\documentclass[%
reprint,
superscriptaddress,
nofootinbib,
amsmath,amssymb,
aps,
prb,
]{revtex4-2}
\usepackage{booktabs}
\usepackage{graphicx}
\usepackage{dcolumn}
\usepackage{bm}
\usepackage{chemformula}
\usepackage{hyperref}
\usepackage{amsfonts,amssymb}

\usepackage{tabularx}
\usepackage{amsmath}
\usepackage{float}

\begin{document}
	
	\title{Magnetic ground state of the Kitaev \ch{Na_2Co_2TeO_6} spin liquid candidate}
	
	\author{Weiliang~Yao}
    \email{wyao4@utk.edu}
	\altaffiliation{Present address: Department of Physics, University of Tennessee, Knoxville, Tennessee 37996, USA}
    \affiliation{International Center for Quantum Materials, School of Physics, Peking University, Beijing 100871, China}
	\author{Yang~Zhao}
	\affiliation{NIST Center for Neutron Research, National Institute of Standards and Technology, Gaithersburg, Maryland 20899, USA}
	\affiliation{Department of Materials Science and Engineering, University of Maryland, College Park, Maryland 20742, USA}
	\author{Yiming~Qiu}
	\affiliation{NIST Center for Neutron Research, National Institute of Standards and Technology, Gaithersburg, Maryland 20899, USA}
	\author{Christian~Balz}
	\affiliation{ISIS Neutron and Muon Source, STFC Rutherford Appleton Laboratory, Didcot OX11 0QX, United Kingdom}
	\author{J.~Ross~Stewart}
	\affiliation{ISIS Neutron and Muon Source, STFC Rutherford Appleton Laboratory, Didcot OX11 0QX, United Kingdom}
	\author{Jeffrey W.~Lynn}
	\affiliation{NIST Center for Neutron Research, National Institute of Standards and Technology, Gaithersburg, Maryland 20899, USA}
	\author{Yuan~Li}
	\email{yuan.li@pku.edu.cn}
	\affiliation{International Center for Quantum Materials, School of Physics, Peking University, Beijing 100871, China}
	\affiliation{Collaborative Innovation Center of Quantum Matter, Beijing 100871, China}
    \date{\today}
	
	\begin{abstract}
    As a candidate Kitaev material, \ch{Na_2Co_2TeO_6} exhibits intriguing magnetism on a honeycomb lattice that is believed to be $C_3$-symmetric. Here we report a neutron diffraction study of high quality single crystals under $a$-axis magnetic fields. Our data support the less common notion of a magnetic ground state that corresponds to a triple-$\mathbf{q}$ magnetic structure with $C_3$ symmetry, rather than the multi-domain zigzag structure typically assumed in prototype Kitaev spin liquid candidates. In particular, we find that the field is unable to repopulate the supposed zigzag domains, where the only alternative explanation is that the domains are strongly pinned by hitherto unidentified structural reasons. If the triple-$\mathbf{q}$ structure is correct then this requires reevaluation of many candidate Kitaev materials. We also find that fields beyond about 10 Tesla suppress the long range antiferromagnetic order, allowing new magnetic behavior to emerge different from that expected for a spin liquid.
	\end{abstract}
	
	\maketitle

The exactly solvable Kitaev model \cite{Kitaev2006} represents a distinct route to quantum many-body entanglement of spins \cite{Anderson1973} and has important potential for topological quantum computing \cite{Kitaev2006,NayakRMP2008}. Pursuit of Kitaev spin liquids (KSLs) \cite{Kitaev2006} in crystalline materials has fueled intense research \cite{Takagi2019,TrebstPR2022}. Among materialization ideas \cite{JackeliPRL2009,ChaloupkaPRL2010,Winter2017,Takagi2019,TrebstPR2022}, several recently proposed cobalt oxides \cite{LiuPRB2018,SanoPRB2018,MotomeJPCM2020,KimJPCM2021_2} are promising, since their 3$d^7$ magnetic electrons are desirable for weakening non-Kitaev interactions \cite{LiuPRB2018,SanoPRB2018}. Moreover, unlike $\alpha$-\ch{RuCl_3} \cite{PlumbPRB2014} and \ch{H_3LiIr_2O_6} \cite{KitagawaNature2018} which are van der Waals materials, the cobaltates can be grown into large single crystals with relatively few imperfections \cite{XiaoCGD2019,YaoPRB2020,ZhongSA2020,HalloranArxiv2022,YanPRM2019,LiPRX2022}.

An important common characteristic of the cobaltates and the 4$d$-electron counterpart $\alpha$-\ch{RuCl_3} is their tunability by magnetic fields. Such external tuning \cite{JanssenPRL2016,Janssen2019,GordonNC2019,HickeyNC2019,LiNC2021} is widely considered necessary for finding (field-driven) spin liquids, because most KSL candidate materials do have magnetic order at low temperature \cite{Winter2017,Takagi2019,TrebstPR2022}. In $\alpha$-\ch{RuCl_3}, a hallmark of the tunability is field suppression of thermodynamic signatures of magnetic order \cite{SearsPRB2017,WolterPRB2017}, which has led to a flurry of studies of excitations in the intermediate and high-field states \cite{BaekPRL2017,ZhengPRL2017,BanerjeeScience2017,DoNP2017,KasaharaNature2018,BanerjeeNPJQM2018,HentrichPRL2018,BalzPRB2019,YokoiScience2021,BruinNPhys2022,LefrancoisPRX2022}. Indeed, similar field suppression of order and unconventional transport behaviors have been found in the cobaltates \cite{YaoPRB2020,LinNC2021,HongPRB2021,LiArxiv2022,YangPRB2022,XiaoJPCM2021,ZhongSA2020,YanPRM2019,TakedaPRR2022,LiPRX2022}, which imply not only chances for finding spin liquids but also an experimental opportunity -- brought by the high crystal quality -- for elucidating the microscopic mechanisms. The latter aspect is significant because microscopic models of essentially \textit{all} KSL candidate materials are currently under debate \cite{RusnackoPRB2019,MaksimovPRR2020,LaurellNPJQM2020,SongvilayPRB2020,SamarakoonPRB2021,LinNC2021,KimJPCM2021,DasPRB2021,SandersPRB2022,YaoPRL2022,WinterJPM2022,MaksimovPRB2022,PandeyPRB2022,LinPreprint2022}. From an optimistic perspective, establishing a concrete case for at least one of them, despite the difficulty of the problem itself, may already provide valuable insight into many of the candidate materials.

\begin{figure}[t!]
		\centering{\includegraphics[clip,width=3.2in]{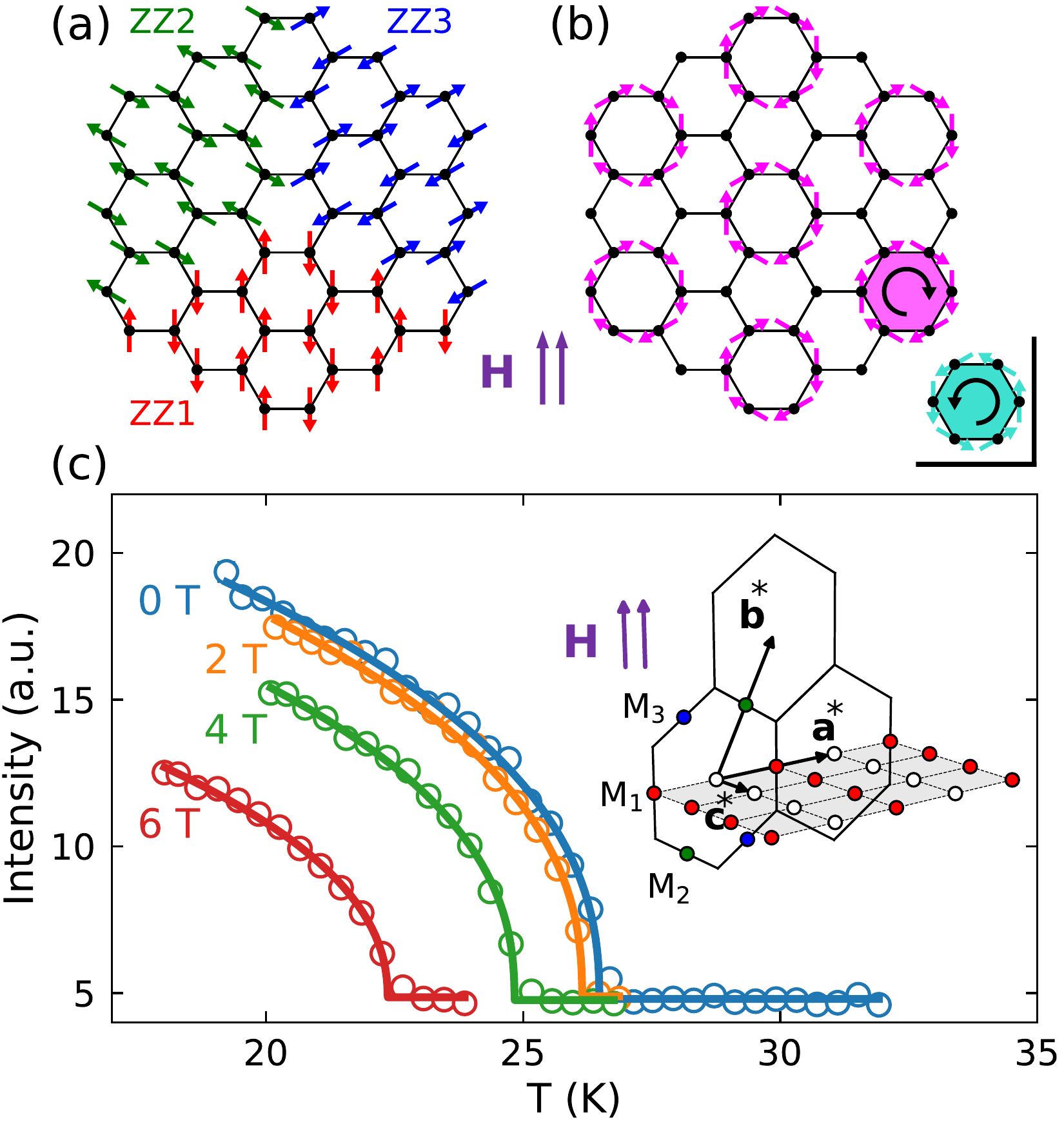}}
		\advance\leftskip-0.3cm
		\caption{(a) The zigzag magnetic structure and its orientational domains. (b) The triple-$\mathbf{q}$ magnetic structure. The moments can be thought of as a vector sum of the three patterns in (a) extended over the whole lattice. When one or three of the ZZ$n$ components are reversed, the chirality is reversed (inset). (c) Temperature dependence of the M$_1$(0.5, 0, 1) reflection in selected fields, where the long range magnetic order is robust (see Fig. \ref{figS3} in \cite{SM}). Solid curves are power-law fits to the data (see text). Inset shows the reciprocal lattice in our setting, where hexagons are boundaries of 2D Brillouin zones, and the shaded ($H$, 0, $L$) plane is perpendicular to the field (purple arrows). Empty circles are structural Brillouin zone centers. Filled circles are magnetic Bragg peaks at the M-points, color-coded with the zigzag domains in (a). }
		\label{fig1}
	\end{figure}

Among the cobaltates, \ch{Na_2Co_2TeO_6} has been studied the most by spectroscopic methods \cite{SongvilayPRB2020,SamarakoonPRB2021,LinNC2021,KimJPCM2021,SandersPRB2022,YaoPRL2022,XPRB2021,LeePRB2021,KikuchiArxiv2022,LinPreprint2022}. Its crystal structure (space group $P6_322$) furthermore stands out among KSL candidate materials for having, at least nominally, three-fold rotational ($C_3$) symmetry about the $c$-axis \cite{ViciuJSSC2007,LefrancoisPRB2016,BeraPRB2017}, whereas many other materials have monoclinic stacking which removes the $C_3$ symmetry. Notably, $C_3$ is a symmetry that becomes broken in the presence of ``zigzag'' magnetic order [Fig.~\ref{fig1}(a)], which is the most commonly considered form of order in KSL candidate materials \cite{Takagi2019,TrebstPR2022}. The magnetic ground state of \ch{Na_2Co_2TeO_6} was initially reported to be zigzag based on neutron diffraction \cite{LefrancoisPRB2016,BeraPRB2017}, which has also been used to identify zigzag order in other KSL candidate materials \cite{YePRB2012,SearsPRB2015,CaoPRB2016,YanPRM2019}. Recently, an alternative novel ``triple-$\mathbf{q}$'' magnetic state [Fig.~\ref{fig1}(b)] was suggested based on a distinct signature in the spin waves \cite{XPRB2021}, which subsequently received indirect support from magnetic resonance \cite{LeePRB2021,KikuchiArxiv2022}. The $C_3$-symmetric triple-$\mathbf{q}$ state can be constructed by adding zigzag components of three different orientations. For this reason, the triple-$\mathbf{q}$ and zigzag orders cannot be distinguished by diffraction \cite{XPRB2021}, unless the $C_3$-symmetry breaking is revealed by observing uneven populations of its orientational domains. The $C_3$ structure of \ch{Na_2Co_2TeO_6} is desirable for this purpose, because a weak external perturbation (\textit{e.g.}, in-plane magnetic field, strain, \textit{etc.}) can be expected to selectively populate the domains if the zigzag ground state is realized. Given the prominence of the zigzag order in KSL research, and since it has not been ruled out in \ch{Na_2Co_2TeO_6} \cite{SongvilayPRB2020,SamarakoonPRB2021,LinNC2021,KimJPCM2021,SandersPRB2022,YaoPRL2022}, such an explicit test is much needed.

	\begin{figure*}[t!]
		\centering{\includegraphics[width=0.9\textwidth]{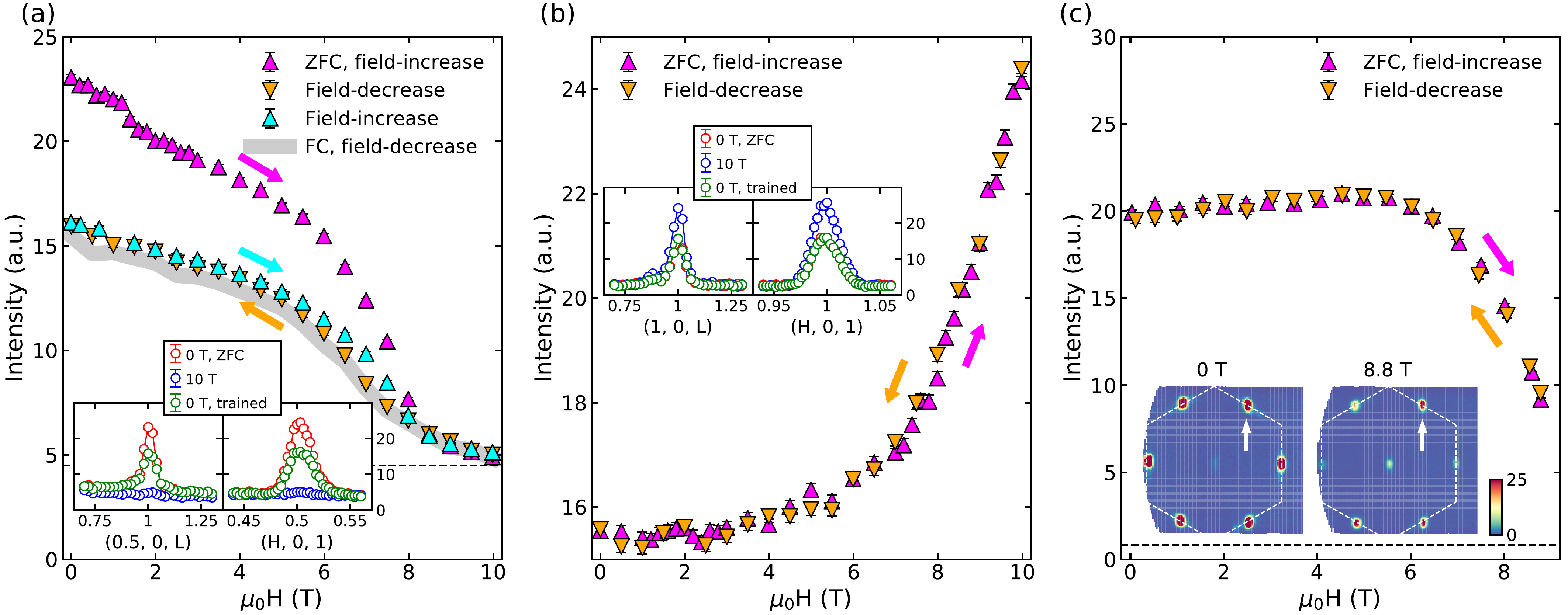}}
		\caption{(a) Field evolution of magnetic Bragg peak M$_1$(0.5, 0, 1) at 2 K, with the sample undergoing a series of field scans after zero-field cooling (ZFC, see text), as well as after field-cooling (FC) in 10 T. (b) Field evolution of Bragg peak (1,~0,~1) at 2 K. In the virgin zero field state, the observed intensity is due to nuclear Bragg scattering, whereas the field-enhanced intensity is a measure of uniform magnetization. (c) Field evolution of magnetic Bragg peak M$_2$(0,~0.5,~1) at 2 K. Measurements displayed in the main panels are performed at the maximum of the peak profiles displayed in the insets. Horizontal dashed lines in (a) and (c) indicate background level. Error bars indicate statistical uncertainty (1 s.d.).}
		\label{fig2}
	\end{figure*}

Here we report our magnetic neutron diffraction study of \ch{Na_2Co_2TeO_6} single crystals in order to test whether in-plane fields along the $a$-axis can selectively populate magnetic domains. We also examine whether magnetic fields (up to 10 Tesla) can drive the system into a spin-disordered state, as has been previously suggested \cite{YaoPRB2020,LinNC2021,HongPRB2021}. Our conclusion is that the fields can do neither of these. While a spin liquid might still be reachable with fields in other directions \cite{LinPreprint2022} and/or greater than 10 T, our results set a definitive constraint on the zero-field magnetic ground state. Namely, unless a lower structural symmetry without $C_3$ has previously been missed, the system prefers a $C_3$-symmetric triple-$\mathbf{q}$ state over the widely considered zigzag order.

Our experimental geometry is shown in the inset of Fig.~\ref{fig1}(c). Magnetic Bragg peaks are expected at the M-points of the two-dimensional (2D) Brillouin zone. They originate either separately from different zigzag domains [ZZ1-ZZ3 in Fig.~\ref{fig1}(a), peaks at M$_1$-M$_3$, respectively], or together from the triple-$\mathbf{q}$ order. Figure~\ref{fig1}(c) displays the temperature ($T$) dependence of the magnetic peak at M$_1$(0.5, 0, 1). In the zigzag scenario, this peak arises from the ZZ1 domain, where the in-plane magnetic moments are collinear with the applied field [Fig.~\ref{fig1}(a)]. The transition temperature ($T_N\sim26.5$ K at 0 T) is gradually suppressed by the field, and the data can be fit with a power-law function: $I=A\,(T_N-T)^{2\beta} +B$, where $A$ and $B$ are scale and background constants, respectively, and $\beta$ is the critical exponent of the order parameter, which changes very little from 0.209(7) to 0.227(13) between 0~T and 6~T. This indicates that the nature of the magnetic transition barely changes with field, and that it is different from the 2D Ising case found in $\alpha$-\ch{RuCl_3} \cite{BanerjeeScience2017}. The deviation might be attributable to the fact that $T_N$ marks three-dimensional ordering, which is preceded by a minor 2D transition at a slightly higher temperature \cite{XPRB2021}. The 2D transition cannot be observed in these data because of the small sample volume \cite{SM}.

Figure~\ref{fig2}(a) displays the system's field evolution as seen from the M$_1$(0.5, 0, 1) magnetic peak at 2 K. Starting from an initial state prepared by zero-field cooling (ZFC), the intensity monotonically decreases with increasing field. Besides a subtle anomaly at about 1.5 T, the main decrease occurs between about 6~T and 8.2~T, and a small but finite intensity remains at the highest field of 10 T, which we will revisit later. At first sight, the intensity decrease could be attributed to two reasons: (1) the antiferromagnetic order is suppressed by the field; (2) the zigzag domain ZZ1 responsible for the measured peak is unfavored by the field and gets transformed into ZZ2 and ZZ3. To test the relevance of (2), we continued our measurement upon removing and then reapplying the field. Intriguingly, the intensity recovers to about 2/3 of the original after the field is removed, and the sample appears to have entered a stable field-trained state -- reapplying the field results in a field-dependent behavior different from the initial field application up to 8.2~T. The data further reveal a hysteretic behavior between 6~T and 8.2~T. A cleaner procedure to prepare the field-trained state involves field-cooling (FC) the sample in a 10 T field and then removing the field.

A central issue here is whether or not the partial intensity loss in the field-trained state is due to domain repopulation. We first note that, with the structural $C_3$ symmetry, a sample prepared by FC can have no ZZ1 domain whatsoever, but this view is defied by the 2/3-recovered intensity. In Fig.~\ref{fig2}(b), we present data measured on an integer-indexed peak (1, 0, 1), which show that the field-trained state is \textit{not} different from the ZFC state as far as uniform magnetization and susceptibility are concerned -- the intensity and its field derivative at 0 T both recover to the original values. Since the zigzag domains have different susceptibility in a given field direction, this result implies that there is no zigzag domain repopulation after the field is removed. As a further test, Fig.~\ref{fig2}(c) displays the field evolution of the M$_2$(0,~0.5,~1) peak, which is associated with domain ZZ2 in the zigzag scenario (see Fig. \ref{figS4} in \cite{SM} for similar result for the M$_3$ peak). While the behavior is qualitatively different from the M$_1$ peak below 6 T, there is no intensity gain on M$_2$ in the field-trained state. We thus conclude that the loss of the M$_1$ peak intensity is unrelated to domain repopulation. In this context, we note that some previous related results in $\alpha$-\ch{RuCl_3} \cite{SearsPRB2017,BanerjeeNPJQM2018} have been attributed to zigzag-domain repopulation by small in-plane fields. Those results are qualitatively similar to our data obtained upon the initial field application in Figs.~\ref{fig2}(a) and (c), and the interpretation was made even in the absence of $C_3$ structural symmetry of $\alpha$-\ch{RuCl_3}. As the structural symmetry is expected to make the zigzag domains energetically unequal, it follows that the magnetization energy must be able to overcome the difference. In this sense, our results in \ch{Na_2Co_2TeO_6} are particularly difficult to comprehend under the zigzag scenario, because the structural $C_3$ symmetry should make the magnetic domains even easier to repopulate than in $\alpha$-\ch{RuCl_3}.

	\begin{figure}[t!]
		\centering{\includegraphics[clip,width=3.2in]{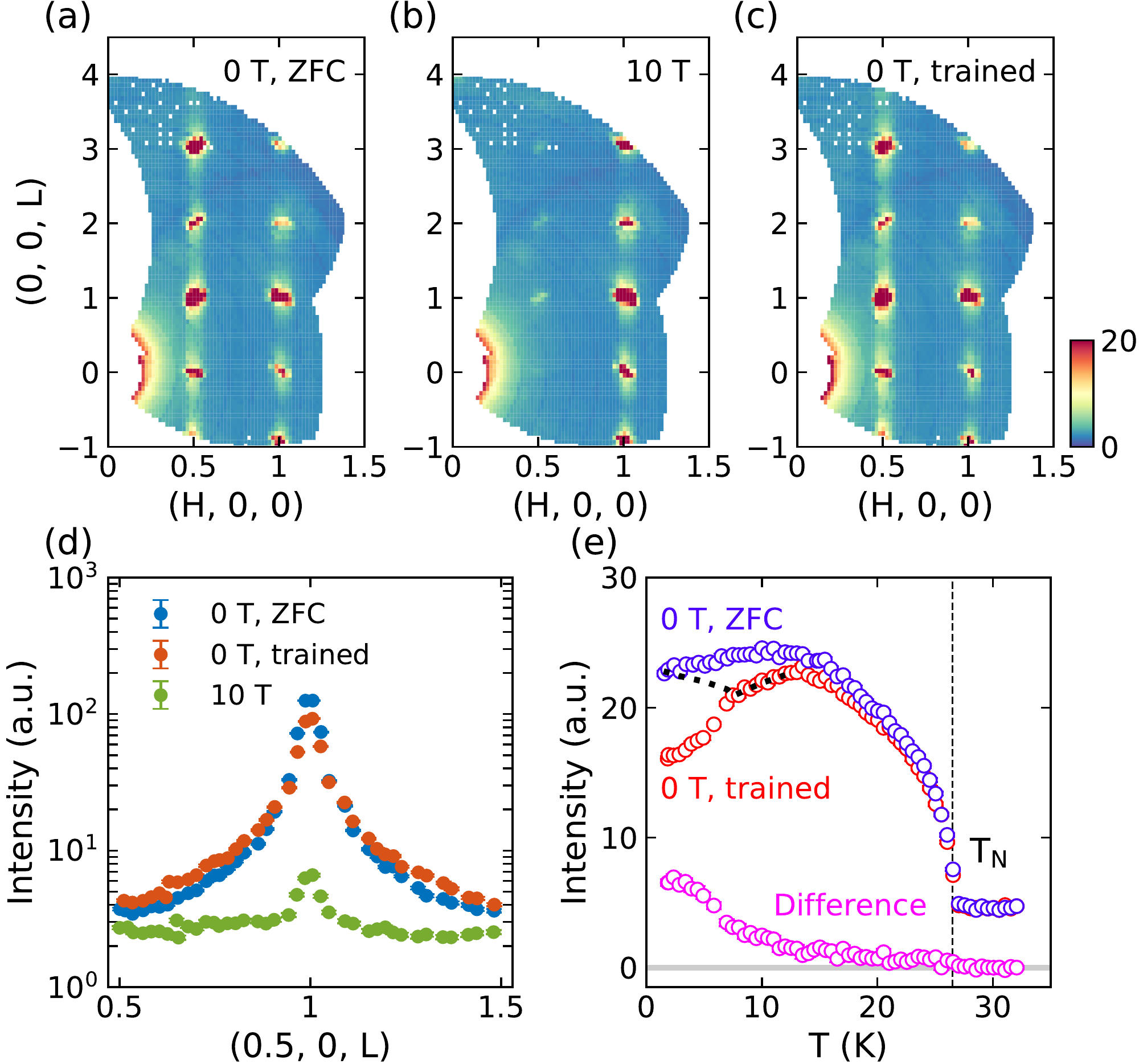}}
		\caption{(a)-(c) Elastic scattering in the ($H$, 0, $L$) plane measured at 0.1 K under the specified field conditions. (d) Line-cuts through the data in (a-c) along $\mathbf{c^*}$ at $H=0.5$. (e) $T$ dependence of the signal at M$_1$(0.5, 0, 1) measured in zero field upon warming the sample, before and after field training. Black dotted curve illustrates expected $T$ dependence of the M$_2$(0,~0.5,~1) and M$_3$(-0.5,~0.5,~1) signals after field training, if the triple-$\mathbf{q}$ scenario is correct (see text).}
		\label{fig3}
	\end{figure}

To reveal where the lost intensity of M$_1$(0.5,~0,~1) has gone in the field-trained state, Fig.~\ref{fig3}(a-c) presents our measurement in an extensive region of the ($H$,~0,~$L$) reciprocal plane. After ZFC, a rod of magnetic scattering running along $\mathbf{c^*}$ is observed at $H=0.5$, in addition to the sharp peaks at integer $L$. It signifies quasi-2D magnetic correlations \cite{XPRB2021}, and the signal becomes noticeably enhanced in the field-trained state [Fig.~\ref{fig3}(c-d)]. The enhancement occurs upon decreasing the field between 8.2 T and 6 T (Fig.~\ref{figS5} in \cite{SM}) and can approximately account for the intensity loss at integer $L$ (Fig.~\ref{figS9} in \cite{SM}). Therefore, field training suppresses $c$-axis correlations, but it leaves the $L$-integrated 2D diffraction signal at M$_1$(0.5,~0) unaffected. This reinforces our conclusion of no zigzag domain repopulation. The field training leaves no significant change in the 2D correlation length [Fig.~\ref{fig2}(a) inset], or in the $c$-axis correlations characterized by M$_{2,3}$ [Fig.~\ref{fig2}(c)]. Moreover, the lost $c$-axis correlations at M$_1$(0.5,~0,~1) can be partially recovered [Fig.~\ref{fig3}(e)] by warming up the field-trained sample. The implication of these observations will be discussed later. We note that the system behaves somewhat differently from  $\alpha$-\ch{RuCl_3}, where a distinct form of magnetic order perpendicular to the honeycomb plane can be stabilized by intermediate in-plane fields \cite{BalzPRB2021}, presumably due to a more significant role of the system's inter-plane coupling \cite{BalzPRB2019}.

Comparing the data in Fig.~\ref{fig3}(a-b), we notice enhanced scattering at integer $H$ and $L$ at 10 T, where no magnetic scattering exists at 0 T (Fig.~\ref{figS7} in \cite{SM}). This additional signal is therefore purely due to field-induced uniform magnetization. An induced moment of about 2.05(3) $\mu_\mathrm{B}$/Co can be estimated from the data \cite{SM}, consistent with previous reports \cite{LinNC2021,XiaoJPCM2021}. While this means that the field suppresses antiferromagnetic order by causing significant spin polarization, the peaks at $H=0.5$ are not fully suppressed [Fig.~\ref{fig3}(b \& d)], and their magnetic nature has been confirmed by comparing to measurements at high temperature (Fig.~\ref{figS8} in \cite{SM}). The system is always in a magnetically ordered state under $a$-axis fields up to 10 T, and is therefore not yet a spin liquid. Nevertheless, recent studies of \ch{Na_2Co_2TeO_6} have revealed unusual thermal transport properties under in-plane fields, which implies that the near-polarized state is distinct from a conventional paramagnet \cite{HongPRB2021,TakedaPRR2022}. Similar behaviors are also observed in \ch{BaCo_2(AsO_4)_2}, where an intriguing state related to Kitaev interactions has been inferred near full polarization \cite{ZhongSA2020,ZhangNM2022}. These studies motivate further searches for exotic magnetism in the cobalt-based Kitaev candidate materials.

\begin{figure}[t!]
		\centering{\includegraphics[clip,width=3.0in]{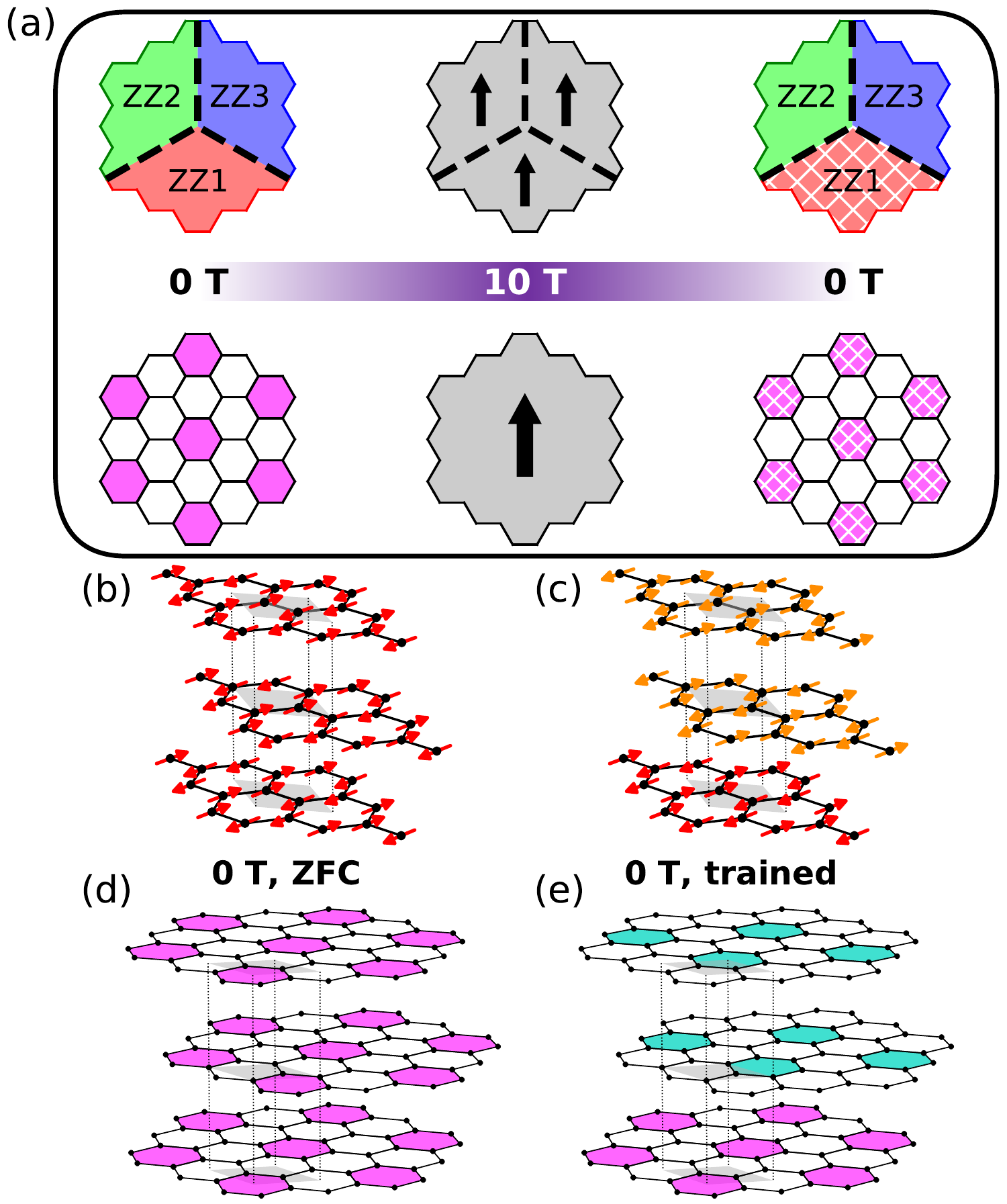}}
		\caption{(a) Schematic field-training processes under the zigzag (upper half) and triple-$\mathbf{q}$ (lower half) scenarios. The 10 T state is approximated as fully spin-polarized. Polygons are color-coded with Fig.~\ref{fig1}(a-b). Dashed lines are boundaries between ``hidden'' low-symmetry structural domains. Hatches indicate suppressed $c$-axis correlations. (b) \& (c) Schematic stacking in the (supposed) ZZ1 domain before and after field training. Yellow arrows indicate randomly reversed layers. (d) \& (e) Schematic stacking in the triple-$\mathbf{q}$ structure before and after field training, color-coded after Fig.~\ref{fig1}(b). Grey rhombuses in (b-e) indicate the structural primitive cell.}
		\label{fig4}
	\end{figure}

We now discuss possible scenarios for the field training to cause no diffraction intensity transfer between the M-points. In the first scenario, the M-points are associated with spatially separated zigzag domains, as we illustrate in the upper half of Fig.~\ref{fig4}(a). In order for the field training not to repopulate the domains, they must be completely pinned by the local crystal lattice regarding their zigzag-chain orientations. Given the high quality of our crystals, we believe that the pinning is not due to defects, and can only be explained by a hitherto unidentified departure from the nominal $C_3$-symmetric structure: On the one hand, a tiny orthorhombic distortion may already produce a strong pinning effect, since the sister compound \ch{Na_3Co_2SbO_6} has demonstrated a large magnetic anisotropy arising from a small lattice distortion \cite{LiPRX2022}. On the other hand, structural orthorhombicity might arise from long-period stacking  \cite{SpitzJACS2022} that can be easily missed in experiments due to the presence of stacking faults. Moreover, the crystal structure of \ch{Na_2Co_2TeO_6} still has some loose ends, including additional weak Bragg peaks previously seen with both neutron and X-ray diffraction \cite{BeraPRB2017,XPRB2021}. The diffraction peaks share the same wave vectors as the magnetic ones seen at low temperature, and may signify a superstructure in the sodium layer \cite{XPRB2021}. If the superstructure breaks the $C_3$ symmetry, which is yet to be clarified such as by high-resolution single-crystal diffraction, it may pin magnetic zigzag domains. In the second scenario, as illustrated in the lower half of Fig.~\ref{fig4}(a), the M-points all belong to the same triple-$\mathbf{q}$ order parameter, which naturally explains the lack of opportunity for orientational domain repopulation.

To this end, the field-training effect on the $c$-axis correlations deserves some thought. Given that the effect is only observed at M$_1$ [Fig.~\ref{fig2}(a \& c)], we illustrate plausible changes caused by the training in Figs.~\ref{fig4}(b-c) and (d-e), respectively, for the zigzag and the triple-$\mathbf{q}$ cases. In the former case, the inter-layer arrangement inside the ZZ1 domain is disturbed by the training, probably because the field causes a spin-flop-like transition between 6~T and 8.2~T, as the hysteretic behavior [Fig.~\ref{fig2}(a)] suggests. Indeed, the first-order nature of the transition is expected to strongly disturb ZZ1, but it would naturally leave ZZ2 and ZZ3 intact. In the latter case, we note that inside a given honeycomb layer, reversing one zigzag component in the triple-$\mathbf{q}$ structure would reverse the layer's spin chirality [Fig.~\ref{fig1}(b)]. Hence, the suppressed $c$-axis correlations at M$_1$, but not at M$_2$ or M$_3$, imply a scrambled arrangement of the chirality across the layers [Figs.~\ref{fig4}(d-e)]. Importantly, the fact that warming up a field-trained sample recovers part of the $c$-axis correlations seen at M$_1$, as shown in Fig.~\ref{fig3}(e), also has very different explanations in the two cases. In the zigzag case, the recovery pertains to only the ZZ1 domain, which means that the diffraction signals at M$_2$ and M$_3$ will not be affected. Since the latter signals are the same in the ZFC and field-trained states [Fig.~\ref{fig2}(c)], upon warming the sample from 2 K, they are expected to simply follow the ZFC curve in Fig.~\ref{fig3}(e). In contrast, in the triple-$\mathbf{q}$ case, the scrambled chirality between the layers is not expected to recover easily by thermal fluctuations. Instead, the pattern in each layer might be able to translate, which corresponds to reversing \textit{two} zigzag components simultaneously (Fig.~\ref{figS10} in \cite{SM}). Mathematically, such a process would partially recover the $c$-axis correlations seen at M$_1$, but at the cost of the correlations at M$_2$ and M$_3$. It means that if one can monitor, \textit{e.g.}, the M$_2$(0,~0.5,~1) peak upon warming the sample from a field-trained state, the measured intensity would be like the dotted lines in Fig.~\ref{fig3}(e). Such a distinct behavior from the zigzag case, if confirmed in future studies, will firmly establish the triple-$\mathbf{q}$ scenario. In fact, we believe that such crosstalk between signals at different wave vectors can be utilized, on very general grounds indeed, for experimental differentiation between single- and multi-$\mathbf{q}$ magnetic orders regardless of the crystal structure. The experiment requires a demanding condition with both a magnet and detector coverage for observing the out-of-horizontal-plane diffraction peaks.

To conclude, we have investigated the $a$-axis field dependence of magnetic order in \ch{Na_2Co_2TeO_6} with neutron diffraction. In spite of the nominal $C_3$ crystal symmetry, we find that an $a$-axis applied field is unable to repopulate $C_3$-breaking magnetic domains -- either such domains exist but are completely pinned by an as yet unknown low-symmetry structure, or the magnetic ground state features the $C_3$-symmetric triple-$\mathbf{q}$ structure. Our study brings unprecedented insight into the crystal and magnetic structures of not only \ch{Na_2Co_2TeO_6}, but also related systems with presumed zigzag order that may actually be triple-$\mathbf{q}$. Finally, we show that \ch{Na_2Co_2TeO_6} is not yet a spin liquid up to 10 T, but its magnetism remains highly intriguing and awaits further elucidation.

\textit{Note added.} A parallel work reports theoretical analyses of triple-$\mathbf{q}$ order in \ch{Na_2Co_2TeO_6}, which are consistent with our results \cite{KruegerPreprint}.

\begin{acknowledgments}
We are grateful for discussions with L. Chen, W. Chen, C. Hess, X. Hong, L. Janssen, X. Jin, D. Khalyavin, C. Kim, V. Kocsis, W. G. F. Kr\"{u}ger, J.-G. Park, L. Taillefer, and A.~U.~B. Wolter. The work at Peking University was supported by the National Basic Research Program of China (Grant No. 2021YFA1401900) and the NSF of China (Grant Nos. 12061131004, and 11888101). Access to MACS was provided by the Center for High Resolution Neutron Scattering, a partnership between the National Institute of Standards and Technology and the National Science Foundation under Agreement No. DMR-1508249. We acknowledge ISIS for beamtime under proposal RB2010025 \cite{LET2021}.

\end{acknowledgments}

	\bibliographystyle{apsrev4-2}
	
	\bibliography{reference_diffraction}

\begin{thebibliography}{78}%
\makeatletter
\providecommand \@ifxundefined [1]{%
 \@ifx{#1\undefined}
}%
\providecommand \@ifnum [1]{%
 \ifnum #1\expandafter \@firstoftwo
 \else \expandafter \@secondoftwo
 \fi
}%
\providecommand \@ifx [1]{%
 \ifx #1\expandafter \@firstoftwo
 \else \expandafter \@secondoftwo
 \fi
}%
\providecommand \natexlab [1]{#1}%
\providecommand \enquote  [1]{``#1''}%
\providecommand \bibnamefont  [1]{#1}%
\providecommand \bibfnamefont [1]{#1}%
\providecommand \citenamefont [1]{#1}%
\providecommand \href@noop [0]{\@secondoftwo}%
\providecommand \href [0]{\begingroup \@sanitize@url \@href}%
\providecommand \@href[1]{\@@startlink{#1}\@@href}%
\providecommand \@@href[1]{\endgroup#1\@@endlink}%
\providecommand \@sanitize@url [0]{\catcode `\\12\catcode `\$12\catcode
  `\&12\catcode `\#12\catcode `\^12\catcode `\_12\catcode `\%12\relax}%
\providecommand \@@startlink[1]{}%
\providecommand \@@endlink[0]{}%
\providecommand \url  [0]{\begingroup\@sanitize@url \@url }%
\providecommand \@url [1]{\endgroup\@href {#1}{\urlprefix }}%
\providecommand \urlprefix  [0]{URL }%
\providecommand \Eprint [0]{\href }%
\providecommand \doibase [0]{https://doi.org/}%
\providecommand \selectlanguage [0]{\@gobble}%
\providecommand \bibinfo  [0]{\@secondoftwo}%
\providecommand \bibfield  [0]{\@secondoftwo}%
\providecommand \translation [1]{[#1]}%
\providecommand \BibitemOpen [0]{}%
\providecommand \bibitemStop [0]{}%
\providecommand \bibitemNoStop [0]{.\EOS\space}%
\providecommand \EOS [0]{\spacefactor3000\relax}%
\providecommand \BibitemShut  [1]{\csname bibitem#1\endcsname}%
\let\auto@bib@innerbib\@empty
\bibitem [{\citenamefont {Kitaev}(2006)}]{Kitaev2006}%
  \BibitemOpen
  \bibfield  {author} {\bibinfo {author} {\bibfnamefont {A.}~\bibnamefont
  {Kitaev}},\ }\href {https://doi.org/10.1016/j.aop.2005.10.005} {\bibfield
  {journal} {\bibinfo  {journal} {Annals of Physics}\ }\textbf {\bibinfo
  {volume} {321}},\ \bibinfo {pages} {2} (\bibinfo {year} {2006})}\BibitemShut
  {NoStop}%
\bibitem [{\citenamefont {Anderson}(1973)}]{Anderson1973}%
  \BibitemOpen
  \bibfield  {author} {\bibinfo {author} {\bibfnamefont {P.~W.}\ \bibnamefont
  {Anderson}},\ }\href {https://doi.org/10.1016/0025-5408(73)90167-0}
  {\bibfield  {journal} {\bibinfo  {journal} {Materials Research Bulletin}\
  }\textbf {\bibinfo {volume} {8}},\ \bibinfo {pages} {153} (\bibinfo {year}
  {1973})}\BibitemShut {NoStop}%
\bibitem [{\citenamefont {Nayak}\ \emph {et~al.}(2008)\citenamefont {Nayak},
  \citenamefont {Simon}, \citenamefont {Stern}, \citenamefont {Freedman},\ and\
  \citenamefont {Das~Sarma}}]{NayakRMP2008}%
  \BibitemOpen
  \bibfield  {author} {\bibinfo {author} {\bibfnamefont {C.}~\bibnamefont
  {Nayak}}, \bibinfo {author} {\bibfnamefont {S.~H.}\ \bibnamefont {Simon}},
  \bibinfo {author} {\bibfnamefont {A.}~\bibnamefont {Stern}}, \bibinfo
  {author} {\bibfnamefont {M.}~\bibnamefont {Freedman}},\ and\ \bibinfo
  {author} {\bibfnamefont {S.}~\bibnamefont {Das~Sarma}},\ }\href
  {https://doi.org/10.1103/RevModPhys.80.1083} {\bibfield  {journal} {\bibinfo
  {journal} {Rev. Mod. Phys.}\ }\textbf {\bibinfo {volume} {80}},\ \bibinfo
  {pages} {1083} (\bibinfo {year} {2008})}\BibitemShut {NoStop}%
\bibitem [{\citenamefont {Takagi}\ \emph {et~al.}(2019)\citenamefont {Takagi},
  \citenamefont {Takayama}, \citenamefont {Jackeli}, \citenamefont
  {Khaliullin},\ and\ \citenamefont {Nagler}}]{Takagi2019}%
  \BibitemOpen
  \bibfield  {author} {\bibinfo {author} {\bibfnamefont {H.}~\bibnamefont
  {Takagi}}, \bibinfo {author} {\bibfnamefont {T.}~\bibnamefont {Takayama}},
  \bibinfo {author} {\bibfnamefont {G.}~\bibnamefont {Jackeli}}, \bibinfo
  {author} {\bibfnamefont {G.}~\bibnamefont {Khaliullin}},\ and\ \bibinfo
  {author} {\bibfnamefont {S.~E.}\ \bibnamefont {Nagler}},\ }\href
  {https://doi.org/10.1038/s42254-019-0038-2} {\bibfield  {journal} {\bibinfo
  {journal} {Nature Reviews Physics}\ }\textbf {\bibinfo {volume} {1}},\
  \bibinfo {pages} {264} (\bibinfo {year} {2019})}\BibitemShut {NoStop}%
\bibitem [{\citenamefont {Trebst}\ and\ \citenamefont
  {Hickey}(2022)}]{TrebstPR2022}%
  \BibitemOpen
  \bibfield  {author} {\bibinfo {author} {\bibfnamefont {S.}~\bibnamefont
  {Trebst}}\ and\ \bibinfo {author} {\bibfnamefont {C.}~\bibnamefont
  {Hickey}},\ }\href
  {https://doi.org/https://doi.org/10.1016/j.physrep.2021.11.003} {\bibfield
  {journal} {\bibinfo  {journal} {Physics Reports}\ }\textbf {\bibinfo {volume}
  {950}},\ \bibinfo {pages} {1} (\bibinfo {year} {2022})}\BibitemShut {NoStop}%
\bibitem [{\citenamefont {Jackeli}\ and\ \citenamefont
  {Khaliullin}(2009)}]{JackeliPRL2009}%
  \BibitemOpen
  \bibfield  {author} {\bibinfo {author} {\bibfnamefont {G.}~\bibnamefont
  {Jackeli}}\ and\ \bibinfo {author} {\bibfnamefont {G.}~\bibnamefont
  {Khaliullin}},\ }\href {https://doi.org/10.1103/PhysRevLett.102.017205}
  {\bibfield  {journal} {\bibinfo  {journal} {Phys. Rev. Lett.}\ }\textbf
  {\bibinfo {volume} {102}},\ \bibinfo {pages} {017205} (\bibinfo {year}
  {2009})}\BibitemShut {NoStop}%
\bibitem [{\citenamefont {Chaloupka}\ \emph {et~al.}(2010)\citenamefont
  {Chaloupka}, \citenamefont {Jackeli},\ and\ \citenamefont
  {Khaliullin}}]{ChaloupkaPRL2010}%
  \BibitemOpen
  \bibfield  {author} {\bibinfo {author} {\bibfnamefont {J.~c.~v.}\
  \bibnamefont {Chaloupka}}, \bibinfo {author} {\bibfnamefont {G.}~\bibnamefont
  {Jackeli}},\ and\ \bibinfo {author} {\bibfnamefont {G.}~\bibnamefont
  {Khaliullin}},\ }\href {https://doi.org/10.1103/PhysRevLett.105.027204}
  {\bibfield  {journal} {\bibinfo  {journal} {Phys. Rev. Lett.}\ }\textbf
  {\bibinfo {volume} {105}},\ \bibinfo {pages} {027204} (\bibinfo {year}
  {2010})}\BibitemShut {NoStop}%
\bibitem [{\citenamefont {Winter}\ \emph {et~al.}(2017)\citenamefont {Winter},
  \citenamefont {Tsirlin}, \citenamefont {Daghofer}, \citenamefont {van~den
  Brink}, \citenamefont {Singh}, \citenamefont {Gegenwart},\ and\ \citenamefont
  {Valent\'{i}}}]{Winter2017}%
  \BibitemOpen
  \bibfield  {author} {\bibinfo {author} {\bibfnamefont {S.~M.}\ \bibnamefont
  {Winter}}, \bibinfo {author} {\bibfnamefont {A.~A.}\ \bibnamefont {Tsirlin}},
  \bibinfo {author} {\bibfnamefont {M.}~\bibnamefont {Daghofer}}, \bibinfo
  {author} {\bibfnamefont {J.}~\bibnamefont {van~den Brink}}, \bibinfo {author}
  {\bibfnamefont {Y.}~\bibnamefont {Singh}}, \bibinfo {author} {\bibfnamefont
  {P.}~\bibnamefont {Gegenwart}},\ and\ \bibinfo {author} {\bibfnamefont
  {R.}~\bibnamefont {Valent\'{i}}},\ }\href
  {https://doi.org/10.1088/1361-648X/aa8cf5} {\bibfield  {journal} {\bibinfo
  {journal} {J. Phys.: Condens. Matter}\ }\textbf {\bibinfo {volume} {29}},\
  \bibinfo {pages} {493002} (\bibinfo {year} {2017})}\BibitemShut {NoStop}%
\bibitem [{\citenamefont {Liu}\ and\ \citenamefont
  {Khaliullin}(2018)}]{LiuPRB2018}%
  \BibitemOpen
  \bibfield  {author} {\bibinfo {author} {\bibfnamefont {H.}~\bibnamefont
  {Liu}}\ and\ \bibinfo {author} {\bibfnamefont {G.}~\bibnamefont
  {Khaliullin}},\ }\href {https://doi.org/10.1103/PhysRevB.97.014407}
  {\bibfield  {journal} {\bibinfo  {journal} {Phys. Rev. B}\ }\textbf {\bibinfo
  {volume} {97}},\ \bibinfo {pages} {014407} (\bibinfo {year}
  {2018})}\BibitemShut {NoStop}%
\bibitem [{\citenamefont {Sano}\ \emph {et~al.}(2018)\citenamefont {Sano},
  \citenamefont {Kato},\ and\ \citenamefont {Motome}}]{SanoPRB2018}%
  \BibitemOpen
  \bibfield  {author} {\bibinfo {author} {\bibfnamefont {R.}~\bibnamefont
  {Sano}}, \bibinfo {author} {\bibfnamefont {Y.}~\bibnamefont {Kato}},\ and\
  \bibinfo {author} {\bibfnamefont {Y.}~\bibnamefont {Motome}},\ }\href
  {https://doi.org/10.1103/PhysRevB.97.014408} {\bibfield  {journal} {\bibinfo
  {journal} {Phys. Rev. B}\ }\textbf {\bibinfo {volume} {97}},\ \bibinfo
  {pages} {014408} (\bibinfo {year} {2018})}\BibitemShut {NoStop}%
\bibitem [{\citenamefont {Motome}\ \emph {et~al.}(2020)\citenamefont {Motome},
  \citenamefont {Sano}, \citenamefont {Jang}, \citenamefont {Sugita},\ and\
  \citenamefont {Kato}}]{MotomeJPCM2020}%
  \BibitemOpen
  \bibfield  {author} {\bibinfo {author} {\bibfnamefont {Y.}~\bibnamefont
  {Motome}}, \bibinfo {author} {\bibfnamefont {R.}~\bibnamefont {Sano}},
  \bibinfo {author} {\bibfnamefont {S.}~\bibnamefont {Jang}}, \bibinfo {author}
  {\bibfnamefont {Y.}~\bibnamefont {Sugita}},\ and\ \bibinfo {author}
  {\bibfnamefont {Y.}~\bibnamefont {Kato}},\ }\href
  {https://doi.org/10.1088/1361-648x/ab8525} {\bibfield  {journal} {\bibinfo
  {journal} {Journal of Physics: Condensed Matter}\ }\textbf {\bibinfo {volume}
  {32}},\ \bibinfo {pages} {404001} (\bibinfo {year} {2020})}\BibitemShut
  {NoStop}%
\bibitem [{\citenamefont {Kim}\ \emph {et~al.}(2021)\citenamefont {Kim},
  \citenamefont {Kim},\ and\ \citenamefont {Park}}]{KimJPCM2021_2}%
  \BibitemOpen
  \bibfield  {author} {\bibinfo {author} {\bibfnamefont {C.}~\bibnamefont
  {Kim}}, \bibinfo {author} {\bibfnamefont {H.-S.}\ \bibnamefont {Kim}},\ and\
  \bibinfo {author} {\bibfnamefont {J.-G.}\ \bibnamefont {Park}},\ }\href
  {https://doi.org/10.1088/1361-648x/ac2d5d} {\bibfield  {journal} {\bibinfo
  {journal} {Journal of Physics: Condensed Matter}\ }\textbf {\bibinfo {volume}
  {34}},\ \bibinfo {pages} {023001} (\bibinfo {year} {2021})}\BibitemShut
  {NoStop}%
\bibitem [{\citenamefont {Plumb}\ \emph {et~al.}(2014)\citenamefont {Plumb},
  \citenamefont {Clancy}, \citenamefont {Sandilands}, \citenamefont {Shankar},
  \citenamefont {Hu}, \citenamefont {Burch}, \citenamefont {Kee},\ and\
  \citenamefont {Kim}}]{PlumbPRB2014}%
  \BibitemOpen
  \bibfield  {author} {\bibinfo {author} {\bibfnamefont {K.~W.}\ \bibnamefont
  {Plumb}}, \bibinfo {author} {\bibfnamefont {J.~P.}\ \bibnamefont {Clancy}},
  \bibinfo {author} {\bibfnamefont {L.~J.}\ \bibnamefont {Sandilands}},
  \bibinfo {author} {\bibfnamefont {V.~V.}\ \bibnamefont {Shankar}}, \bibinfo
  {author} {\bibfnamefont {Y.~F.}\ \bibnamefont {Hu}}, \bibinfo {author}
  {\bibfnamefont {K.~S.}\ \bibnamefont {Burch}}, \bibinfo {author}
  {\bibfnamefont {H.-Y.}\ \bibnamefont {Kee}},\ and\ \bibinfo {author}
  {\bibfnamefont {Y.-J.}\ \bibnamefont {Kim}},\ }\href
  {https://doi.org/10.1103/PhysRevB.90.041112} {\bibfield  {journal} {\bibinfo
  {journal} {Phys. Rev. B}\ }\textbf {\bibinfo {volume} {90}},\ \bibinfo
  {pages} {041112} (\bibinfo {year} {2014})}\BibitemShut {NoStop}%
\bibitem [{\citenamefont {Kitagawa}\ \emph {et~al.}(2018)\citenamefont
  {Kitagawa}, \citenamefont {Takayama}, \citenamefont {Matsumoto},
  \citenamefont {Kato}, \citenamefont {Takano}, \citenamefont {Kishimoto},
  \citenamefont {Bette}, \citenamefont {Dinnebier}, \citenamefont {Jackeli},\
  and\ \citenamefont {Takagi}}]{KitagawaNature2018}%
  \BibitemOpen
  \bibfield  {author} {\bibinfo {author} {\bibfnamefont {K.}~\bibnamefont
  {Kitagawa}}, \bibinfo {author} {\bibfnamefont {T.}~\bibnamefont {Takayama}},
  \bibinfo {author} {\bibfnamefont {Y.}~\bibnamefont {Matsumoto}}, \bibinfo
  {author} {\bibfnamefont {A.}~\bibnamefont {Kato}}, \bibinfo {author}
  {\bibfnamefont {R.}~\bibnamefont {Takano}}, \bibinfo {author} {\bibfnamefont
  {Y.}~\bibnamefont {Kishimoto}}, \bibinfo {author} {\bibfnamefont
  {S.}~\bibnamefont {Bette}}, \bibinfo {author} {\bibfnamefont
  {R.}~\bibnamefont {Dinnebier}}, \bibinfo {author} {\bibfnamefont
  {G.}~\bibnamefont {Jackeli}},\ and\ \bibinfo {author} {\bibfnamefont
  {H.}~\bibnamefont {Takagi}},\ }\href {https://doi.org/10.1038/nature25482}
  {\bibfield  {journal} {\bibinfo  {journal} {Nature}\ }\textbf {\bibinfo
  {volume} {554}},\ \bibinfo {pages} {341} (\bibinfo {year}
  {2018})}\BibitemShut {NoStop}%
\bibitem [{\citenamefont {Xiao}\ \emph {et~al.}(2019)\citenamefont {Xiao},
  \citenamefont {Xia}, \citenamefont {Zhang}, \citenamefont {Yue},
  \citenamefont {Huang}, \citenamefont {Zhang}, \citenamefont {Yang},
  \citenamefont {Song}, \citenamefont {Wei}, \citenamefont {Deng},\ and\
  \citenamefont {Jiang}}]{XiaoCGD2019}%
  \BibitemOpen
  \bibfield  {author} {\bibinfo {author} {\bibfnamefont {G.}~\bibnamefont
  {Xiao}}, \bibinfo {author} {\bibfnamefont {Z.}~\bibnamefont {Xia}}, \bibinfo
  {author} {\bibfnamefont {W.}~\bibnamefont {Zhang}}, \bibinfo {author}
  {\bibfnamefont {X.}~\bibnamefont {Yue}}, \bibinfo {author} {\bibfnamefont
  {S.}~\bibnamefont {Huang}}, \bibinfo {author} {\bibfnamefont
  {X.}~\bibnamefont {Zhang}}, \bibinfo {author} {\bibfnamefont
  {F.}~\bibnamefont {Yang}}, \bibinfo {author} {\bibfnamefont {Y.}~\bibnamefont
  {Song}}, \bibinfo {author} {\bibfnamefont {M.}~\bibnamefont {Wei}}, \bibinfo
  {author} {\bibfnamefont {H.}~\bibnamefont {Deng}},\ and\ \bibinfo {author}
  {\bibfnamefont {D.}~\bibnamefont {Jiang}},\ }\href
  {https://doi.org/10.1021/acs.cgd.8b01770} {\bibfield  {journal} {\bibinfo
  {journal} {Cryst. Growth Des.}\ }\textbf {\bibinfo {volume} {19}},\ \bibinfo
  {pages} {2658} (\bibinfo {year} {2019})}\BibitemShut {NoStop}%
\bibitem [{\citenamefont {Yao}\ and\ \citenamefont {Li}(2020)}]{YaoPRB2020}%
  \BibitemOpen
  \bibfield  {author} {\bibinfo {author} {\bibfnamefont {W.}~\bibnamefont
  {Yao}}\ and\ \bibinfo {author} {\bibfnamefont {Y.}~\bibnamefont {Li}},\
  }\href {https://doi.org/10.1103/PhysRevB.101.085120} {\bibfield  {journal}
  {\bibinfo  {journal} {Phys. Rev. B}\ }\textbf {\bibinfo {volume} {101}},\
  \bibinfo {pages} {085120} (\bibinfo {year} {2020})}\BibitemShut {NoStop}%
\bibitem [{\citenamefont {Zhong}\ \emph {et~al.}(2020)\citenamefont {Zhong},
  \citenamefont {Gao}, \citenamefont {Ong},\ and\ \citenamefont
  {Cava}}]{ZhongSA2020}%
  \BibitemOpen
  \bibfield  {author} {\bibinfo {author} {\bibfnamefont {R.}~\bibnamefont
  {Zhong}}, \bibinfo {author} {\bibfnamefont {T.}~\bibnamefont {Gao}}, \bibinfo
  {author} {\bibfnamefont {N.~P.}\ \bibnamefont {Ong}},\ and\ \bibinfo {author}
  {\bibfnamefont {R.~J.}\ \bibnamefont {Cava}},\ }\href
  {https://doi.org/10.1126/sciadv.aay6953} {\bibfield  {journal} {\bibinfo
  {journal} {Science Advances}\ }\textbf {\bibinfo {volume} {6}},\ \bibinfo
  {pages} {eaay6953} (\bibinfo {year} {2020})}\BibitemShut {NoStop}%
\bibitem [{\citenamefont {Halloran}\ \emph {et~al.}(2022)\citenamefont
  {Halloran}, \citenamefont {Desrochers}, \citenamefont {Zhang}, \citenamefont
  {Chen}, \citenamefont {Chern}, \citenamefont {Xu}, \citenamefont {Winn},
  \citenamefont {Graves-Brook}, \citenamefont {Stone}, \citenamefont
  {Kolesnikov} \emph {et~al.}}]{HalloranArxiv2022}%
  \BibitemOpen
  \bibfield  {author} {\bibinfo {author} {\bibfnamefont {T.}~\bibnamefont
  {Halloran}}, \bibinfo {author} {\bibfnamefont {F.}~\bibnamefont
  {Desrochers}}, \bibinfo {author} {\bibfnamefont {E.~Z.}\ \bibnamefont
  {Zhang}}, \bibinfo {author} {\bibfnamefont {T.}~\bibnamefont {Chen}},
  \bibinfo {author} {\bibfnamefont {L.~E.}\ \bibnamefont {Chern}}, \bibinfo
  {author} {\bibfnamefont {Z.}~\bibnamefont {Xu}}, \bibinfo {author}
  {\bibfnamefont {B.}~\bibnamefont {Winn}}, \bibinfo {author} {\bibfnamefont
  {M.}~\bibnamefont {Graves-Brook}}, \bibinfo {author} {\bibfnamefont
  {M.}~\bibnamefont {Stone}}, \bibinfo {author} {\bibfnamefont {A.~I.}\
  \bibnamefont {Kolesnikov}}, \emph {et~al.},\ }\href
  {https://doi.org/10.48550/arXiv.2205.15262} {\bibfield  {journal} {\bibinfo
  {journal} {arXiv preprint arXiv:2205.15262}\ } (\bibinfo {year}
  {2022})}\BibitemShut {NoStop}%
\bibitem [{\citenamefont {Yan}\ \emph {et~al.}(2019)\citenamefont {Yan},
  \citenamefont {Okamoto}, \citenamefont {Wu}, \citenamefont {Zheng},
  \citenamefont {Zhou}, \citenamefont {Cao},\ and\ \citenamefont
  {McGuire}}]{YanPRM2019}%
  \BibitemOpen
  \bibfield  {author} {\bibinfo {author} {\bibfnamefont {J.-Q.}\ \bibnamefont
  {Yan}}, \bibinfo {author} {\bibfnamefont {S.}~\bibnamefont {Okamoto}},
  \bibinfo {author} {\bibfnamefont {Y.}~\bibnamefont {Wu}}, \bibinfo {author}
  {\bibfnamefont {Q.}~\bibnamefont {Zheng}}, \bibinfo {author} {\bibfnamefont
  {H.~D.}\ \bibnamefont {Zhou}}, \bibinfo {author} {\bibfnamefont {H.~B.}\
  \bibnamefont {Cao}},\ and\ \bibinfo {author} {\bibfnamefont {M.~A.}\
  \bibnamefont {McGuire}},\ }\href
  {https://doi.org/10.1103/PhysRevMaterials.3.074405} {\bibfield  {journal}
  {\bibinfo  {journal} {Phys. Rev. Materials}\ }\textbf {\bibinfo {volume}
  {3}},\ \bibinfo {pages} {074405} (\bibinfo {year} {2019})}\BibitemShut
  {NoStop}%
\bibitem [{\citenamefont {Li}\ \emph {et~al.}(2022{\natexlab{a}})\citenamefont
  {Li}, \citenamefont {Gu}, \citenamefont {Chen}, \citenamefont {Garlea},
  \citenamefont {Iida}, \citenamefont {Kamazawa}, \citenamefont {Li},
  \citenamefont {Deng}, \citenamefont {Xiao}, \citenamefont {Zheng},
  \citenamefont {Ye}, \citenamefont {Peng}, \citenamefont {Zaliznyak},
  \citenamefont {Tranquada},\ and\ \citenamefont {Li}}]{LiPRX2022}%
  \BibitemOpen
  \bibfield  {author} {\bibinfo {author} {\bibfnamefont {X.}~\bibnamefont
  {Li}}, \bibinfo {author} {\bibfnamefont {Y.}~\bibnamefont {Gu}}, \bibinfo
  {author} {\bibfnamefont {Y.}~\bibnamefont {Chen}}, \bibinfo {author}
  {\bibfnamefont {V.~O.}\ \bibnamefont {Garlea}}, \bibinfo {author}
  {\bibfnamefont {K.}~\bibnamefont {Iida}}, \bibinfo {author} {\bibfnamefont
  {K.}~\bibnamefont {Kamazawa}}, \bibinfo {author} {\bibfnamefont
  {Y.}~\bibnamefont {Li}}, \bibinfo {author} {\bibfnamefont {G.}~\bibnamefont
  {Deng}}, \bibinfo {author} {\bibfnamefont {Q.}~\bibnamefont {Xiao}}, \bibinfo
  {author} {\bibfnamefont {X.}~\bibnamefont {Zheng}}, \bibinfo {author}
  {\bibfnamefont {Z.}~\bibnamefont {Ye}}, \bibinfo {author} {\bibfnamefont
  {Y.}~\bibnamefont {Peng}}, \bibinfo {author} {\bibfnamefont {I.~A.}\
  \bibnamefont {Zaliznyak}}, \bibinfo {author} {\bibfnamefont {J.~M.}\
  \bibnamefont {Tranquada}},\ and\ \bibinfo {author} {\bibfnamefont
  {Y.}~\bibnamefont {Li}},\ }\href {https://doi.org/10.48550/arXiv.2204.04593}
  {\bibfield  {journal} {\bibinfo  {journal} {arXiv preprint arXiv:2204.04593}\
  } (\bibinfo {year} {2022}{\natexlab{a}})}\BibitemShut {NoStop}%
\bibitem [{\citenamefont {Janssen}\ \emph {et~al.}(2016)\citenamefont
  {Janssen}, \citenamefont {Andrade},\ and\ \citenamefont
  {Vojta}}]{JanssenPRL2016}%
  \BibitemOpen
  \bibfield  {author} {\bibinfo {author} {\bibfnamefont {L.}~\bibnamefont
  {Janssen}}, \bibinfo {author} {\bibfnamefont {E.~C.}\ \bibnamefont
  {Andrade}},\ and\ \bibinfo {author} {\bibfnamefont {M.}~\bibnamefont
  {Vojta}},\ }\href {https://doi.org/10.1103/PhysRevLett.117.277202} {\bibfield
   {journal} {\bibinfo  {journal} {Phys. Rev. Lett.}\ }\textbf {\bibinfo
  {volume} {117}},\ \bibinfo {pages} {277202} (\bibinfo {year}
  {2016})}\BibitemShut {NoStop}%
\bibitem [{\citenamefont {Janssen}\ and\ \citenamefont
  {Vojta}(2019)}]{Janssen2019}%
  \BibitemOpen
  \bibfield  {author} {\bibinfo {author} {\bibfnamefont {L.}~\bibnamefont
  {Janssen}}\ and\ \bibinfo {author} {\bibfnamefont {M.}~\bibnamefont
  {Vojta}},\ }\href {https://doi.org/10.1088/1361-648x/ab283e} {\bibfield
  {journal} {\bibinfo  {journal} {Journal of Physics: Condensed Matter}\
  }\textbf {\bibinfo {volume} {31}},\ \bibinfo {pages} {423002} (\bibinfo
  {year} {2019})}\BibitemShut {NoStop}%
\bibitem [{\citenamefont {Gordon}\ \emph {et~al.}(2019)\citenamefont {Gordon},
  \citenamefont {Catuneanu}, \citenamefont {S$\mathrm{\o}$rensen},\ and\
  \citenamefont {Kee}}]{GordonNC2019}%
  \BibitemOpen
  \bibfield  {author} {\bibinfo {author} {\bibfnamefont {J.~S.}\ \bibnamefont
  {Gordon}}, \bibinfo {author} {\bibfnamefont {A.}~\bibnamefont {Catuneanu}},
  \bibinfo {author} {\bibfnamefont {E.~S.}\ \bibnamefont
  {S$\mathrm{\o}$rensen}},\ and\ \bibinfo {author} {\bibfnamefont {H.-Y.}\
  \bibnamefont {Kee}},\ }\href {https://doi.org/10.1038/s41467-019-10405-8}
  {\bibfield  {journal} {\bibinfo  {journal} {Nat. Commun.}\ }\textbf {\bibinfo
  {volume} {10}},\ \bibinfo {pages} {2470} (\bibinfo {year}
  {2019})}\BibitemShut {NoStop}%
\bibitem [{\citenamefont {Hickey}\ and\ \citenamefont
  {Trebst}(2019)}]{HickeyNC2019}%
  \BibitemOpen
  \bibfield  {author} {\bibinfo {author} {\bibfnamefont {C.}~\bibnamefont
  {Hickey}}\ and\ \bibinfo {author} {\bibfnamefont {S.}~\bibnamefont
  {Trebst}},\ }\href {https://doi.org/10.1038/s41467-019-08459-9} {\bibfield
  {journal} {\bibinfo  {journal} {Nat. Commun.}\ }\textbf {\bibinfo {volume}
  {10}},\ \bibinfo {pages} {530} (\bibinfo {year} {2019})}\BibitemShut
  {NoStop}%
\bibitem [{\citenamefont {Li}\ \emph {et~al.}(2021)\citenamefont {Li},
  \citenamefont {Zhang}, \citenamefont {Wang}, \citenamefont {Wu},
  \citenamefont {Gao}, \citenamefont {Qu}, \citenamefont {Liu}, \citenamefont
  {Gong},\ and\ \citenamefont {Li}}]{LiNC2021}%
  \BibitemOpen
  \bibfield  {author} {\bibinfo {author} {\bibfnamefont {H.}~\bibnamefont
  {Li}}, \bibinfo {author} {\bibfnamefont {H.-K.}\ \bibnamefont {Zhang}},
  \bibinfo {author} {\bibfnamefont {J.}~\bibnamefont {Wang}}, \bibinfo {author}
  {\bibfnamefont {H.-Q.}\ \bibnamefont {Wu}}, \bibinfo {author} {\bibfnamefont
  {Y.}~\bibnamefont {Gao}}, \bibinfo {author} {\bibfnamefont {D.-W.}\
  \bibnamefont {Qu}}, \bibinfo {author} {\bibfnamefont {Z.-X.}\ \bibnamefont
  {Liu}}, \bibinfo {author} {\bibfnamefont {S.-S.}\ \bibnamefont {Gong}},\ and\
  \bibinfo {author} {\bibfnamefont {W.}~\bibnamefont {Li}},\ }\href
  {https://doi.org/10.1038/s41467-021-24257-8} {\bibfield  {journal} {\bibinfo
  {journal} {Nature Communications}\ }\textbf {\bibinfo {volume} {12}},\
  \bibinfo {pages} {4007} (\bibinfo {year} {2021})}\BibitemShut {NoStop}%
\bibitem [{\citenamefont {Sears}\ \emph {et~al.}(2017)\citenamefont {Sears},
  \citenamefont {Zhao}, \citenamefont {Xu}, \citenamefont {Lynn},\ and\
  \citenamefont {Kim}}]{SearsPRB2017}%
  \BibitemOpen
  \bibfield  {author} {\bibinfo {author} {\bibfnamefont {J.~A.}\ \bibnamefont
  {Sears}}, \bibinfo {author} {\bibfnamefont {Y.}~\bibnamefont {Zhao}},
  \bibinfo {author} {\bibfnamefont {Z.}~\bibnamefont {Xu}}, \bibinfo {author}
  {\bibfnamefont {J.~W.}\ \bibnamefont {Lynn}},\ and\ \bibinfo {author}
  {\bibfnamefont {Y.-J.}\ \bibnamefont {Kim}},\ }\href
  {https://doi.org/10.1103/PhysRevB.95.180411} {\bibfield  {journal} {\bibinfo
  {journal} {Phys. Rev. B}\ }\textbf {\bibinfo {volume} {95}},\ \bibinfo
  {pages} {180411} (\bibinfo {year} {2017})}\BibitemShut {NoStop}%
\bibitem [{\citenamefont {Wolter}\ \emph {et~al.}(2017)\citenamefont {Wolter},
  \citenamefont {Corredor}, \citenamefont {Janssen}, \citenamefont {Nenkov},
  \citenamefont {Sch\"onecker}, \citenamefont {Do}, \citenamefont {Choi},
  \citenamefont {Albrecht}, \citenamefont {Hunger}, \citenamefont {Doert},
  \citenamefont {Vojta},\ and\ \citenamefont {B\"uchner}}]{WolterPRB2017}%
  \BibitemOpen
  \bibfield  {author} {\bibinfo {author} {\bibfnamefont {A.~U.~B.}\
  \bibnamefont {Wolter}}, \bibinfo {author} {\bibfnamefont {L.~T.}\
  \bibnamefont {Corredor}}, \bibinfo {author} {\bibfnamefont {L.}~\bibnamefont
  {Janssen}}, \bibinfo {author} {\bibfnamefont {K.}~\bibnamefont {Nenkov}},
  \bibinfo {author} {\bibfnamefont {S.}~\bibnamefont {Sch\"onecker}}, \bibinfo
  {author} {\bibfnamefont {S.-H.}\ \bibnamefont {Do}}, \bibinfo {author}
  {\bibfnamefont {K.-Y.}\ \bibnamefont {Choi}}, \bibinfo {author}
  {\bibfnamefont {R.}~\bibnamefont {Albrecht}}, \bibinfo {author}
  {\bibfnamefont {J.}~\bibnamefont {Hunger}}, \bibinfo {author} {\bibfnamefont
  {T.}~\bibnamefont {Doert}}, \bibinfo {author} {\bibfnamefont
  {M.}~\bibnamefont {Vojta}},\ and\ \bibinfo {author} {\bibfnamefont
  {B.}~\bibnamefont {B\"uchner}},\ }\href
  {https://doi.org/10.1103/PhysRevB.96.041405} {\bibfield  {journal} {\bibinfo
  {journal} {Phys. Rev. B}\ }\textbf {\bibinfo {volume} {96}},\ \bibinfo
  {pages} {041405} (\bibinfo {year} {2017})}\BibitemShut {NoStop}%
\bibitem [{\citenamefont {Baek}\ \emph {et~al.}(2017)\citenamefont {Baek},
  \citenamefont {Do}, \citenamefont {Choi}, \citenamefont {Kwon}, \citenamefont
  {Wolter}, \citenamefont {Nishimoto}, \citenamefont {van~den Brink},\ and\
  \citenamefont {B\"uchner}}]{BaekPRL2017}%
  \BibitemOpen
  \bibfield  {author} {\bibinfo {author} {\bibfnamefont {S.-H.}\ \bibnamefont
  {Baek}}, \bibinfo {author} {\bibfnamefont {S.-H.}\ \bibnamefont {Do}},
  \bibinfo {author} {\bibfnamefont {K.-Y.}\ \bibnamefont {Choi}}, \bibinfo
  {author} {\bibfnamefont {Y.~S.}\ \bibnamefont {Kwon}}, \bibinfo {author}
  {\bibfnamefont {A.~U.~B.}\ \bibnamefont {Wolter}}, \bibinfo {author}
  {\bibfnamefont {S.}~\bibnamefont {Nishimoto}}, \bibinfo {author}
  {\bibfnamefont {J.}~\bibnamefont {van~den Brink}},\ and\ \bibinfo {author}
  {\bibfnamefont {B.}~\bibnamefont {B\"uchner}},\ }\href
  {https://doi.org/10.1103/PhysRevLett.119.037201} {\bibfield  {journal}
  {\bibinfo  {journal} {Phys. Rev. Lett.}\ }\textbf {\bibinfo {volume} {119}},\
  \bibinfo {pages} {037201} (\bibinfo {year} {2017})}\BibitemShut {NoStop}%
\bibitem [{\citenamefont {Zheng}\ \emph {et~al.}(2017)\citenamefont {Zheng},
  \citenamefont {Ran}, \citenamefont {Li}, \citenamefont {Wang}, \citenamefont
  {Wang}, \citenamefont {Liu}, \citenamefont {Liu}, \citenamefont {Normand},
  \citenamefont {Wen},\ and\ \citenamefont {Yu}}]{ZhengPRL2017}%
  \BibitemOpen
  \bibfield  {author} {\bibinfo {author} {\bibfnamefont {J.}~\bibnamefont
  {Zheng}}, \bibinfo {author} {\bibfnamefont {K.}~\bibnamefont {Ran}}, \bibinfo
  {author} {\bibfnamefont {T.}~\bibnamefont {Li}}, \bibinfo {author}
  {\bibfnamefont {J.}~\bibnamefont {Wang}}, \bibinfo {author} {\bibfnamefont
  {P.}~\bibnamefont {Wang}}, \bibinfo {author} {\bibfnamefont {B.}~\bibnamefont
  {Liu}}, \bibinfo {author} {\bibfnamefont {Z.-X.}\ \bibnamefont {Liu}},
  \bibinfo {author} {\bibfnamefont {B.}~\bibnamefont {Normand}}, \bibinfo
  {author} {\bibfnamefont {J.}~\bibnamefont {Wen}},\ and\ \bibinfo {author}
  {\bibfnamefont {W.}~\bibnamefont {Yu}},\ }\href
  {https://doi.org/10.1103/PhysRevLett.119.227208} {\bibfield  {journal}
  {\bibinfo  {journal} {Phys. Rev. Lett.}\ }\textbf {\bibinfo {volume} {119}},\
  \bibinfo {pages} {227208} (\bibinfo {year} {2017})}\BibitemShut {NoStop}%
\bibitem [{\citenamefont {Banerjee}\ \emph {et~al.}(2017)\citenamefont
  {Banerjee}, \citenamefont {Yan}, \citenamefont {Knolle}, \citenamefont
  {Bridges}, \citenamefont {Stone}, \citenamefont {Lumsden}, \citenamefont
  {Mandrus}, \citenamefont {Tennant}, \citenamefont {Moessner},\ and\
  \citenamefont {Nagler}}]{BanerjeeScience2017}%
  \BibitemOpen
  \bibfield  {author} {\bibinfo {author} {\bibfnamefont {A.}~\bibnamefont
  {Banerjee}}, \bibinfo {author} {\bibfnamefont {J.}~\bibnamefont {Yan}},
  \bibinfo {author} {\bibfnamefont {J.}~\bibnamefont {Knolle}}, \bibinfo
  {author} {\bibfnamefont {C.~A.}\ \bibnamefont {Bridges}}, \bibinfo {author}
  {\bibfnamefont {M.~B.}\ \bibnamefont {Stone}}, \bibinfo {author}
  {\bibfnamefont {M.~D.}\ \bibnamefont {Lumsden}}, \bibinfo {author}
  {\bibfnamefont {D.~G.}\ \bibnamefont {Mandrus}}, \bibinfo {author}
  {\bibfnamefont {D.~A.}\ \bibnamefont {Tennant}}, \bibinfo {author}
  {\bibfnamefont {R.}~\bibnamefont {Moessner}},\ and\ \bibinfo {author}
  {\bibfnamefont {S.~E.}\ \bibnamefont {Nagler}},\ }\href
  {https://doi.org/10.1126/science.aah6015} {\bibfield  {journal} {\bibinfo
  {journal} {Science}\ }\textbf {\bibinfo {volume} {356}},\ \bibinfo {pages}
  {1055} (\bibinfo {year} {2017})}\BibitemShut {NoStop}%
\bibitem [{\citenamefont {Do}\ \emph {et~al.}(2017)\citenamefont {Do},
  \citenamefont {Park}, \citenamefont {Yoshitake}, \citenamefont {Nasu},
  \citenamefont {Motome}, \citenamefont {Kwon}, \citenamefont {Adroja},
  \citenamefont {Voneshen}, \citenamefont {Kim}, \citenamefont {Jang},
  \citenamefont {Park}, \citenamefont {Choi},\ and\ \citenamefont
  {Ji}}]{DoNP2017}%
  \BibitemOpen
  \bibfield  {author} {\bibinfo {author} {\bibfnamefont {S.-H.}\ \bibnamefont
  {Do}}, \bibinfo {author} {\bibfnamefont {S.-Y.}\ \bibnamefont {Park}},
  \bibinfo {author} {\bibfnamefont {J.}~\bibnamefont {Yoshitake}}, \bibinfo
  {author} {\bibfnamefont {J.}~\bibnamefont {Nasu}}, \bibinfo {author}
  {\bibfnamefont {Y.}~\bibnamefont {Motome}}, \bibinfo {author} {\bibfnamefont
  {Y.~S.}\ \bibnamefont {Kwon}}, \bibinfo {author} {\bibfnamefont
  {D.}~\bibnamefont {Adroja}}, \bibinfo {author} {\bibfnamefont
  {D.}~\bibnamefont {Voneshen}}, \bibinfo {author} {\bibfnamefont
  {K.}~\bibnamefont {Kim}}, \bibinfo {author} {\bibfnamefont {T.-H.}\
  \bibnamefont {Jang}}, \bibinfo {author} {\bibfnamefont {J.-H.}\ \bibnamefont
  {Park}}, \bibinfo {author} {\bibfnamefont {K.-Y.}\ \bibnamefont {Choi}},\
  and\ \bibinfo {author} {\bibfnamefont {S.}~\bibnamefont {Ji}},\ }\href
  {https://doi.org/10.1038/nphys4264} {\bibfield  {journal} {\bibinfo
  {journal} {Nature Physics}\ }\textbf {\bibinfo {volume} {13}},\ \bibinfo
  {pages} {1079} (\bibinfo {year} {2017})}\BibitemShut {NoStop}%
\bibitem [{\citenamefont {Kasahara}\ \emph {et~al.}(2018)\citenamefont
  {Kasahara}, \citenamefont {Ohnishi}, \citenamefont {Mizukami}, \citenamefont
  {Tanaka}, \citenamefont {Ma}, \citenamefont {Sugii}, \citenamefont {Kurita},
  \citenamefont {Tanaka}, \citenamefont {Nasu}, \citenamefont {Motome},
  \citenamefont {Shibauchi},\ and\ \citenamefont
  {Matsuda}}]{KasaharaNature2018}%
  \BibitemOpen
  \bibfield  {author} {\bibinfo {author} {\bibfnamefont {Y.}~\bibnamefont
  {Kasahara}}, \bibinfo {author} {\bibfnamefont {T.}~\bibnamefont {Ohnishi}},
  \bibinfo {author} {\bibfnamefont {Y.}~\bibnamefont {Mizukami}}, \bibinfo
  {author} {\bibfnamefont {O.}~\bibnamefont {Tanaka}}, \bibinfo {author}
  {\bibfnamefont {S.}~\bibnamefont {Ma}}, \bibinfo {author} {\bibfnamefont
  {K.}~\bibnamefont {Sugii}}, \bibinfo {author} {\bibfnamefont
  {N.}~\bibnamefont {Kurita}}, \bibinfo {author} {\bibfnamefont
  {H.}~\bibnamefont {Tanaka}}, \bibinfo {author} {\bibfnamefont
  {J.}~\bibnamefont {Nasu}}, \bibinfo {author} {\bibfnamefont {Y.}~\bibnamefont
  {Motome}}, \bibinfo {author} {\bibfnamefont {T.}~\bibnamefont {Shibauchi}},\
  and\ \bibinfo {author} {\bibfnamefont {Y.}~\bibnamefont {Matsuda}},\ }\href
  {https://doi.org/10.1038/s41586-018-0274-0} {\bibfield  {journal} {\bibinfo
  {journal} {Nature}\ }\textbf {\bibinfo {volume} {559}},\ \bibinfo {pages}
  {227} (\bibinfo {year} {2018})}\BibitemShut {NoStop}%
\bibitem [{\citenamefont {Banerjee}\ \emph {et~al.}(2018)\citenamefont
  {Banerjee}, \citenamefont {Lampen-Kelley}, \citenamefont {Knolle},
  \citenamefont {Balz}, \citenamefont {Aczel}, \citenamefont {Winn},
  \citenamefont {Liu}, \citenamefont {Pajerowski}, \citenamefont {Yan},
  \citenamefont {Bridges}, \citenamefont {Savici}, \citenamefont {Chakoumakos},
  \citenamefont {Lumsden}, \citenamefont {Tennant}, \citenamefont {Moessner},
  \citenamefont {Mandrus},\ and\ \citenamefont {Nagler}}]{BanerjeeNPJQM2018}%
  \BibitemOpen
  \bibfield  {author} {\bibinfo {author} {\bibfnamefont {A.}~\bibnamefont
  {Banerjee}}, \bibinfo {author} {\bibfnamefont {P.}~\bibnamefont
  {Lampen-Kelley}}, \bibinfo {author} {\bibfnamefont {J.}~\bibnamefont
  {Knolle}}, \bibinfo {author} {\bibfnamefont {C.}~\bibnamefont {Balz}},
  \bibinfo {author} {\bibfnamefont {A.~A.}\ \bibnamefont {Aczel}}, \bibinfo
  {author} {\bibfnamefont {B.}~\bibnamefont {Winn}}, \bibinfo {author}
  {\bibfnamefont {Y.}~\bibnamefont {Liu}}, \bibinfo {author} {\bibfnamefont
  {D.}~\bibnamefont {Pajerowski}}, \bibinfo {author} {\bibfnamefont
  {J.}~\bibnamefont {Yan}}, \bibinfo {author} {\bibfnamefont {C.~A.}\
  \bibnamefont {Bridges}}, \bibinfo {author} {\bibfnamefont {A.~T.}\
  \bibnamefont {Savici}}, \bibinfo {author} {\bibfnamefont {B.~C.}\
  \bibnamefont {Chakoumakos}}, \bibinfo {author} {\bibfnamefont {M.~D.}\
  \bibnamefont {Lumsden}}, \bibinfo {author} {\bibfnamefont {D.~A.}\
  \bibnamefont {Tennant}}, \bibinfo {author} {\bibfnamefont {R.}~\bibnamefont
  {Moessner}}, \bibinfo {author} {\bibfnamefont {D.~G.}\ \bibnamefont
  {Mandrus}},\ and\ \bibinfo {author} {\bibfnamefont {S.~E.}\ \bibnamefont
  {Nagler}},\ }\href {https://doi.org/10.1038/s41535-018-0079-2} {\bibfield
  {journal} {\bibinfo  {journal} {npj Quantum Materials}\ }\textbf {\bibinfo
  {volume} {3}},\ \bibinfo {pages} {8} (\bibinfo {year} {2018})}\BibitemShut
  {NoStop}%
\bibitem [{\citenamefont {Hentrich}\ \emph {et~al.}(2018)\citenamefont
  {Hentrich}, \citenamefont {Wolter}, \citenamefont {Zotos}, \citenamefont
  {Brenig}, \citenamefont {Nowak}, \citenamefont {Isaeva}, \citenamefont
  {Doert}, \citenamefont {Banerjee}, \citenamefont {Lampen-Kelley},
  \citenamefont {Mandrus}, \citenamefont {Nagler}, \citenamefont {Sears},
  \citenamefont {Kim}, \citenamefont {B\"uchner},\ and\ \citenamefont
  {Hess}}]{HentrichPRL2018}%
  \BibitemOpen
  \bibfield  {author} {\bibinfo {author} {\bibfnamefont {R.}~\bibnamefont
  {Hentrich}}, \bibinfo {author} {\bibfnamefont {A.~U.~B.}\ \bibnamefont
  {Wolter}}, \bibinfo {author} {\bibfnamefont {X.}~\bibnamefont {Zotos}},
  \bibinfo {author} {\bibfnamefont {W.}~\bibnamefont {Brenig}}, \bibinfo
  {author} {\bibfnamefont {D.}~\bibnamefont {Nowak}}, \bibinfo {author}
  {\bibfnamefont {A.}~\bibnamefont {Isaeva}}, \bibinfo {author} {\bibfnamefont
  {T.}~\bibnamefont {Doert}}, \bibinfo {author} {\bibfnamefont
  {A.}~\bibnamefont {Banerjee}}, \bibinfo {author} {\bibfnamefont
  {P.}~\bibnamefont {Lampen-Kelley}}, \bibinfo {author} {\bibfnamefont {D.~G.}\
  \bibnamefont {Mandrus}}, \bibinfo {author} {\bibfnamefont {S.~E.}\
  \bibnamefont {Nagler}}, \bibinfo {author} {\bibfnamefont {J.}~\bibnamefont
  {Sears}}, \bibinfo {author} {\bibfnamefont {Y.-J.}\ \bibnamefont {Kim}},
  \bibinfo {author} {\bibfnamefont {B.}~\bibnamefont {B\"uchner}},\ and\
  \bibinfo {author} {\bibfnamefont {C.}~\bibnamefont {Hess}},\ }\href
  {https://doi.org/10.1103/PhysRevLett.120.117204} {\bibfield  {journal}
  {\bibinfo  {journal} {Phys. Rev. Lett.}\ }\textbf {\bibinfo {volume} {120}},\
  \bibinfo {pages} {117204} (\bibinfo {year} {2018})}\BibitemShut {NoStop}%
\bibitem [{\citenamefont {Balz}\ \emph {et~al.}(2019)\citenamefont {Balz},
  \citenamefont {Lampen-Kelley}, \citenamefont {Banerjee}, \citenamefont {Yan},
  \citenamefont {Lu}, \citenamefont {Hu}, \citenamefont {Yadav}, \citenamefont
  {Takano}, \citenamefont {Liu}, \citenamefont {Tennant}, \citenamefont
  {Lumsden}, \citenamefont {Mandrus},\ and\ \citenamefont
  {Nagler}}]{BalzPRB2019}%
  \BibitemOpen
  \bibfield  {author} {\bibinfo {author} {\bibfnamefont {C.}~\bibnamefont
  {Balz}}, \bibinfo {author} {\bibfnamefont {P.}~\bibnamefont {Lampen-Kelley}},
  \bibinfo {author} {\bibfnamefont {A.}~\bibnamefont {Banerjee}}, \bibinfo
  {author} {\bibfnamefont {J.}~\bibnamefont {Yan}}, \bibinfo {author}
  {\bibfnamefont {Z.}~\bibnamefont {Lu}}, \bibinfo {author} {\bibfnamefont
  {X.}~\bibnamefont {Hu}}, \bibinfo {author} {\bibfnamefont {S.~M.}\
  \bibnamefont {Yadav}}, \bibinfo {author} {\bibfnamefont {Y.}~\bibnamefont
  {Takano}}, \bibinfo {author} {\bibfnamefont {Y.}~\bibnamefont {Liu}},
  \bibinfo {author} {\bibfnamefont {D.~A.}\ \bibnamefont {Tennant}}, \bibinfo
  {author} {\bibfnamefont {M.~D.}\ \bibnamefont {Lumsden}}, \bibinfo {author}
  {\bibfnamefont {D.}~\bibnamefont {Mandrus}},\ and\ \bibinfo {author}
  {\bibfnamefont {S.~E.}\ \bibnamefont {Nagler}},\ }\href
  {https://doi.org/10.1103/PhysRevB.100.060405} {\bibfield  {journal} {\bibinfo
   {journal} {Phys. Rev. B}\ }\textbf {\bibinfo {volume} {100}},\ \bibinfo
  {pages} {060405} (\bibinfo {year} {2019})}\BibitemShut {NoStop}%
\bibitem [{\citenamefont {Yokoi}\ \emph {et~al.}(2021)\citenamefont {Yokoi},
  \citenamefont {Ma}, \citenamefont {Kasahara}, \citenamefont {Kasahara},
  \citenamefont {Shibauchi}, \citenamefont {Kurita}, \citenamefont {Tanaka},
  \citenamefont {Nasu}, \citenamefont {Motome}, \citenamefont {Hickey},
  \citenamefont {Trebst},\ and\ \citenamefont {Matsuda}}]{YokoiScience2021}%
  \BibitemOpen
  \bibfield  {author} {\bibinfo {author} {\bibfnamefont {T.}~\bibnamefont
  {Yokoi}}, \bibinfo {author} {\bibfnamefont {S.}~\bibnamefont {Ma}}, \bibinfo
  {author} {\bibfnamefont {Y.}~\bibnamefont {Kasahara}}, \bibinfo {author}
  {\bibfnamefont {S.}~\bibnamefont {Kasahara}}, \bibinfo {author}
  {\bibfnamefont {T.}~\bibnamefont {Shibauchi}}, \bibinfo {author}
  {\bibfnamefont {N.}~\bibnamefont {Kurita}}, \bibinfo {author} {\bibfnamefont
  {H.}~\bibnamefont {Tanaka}}, \bibinfo {author} {\bibfnamefont
  {J.}~\bibnamefont {Nasu}}, \bibinfo {author} {\bibfnamefont {Y.}~\bibnamefont
  {Motome}}, \bibinfo {author} {\bibfnamefont {C.}~\bibnamefont {Hickey}},
  \bibinfo {author} {\bibfnamefont {S.}~\bibnamefont {Trebst}},\ and\ \bibinfo
  {author} {\bibfnamefont {Y.}~\bibnamefont {Matsuda}},\ }\href
  {https://doi.org/10.1126/science.aay5551} {\bibfield  {journal} {\bibinfo
  {journal} {Science}\ }\textbf {\bibinfo {volume} {373}},\ \bibinfo {pages}
  {568} (\bibinfo {year} {2021})}\BibitemShut {NoStop}%
\bibitem [{\citenamefont {Bruin}\ \emph {et~al.}(2022)\citenamefont {Bruin},
  \citenamefont {Claus}, \citenamefont {Matsumoto}, \citenamefont {Kurita},
  \citenamefont {Tanaka},\ and\ \citenamefont {Takagi}}]{BruinNPhys2022}%
  \BibitemOpen
  \bibfield  {author} {\bibinfo {author} {\bibfnamefont {J.~A.~N.}\
  \bibnamefont {Bruin}}, \bibinfo {author} {\bibfnamefont {R.~R.}\ \bibnamefont
  {Claus}}, \bibinfo {author} {\bibfnamefont {Y.}~\bibnamefont {Matsumoto}},
  \bibinfo {author} {\bibfnamefont {N.}~\bibnamefont {Kurita}}, \bibinfo
  {author} {\bibfnamefont {H.}~\bibnamefont {Tanaka}},\ and\ \bibinfo {author}
  {\bibfnamefont {H.}~\bibnamefont {Takagi}},\ }\href
  {https://doi.org/10.1038/s41567-021-01501-y} {\bibfield  {journal} {\bibinfo
  {journal} {Nat. Phys.}\ }\textbf {\bibinfo {volume} {18}},\ \bibinfo {pages}
  {410} (\bibinfo {year} {2022})}\BibitemShut {NoStop}%
\bibitem [{\citenamefont {Lefran\ifmmode~\mbox{\c{c}}\else \c{c}\fi{}ois}\
  \emph {et~al.}(2022)\citenamefont {Lefran\ifmmode~\mbox{\c{c}}\else
  \c{c}\fi{}ois}, \citenamefont {Grissonnanche}, \citenamefont {Baglo},
  \citenamefont {Lampen-Kelley}, \citenamefont {Yan}, \citenamefont {Balz},
  \citenamefont {Mandrus}, \citenamefont {Nagler}, \citenamefont {Kim},
  \citenamefont {Kim}, \citenamefont {Doiron-Leyraud},\ and\ \citenamefont
  {Taillefer}}]{LefrancoisPRX2022}%
  \BibitemOpen
  \bibfield  {author} {\bibinfo {author} {\bibfnamefont {E.}~\bibnamefont
  {Lefran\ifmmode~\mbox{\c{c}}\else \c{c}\fi{}ois}}, \bibinfo {author}
  {\bibfnamefont {G.}~\bibnamefont {Grissonnanche}}, \bibinfo {author}
  {\bibfnamefont {J.}~\bibnamefont {Baglo}}, \bibinfo {author} {\bibfnamefont
  {P.}~\bibnamefont {Lampen-Kelley}}, \bibinfo {author} {\bibfnamefont {J.-Q.}\
  \bibnamefont {Yan}}, \bibinfo {author} {\bibfnamefont {C.}~\bibnamefont
  {Balz}}, \bibinfo {author} {\bibfnamefont {D.}~\bibnamefont {Mandrus}},
  \bibinfo {author} {\bibfnamefont {S.~E.}\ \bibnamefont {Nagler}}, \bibinfo
  {author} {\bibfnamefont {S.}~\bibnamefont {Kim}}, \bibinfo {author}
  {\bibfnamefont {Y.-J.}\ \bibnamefont {Kim}}, \bibinfo {author} {\bibfnamefont
  {N.}~\bibnamefont {Doiron-Leyraud}},\ and\ \bibinfo {author} {\bibfnamefont
  {L.}~\bibnamefont {Taillefer}},\ }\href
  {https://doi.org/10.1103/PhysRevX.12.021025} {\bibfield  {journal} {\bibinfo
  {journal} {Phys. Rev. X}\ }\textbf {\bibinfo {volume} {12}},\ \bibinfo
  {pages} {021025} (\bibinfo {year} {2022})}\BibitemShut {NoStop}%
\bibitem [{\citenamefont {Lin}\ \emph {et~al.}(2021)\citenamefont {Lin},
  \citenamefont {Jeong}, \citenamefont {Kim}, \citenamefont {Wang},
  \citenamefont {Huang}, \citenamefont {Masuda}, \citenamefont {Asai},
  \citenamefont {Itoh}, \citenamefont {G{\"u}nther}, \citenamefont {Russina},
  \citenamefont {Lu}, \citenamefont {Sheng}, \citenamefont {Wang},
  \citenamefont {Wang}, \citenamefont {Wang}, \citenamefont {Ren},
  \citenamefont {Xi}, \citenamefont {Tong}, \citenamefont {Ling}, \citenamefont
  {Liu}, \citenamefont {Wu}, \citenamefont {Mei}, \citenamefont {Qu},
  \citenamefont {Zhou}, \citenamefont {Park},\ and\ \citenamefont
  {Ma}}]{LinNC2021}%
  \BibitemOpen
  \bibfield  {author} {\bibinfo {author} {\bibfnamefont {G.}~\bibnamefont
  {Lin}}, \bibinfo {author} {\bibfnamefont {J.}~\bibnamefont {Jeong}}, \bibinfo
  {author} {\bibfnamefont {C.}~\bibnamefont {Kim}}, \bibinfo {author}
  {\bibfnamefont {Y.}~\bibnamefont {Wang}}, \bibinfo {author} {\bibfnamefont
  {Q.}~\bibnamefont {Huang}}, \bibinfo {author} {\bibfnamefont
  {T.}~\bibnamefont {Masuda}}, \bibinfo {author} {\bibfnamefont
  {S.}~\bibnamefont {Asai}}, \bibinfo {author} {\bibfnamefont {S.}~\bibnamefont
  {Itoh}}, \bibinfo {author} {\bibfnamefont {G.}~\bibnamefont {G{\"u}nther}},
  \bibinfo {author} {\bibfnamefont {M.}~\bibnamefont {Russina}}, \bibinfo
  {author} {\bibfnamefont {Z.}~\bibnamefont {Lu}}, \bibinfo {author}
  {\bibfnamefont {J.}~\bibnamefont {Sheng}}, \bibinfo {author} {\bibfnamefont
  {L.}~\bibnamefont {Wang}}, \bibinfo {author} {\bibfnamefont {J.}~\bibnamefont
  {Wang}}, \bibinfo {author} {\bibfnamefont {G.}~\bibnamefont {Wang}}, \bibinfo
  {author} {\bibfnamefont {Q.}~\bibnamefont {Ren}}, \bibinfo {author}
  {\bibfnamefont {C.}~\bibnamefont {Xi}}, \bibinfo {author} {\bibfnamefont
  {W.}~\bibnamefont {Tong}}, \bibinfo {author} {\bibfnamefont {L.}~\bibnamefont
  {Ling}}, \bibinfo {author} {\bibfnamefont {Z.}~\bibnamefont {Liu}}, \bibinfo
  {author} {\bibfnamefont {L.}~\bibnamefont {Wu}}, \bibinfo {author}
  {\bibfnamefont {J.}~\bibnamefont {Mei}}, \bibinfo {author} {\bibfnamefont
  {Z.}~\bibnamefont {Qu}}, \bibinfo {author} {\bibfnamefont {H.}~\bibnamefont
  {Zhou}}, \bibinfo {author} {\bibfnamefont {J.-G.}\ \bibnamefont {Park}},\
  and\ \bibinfo {author} {\bibfnamefont {J.}~\bibnamefont {Ma}},\ }\href
  {https://doi.org/10.1038/s41467-021-25567-7} {\bibfield  {journal} {\bibinfo
  {journal} {Nature Communications}\ }\textbf {\bibinfo {volume} {12}},\
  \bibinfo {pages} {5559} (\bibinfo {year} {2021})}\BibitemShut {NoStop}%
\bibitem [{\citenamefont {Hong}\ \emph {et~al.}(2021)\citenamefont {Hong},
  \citenamefont {Gillig}, \citenamefont {Hentrich}, \citenamefont {Yao},
  \citenamefont {Kocsis}, \citenamefont {Witte}, \citenamefont {Schreiner},
  \citenamefont {Baumann}, \citenamefont {P\'erez}, \citenamefont {Wolter},
  \citenamefont {Li}, \citenamefont {B\"uchner},\ and\ \citenamefont
  {Hess}}]{HongPRB2021}%
  \BibitemOpen
  \bibfield  {author} {\bibinfo {author} {\bibfnamefont {X.}~\bibnamefont
  {Hong}}, \bibinfo {author} {\bibfnamefont {M.}~\bibnamefont {Gillig}},
  \bibinfo {author} {\bibfnamefont {R.}~\bibnamefont {Hentrich}}, \bibinfo
  {author} {\bibfnamefont {W.}~\bibnamefont {Yao}}, \bibinfo {author}
  {\bibfnamefont {V.}~\bibnamefont {Kocsis}}, \bibinfo {author} {\bibfnamefont
  {A.~R.}\ \bibnamefont {Witte}}, \bibinfo {author} {\bibfnamefont
  {T.}~\bibnamefont {Schreiner}}, \bibinfo {author} {\bibfnamefont
  {D.}~\bibnamefont {Baumann}}, \bibinfo {author} {\bibfnamefont
  {N.}~\bibnamefont {P\'erez}}, \bibinfo {author} {\bibfnamefont {A.~U.~B.}\
  \bibnamefont {Wolter}}, \bibinfo {author} {\bibfnamefont {Y.}~\bibnamefont
  {Li}}, \bibinfo {author} {\bibfnamefont {B.}~\bibnamefont {B\"uchner}},\ and\
  \bibinfo {author} {\bibfnamefont {C.}~\bibnamefont {Hess}},\ }\href
  {https://doi.org/10.1103/PhysRevB.104.144426} {\bibfield  {journal} {\bibinfo
   {journal} {Phys. Rev. B}\ }\textbf {\bibinfo {volume} {104}},\ \bibinfo
  {pages} {144426} (\bibinfo {year} {2021})}\BibitemShut {NoStop}%
\bibitem [{\citenamefont {Li}\ \emph {et~al.}(2022{\natexlab{b}})\citenamefont
  {Li}, \citenamefont {Guang}, \citenamefont {Chu}, \citenamefont {Huang},
  \citenamefont {Liu}, \citenamefont {Xia}, \citenamefont {Zhou}, \citenamefont
  {Yue}, \citenamefont {Sun}, \citenamefont {Wang}, \citenamefont {Li},
  \citenamefont {Lin}, \citenamefont {Ma}, \citenamefont {Zhao}, \citenamefont
  {Zhou},\ and\ \citenamefont {Sun}}]{LiArxiv2022}%
  \BibitemOpen
  \bibfield  {author} {\bibinfo {author} {\bibfnamefont {N.}~\bibnamefont
  {Li}}, \bibinfo {author} {\bibfnamefont {S.}~\bibnamefont {Guang}}, \bibinfo
  {author} {\bibfnamefont {W.}~\bibnamefont {Chu}}, \bibinfo {author}
  {\bibfnamefont {Q.}~\bibnamefont {Huang}}, \bibinfo {author} {\bibfnamefont
  {J.}~\bibnamefont {Liu}}, \bibinfo {author} {\bibfnamefont {K.}~\bibnamefont
  {Xia}}, \bibinfo {author} {\bibfnamefont {X.}~\bibnamefont {Zhou}}, \bibinfo
  {author} {\bibfnamefont {X.}~\bibnamefont {Yue}}, \bibinfo {author}
  {\bibfnamefont {Y.}~\bibnamefont {Sun}}, \bibinfo {author} {\bibfnamefont
  {Y.}~\bibnamefont {Wang}}, \bibinfo {author} {\bibfnamefont {Q.}~\bibnamefont
  {Li}}, \bibinfo {author} {\bibfnamefont {G.}~\bibnamefont {Lin}}, \bibinfo
  {author} {\bibfnamefont {J.}~\bibnamefont {Ma}}, \bibinfo {author}
  {\bibfnamefont {X.}~\bibnamefont {Zhao}}, \bibinfo {author} {\bibfnamefont
  {H.}~\bibnamefont {Zhou}},\ and\ \bibinfo {author} {\bibfnamefont
  {X.}~\bibnamefont {Sun}},\ }\bibfield  {journal} {\bibinfo  {journal} {arXiv
  preprint arXiv:2201.11396}\ }\href
  {https://doi.org/10.48550/arXiv.2201.11396} {10.48550/arXiv.2201.11396}
  (\bibinfo {year} {2022}{\natexlab{b}})\BibitemShut {NoStop}%
\bibitem [{\citenamefont {Yang}\ \emph {et~al.}(2022)\citenamefont {Yang},
  \citenamefont {Kim}, \citenamefont {Choi}, \citenamefont {Lee}, \citenamefont
  {Lin}, \citenamefont {Ma}, \citenamefont {Kratochv\'{\i}lov\'a},
  \citenamefont {Proschek}, \citenamefont {Moon}, \citenamefont {Lee},
  \citenamefont {Oh},\ and\ \citenamefont {Park}}]{YangPRB2022}%
  \BibitemOpen
  \bibfield  {author} {\bibinfo {author} {\bibfnamefont {H.}~\bibnamefont
  {Yang}}, \bibinfo {author} {\bibfnamefont {C.}~\bibnamefont {Kim}}, \bibinfo
  {author} {\bibfnamefont {Y.}~\bibnamefont {Choi}}, \bibinfo {author}
  {\bibfnamefont {J.~H.}\ \bibnamefont {Lee}}, \bibinfo {author} {\bibfnamefont
  {G.}~\bibnamefont {Lin}}, \bibinfo {author} {\bibfnamefont {J.}~\bibnamefont
  {Ma}}, \bibinfo {author} {\bibfnamefont {M.}~\bibnamefont
  {Kratochv\'{\i}lov\'a}}, \bibinfo {author} {\bibfnamefont {P.}~\bibnamefont
  {Proschek}}, \bibinfo {author} {\bibfnamefont {E.-G.}\ \bibnamefont {Moon}},
  \bibinfo {author} {\bibfnamefont {K.~H.}\ \bibnamefont {Lee}}, \bibinfo
  {author} {\bibfnamefont {Y.~S.}\ \bibnamefont {Oh}},\ and\ \bibinfo {author}
  {\bibfnamefont {J.-G.}\ \bibnamefont {Park}},\ }\href
  {https://doi.org/10.1103/PhysRevB.106.L081116} {\bibfield  {journal}
  {\bibinfo  {journal} {Phys. Rev. B}\ }\textbf {\bibinfo {volume} {106}},\
  \bibinfo {pages} {L081116} (\bibinfo {year} {2022})}\BibitemShut {NoStop}%
\bibitem [{\citenamefont {Xiao}\ \emph {et~al.}(2021)\citenamefont {Xiao},
  \citenamefont {Xia}, \citenamefont {Song},\ and\ \citenamefont
  {Xiao}}]{XiaoJPCM2021}%
  \BibitemOpen
  \bibfield  {author} {\bibinfo {author} {\bibfnamefont {G.}~\bibnamefont
  {Xiao}}, \bibinfo {author} {\bibfnamefont {Z.}~\bibnamefont {Xia}}, \bibinfo
  {author} {\bibfnamefont {Y.}~\bibnamefont {Song}},\ and\ \bibinfo {author}
  {\bibfnamefont {L.}~\bibnamefont {Xiao}},\ }\href
  {https://doi.org/10.1088/1361-648x/ac3869} {\bibfield  {journal} {\bibinfo
  {journal} {Journal of Physics: Condensed Matter}\ }\textbf {\bibinfo {volume}
  {34}},\ \bibinfo {pages} {075801} (\bibinfo {year} {2021})}\BibitemShut
  {NoStop}%
\bibitem [{\citenamefont {Takeda}\ \emph {et~al.}(2022)\citenamefont {Takeda},
  \citenamefont {Mai}, \citenamefont {Akazawa}, \citenamefont {Tamura},
  \citenamefont {Yan}, \citenamefont {Moovendaran}, \citenamefont {Raju},
  \citenamefont {Sankar}, \citenamefont {Choi},\ and\ \citenamefont
  {Yamashita}}]{TakedaPRR2022}%
  \BibitemOpen
  \bibfield  {author} {\bibinfo {author} {\bibfnamefont {H.}~\bibnamefont
  {Takeda}}, \bibinfo {author} {\bibfnamefont {J.}~\bibnamefont {Mai}},
  \bibinfo {author} {\bibfnamefont {M.}~\bibnamefont {Akazawa}}, \bibinfo
  {author} {\bibfnamefont {K.}~\bibnamefont {Tamura}}, \bibinfo {author}
  {\bibfnamefont {J.}~\bibnamefont {Yan}}, \bibinfo {author} {\bibfnamefont
  {K.}~\bibnamefont {Moovendaran}}, \bibinfo {author} {\bibfnamefont
  {K.}~\bibnamefont {Raju}}, \bibinfo {author} {\bibfnamefont {R.}~\bibnamefont
  {Sankar}}, \bibinfo {author} {\bibfnamefont {K.-Y.}\ \bibnamefont {Choi}},\
  and\ \bibinfo {author} {\bibfnamefont {M.}~\bibnamefont {Yamashita}},\ }\href
  {https://doi.org/10.1103/PhysRevResearch.4.L042035} {\bibfield  {journal}
  {\bibinfo  {journal} {Phys. Rev. Research}\ }\textbf {\bibinfo {volume}
  {4}},\ \bibinfo {pages} {L042035} (\bibinfo {year} {2022})}\BibitemShut
  {NoStop}%
\bibitem [{\citenamefont {Rusna\ifmmode~\check{c}\else \v{c}\fi{}ko}\ \emph
  {et~al.}(2019)\citenamefont {Rusna\ifmmode~\check{c}\else \v{c}\fi{}ko},
  \citenamefont {Gotfryd},\ and\ \citenamefont {Chaloupka}}]{RusnackoPRB2019}%
  \BibitemOpen
  \bibfield  {author} {\bibinfo {author} {\bibfnamefont {J.}~\bibnamefont
  {Rusna\ifmmode~\check{c}\else \v{c}\fi{}ko}}, \bibinfo {author}
  {\bibfnamefont {D.}~\bibnamefont {Gotfryd}},\ and\ \bibinfo {author}
  {\bibfnamefont {J.}~\bibnamefont {Chaloupka}},\ }\href
  {https://doi.org/10.1103/PhysRevB.99.064425} {\bibfield  {journal} {\bibinfo
  {journal} {Phys. Rev. B}\ }\textbf {\bibinfo {volume} {99}},\ \bibinfo
  {pages} {064425} (\bibinfo {year} {2019})}\BibitemShut {NoStop}%
\bibitem [{\citenamefont {Maksimov}\ and\ \citenamefont
  {Chernyshev}(2020)}]{MaksimovPRR2020}%
  \BibitemOpen
  \bibfield  {author} {\bibinfo {author} {\bibfnamefont {P.~A.}\ \bibnamefont
  {Maksimov}}\ and\ \bibinfo {author} {\bibfnamefont {A.~L.}\ \bibnamefont
  {Chernyshev}},\ }\href {https://doi.org/10.1103/PhysRevResearch.2.033011}
  {\bibfield  {journal} {\bibinfo  {journal} {Phys. Rev. Research}\ }\textbf
  {\bibinfo {volume} {2}},\ \bibinfo {pages} {033011} (\bibinfo {year}
  {2020})}\BibitemShut {NoStop}%
\bibitem [{\citenamefont {Laurell}\ and\ \citenamefont
  {Okamoto}(2020)}]{LaurellNPJQM2020}%
  \BibitemOpen
  \bibfield  {author} {\bibinfo {author} {\bibfnamefont {P.}~\bibnamefont
  {Laurell}}\ and\ \bibinfo {author} {\bibfnamefont {S.}~\bibnamefont
  {Okamoto}},\ }\href {https://doi.org/10.1038/s41535-019-0203-y} {\bibfield
  {journal} {\bibinfo  {journal} {npj Quantum Materials}\ }\textbf {\bibinfo
  {volume} {5}},\ \bibinfo {pages} {1} (\bibinfo {year} {2020})}\BibitemShut
  {NoStop}%
\bibitem [{\citenamefont {Songvilay}\ \emph {et~al.}(2020)\citenamefont
  {Songvilay}, \citenamefont {Robert}, \citenamefont {Petit}, \citenamefont
  {Rodriguez-Rivera}, \citenamefont {Ratcliff}, \citenamefont {Damay},
  \citenamefont {Bal\'edent}, \citenamefont {Jim\'enez-Ruiz}, \citenamefont
  {Lejay}, \citenamefont {Pachoud}, \citenamefont {Hadj-Azzem}, \citenamefont
  {Simonet},\ and\ \citenamefont {Stock}}]{SongvilayPRB2020}%
  \BibitemOpen
  \bibfield  {author} {\bibinfo {author} {\bibfnamefont {M.}~\bibnamefont
  {Songvilay}}, \bibinfo {author} {\bibfnamefont {J.}~\bibnamefont {Robert}},
  \bibinfo {author} {\bibfnamefont {S.}~\bibnamefont {Petit}}, \bibinfo
  {author} {\bibfnamefont {J.~A.}\ \bibnamefont {Rodriguez-Rivera}}, \bibinfo
  {author} {\bibfnamefont {W.~D.}\ \bibnamefont {Ratcliff}}, \bibinfo {author}
  {\bibfnamefont {F.}~\bibnamefont {Damay}}, \bibinfo {author} {\bibfnamefont
  {V.}~\bibnamefont {Bal\'edent}}, \bibinfo {author} {\bibfnamefont
  {M.}~\bibnamefont {Jim\'enez-Ruiz}}, \bibinfo {author} {\bibfnamefont
  {P.}~\bibnamefont {Lejay}}, \bibinfo {author} {\bibfnamefont
  {E.}~\bibnamefont {Pachoud}}, \bibinfo {author} {\bibfnamefont
  {A.}~\bibnamefont {Hadj-Azzem}}, \bibinfo {author} {\bibfnamefont
  {V.}~\bibnamefont {Simonet}},\ and\ \bibinfo {author} {\bibfnamefont
  {C.}~\bibnamefont {Stock}},\ }\href
  {https://doi.org/10.1103/PhysRevB.102.224429} {\bibfield  {journal} {\bibinfo
   {journal} {Phys. Rev. B}\ }\textbf {\bibinfo {volume} {102}},\ \bibinfo
  {pages} {224429} (\bibinfo {year} {2020})}\BibitemShut {NoStop}%
\bibitem [{\citenamefont {Samarakoon}\ \emph {et~al.}(2021)\citenamefont
  {Samarakoon}, \citenamefont {Chen}, \citenamefont {Zhou},\ and\ \citenamefont
  {Garlea}}]{SamarakoonPRB2021}%
  \BibitemOpen
  \bibfield  {author} {\bibinfo {author} {\bibfnamefont {A.~M.}\ \bibnamefont
  {Samarakoon}}, \bibinfo {author} {\bibfnamefont {Q.}~\bibnamefont {Chen}},
  \bibinfo {author} {\bibfnamefont {H.}~\bibnamefont {Zhou}},\ and\ \bibinfo
  {author} {\bibfnamefont {V.~O.}\ \bibnamefont {Garlea}},\ }\href
  {https://doi.org/10.1103/PhysRevB.104.184415} {\bibfield  {journal} {\bibinfo
   {journal} {Phys. Rev. B}\ }\textbf {\bibinfo {volume} {104}},\ \bibinfo
  {pages} {184415} (\bibinfo {year} {2021})}\BibitemShut {NoStop}%
\bibitem [{\citenamefont {Kim}\ \emph {et~al.}(2022)\citenamefont {Kim},
  \citenamefont {Jeong}, \citenamefont {Lin}, \citenamefont {Park},
  \citenamefont {Masuda}, \citenamefont {Asai}, \citenamefont {Itoh},
  \citenamefont {Kim}, \citenamefont {Zhou}, \citenamefont {Ma},\ and\
  \citenamefont {Park}}]{KimJPCM2021}%
  \BibitemOpen
  \bibfield  {author} {\bibinfo {author} {\bibfnamefont {C.}~\bibnamefont
  {Kim}}, \bibinfo {author} {\bibfnamefont {J.}~\bibnamefont {Jeong}}, \bibinfo
  {author} {\bibfnamefont {G.}~\bibnamefont {Lin}}, \bibinfo {author}
  {\bibfnamefont {P.}~\bibnamefont {Park}}, \bibinfo {author} {\bibfnamefont
  {T.}~\bibnamefont {Masuda}}, \bibinfo {author} {\bibfnamefont
  {S.}~\bibnamefont {Asai}}, \bibinfo {author} {\bibfnamefont {S.}~\bibnamefont
  {Itoh}}, \bibinfo {author} {\bibfnamefont {H.-S.}\ \bibnamefont {Kim}},
  \bibinfo {author} {\bibfnamefont {H.}~\bibnamefont {Zhou}}, \bibinfo {author}
  {\bibfnamefont {J.}~\bibnamefont {Ma}},\ and\ \bibinfo {author}
  {\bibfnamefont {J.-G.}\ \bibnamefont {Park}},\ }\href
  {https://doi.org/10.1088/1361-648x/ac2644} {\bibfield  {journal} {\bibinfo
  {journal} {Journal of Physics: Condensed Matter}\ }\textbf {\bibinfo {volume}
  {34}},\ \bibinfo {pages} {045802} (\bibinfo {year} {2022})}\BibitemShut
  {NoStop}%
\bibitem [{\citenamefont {Das}\ \emph {et~al.}(2021)\citenamefont {Das},
  \citenamefont {Voleti}, \citenamefont {Saha-Dasgupta},\ and\ \citenamefont
  {Paramekanti}}]{DasPRB2021}%
  \BibitemOpen
  \bibfield  {author} {\bibinfo {author} {\bibfnamefont {S.}~\bibnamefont
  {Das}}, \bibinfo {author} {\bibfnamefont {S.}~\bibnamefont {Voleti}},
  \bibinfo {author} {\bibfnamefont {T.}~\bibnamefont {Saha-Dasgupta}},\ and\
  \bibinfo {author} {\bibfnamefont {A.}~\bibnamefont {Paramekanti}},\ }\href
  {https://doi.org/10.1103/PhysRevB.104.134425} {\bibfield  {journal} {\bibinfo
   {journal} {Phys. Rev. B}\ }\textbf {\bibinfo {volume} {104}},\ \bibinfo
  {pages} {134425} (\bibinfo {year} {2021})}\BibitemShut {NoStop}%
\bibitem [{\citenamefont {Sanders}\ \emph {et~al.}(2022)\citenamefont
  {Sanders}, \citenamefont {Mole}, \citenamefont {Liu}, \citenamefont {Brown},
  \citenamefont {Yu}, \citenamefont {Ling},\ and\ \citenamefont
  {Rachel}}]{SandersPRB2022}%
  \BibitemOpen
  \bibfield  {author} {\bibinfo {author} {\bibfnamefont {A.~L.}\ \bibnamefont
  {Sanders}}, \bibinfo {author} {\bibfnamefont {R.~A.}\ \bibnamefont {Mole}},
  \bibinfo {author} {\bibfnamefont {J.}~\bibnamefont {Liu}}, \bibinfo {author}
  {\bibfnamefont {A.~J.}\ \bibnamefont {Brown}}, \bibinfo {author}
  {\bibfnamefont {D.}~\bibnamefont {Yu}}, \bibinfo {author} {\bibfnamefont
  {C.~D.}\ \bibnamefont {Ling}},\ and\ \bibinfo {author} {\bibfnamefont
  {S.}~\bibnamefont {Rachel}},\ }\href
  {https://doi.org/10.1103/PhysRevB.106.014413} {\bibfield  {journal} {\bibinfo
   {journal} {Phys. Rev. B}\ }\textbf {\bibinfo {volume} {106}},\ \bibinfo
  {pages} {014413} (\bibinfo {year} {2022})}\BibitemShut {NoStop}%
\bibitem [{\citenamefont {Yao}\ \emph {et~al.}(2022)\citenamefont {Yao},
  \citenamefont {Iida}, \citenamefont {Kamazawa},\ and\ \citenamefont
  {Li}}]{YaoPRL2022}%
  \BibitemOpen
  \bibfield  {author} {\bibinfo {author} {\bibfnamefont {W.}~\bibnamefont
  {Yao}}, \bibinfo {author} {\bibfnamefont {K.}~\bibnamefont {Iida}}, \bibinfo
  {author} {\bibfnamefont {K.}~\bibnamefont {Kamazawa}},\ and\ \bibinfo
  {author} {\bibfnamefont {Y.}~\bibnamefont {Li}},\ }\href
  {https://doi.org/10.1103/PhysRevLett.129.147202} {\bibfield  {journal}
  {\bibinfo  {journal} {Phys. Rev. Lett.}\ }\textbf {\bibinfo {volume} {129}},\
  \bibinfo {pages} {147202} (\bibinfo {year} {2022})}\BibitemShut {NoStop}%
\bibitem [{\citenamefont {Winter}(2022)}]{WinterJPM2022}%
  \BibitemOpen
  \bibfield  {author} {\bibinfo {author} {\bibfnamefont {S.~M.}\ \bibnamefont
  {Winter}},\ }\href {https://doi.org/10.1088/2515-7639/ac94f8} {\bibfield
  {journal} {\bibinfo  {journal} {Journal of Physics: Materials}\ }\textbf
  {\bibinfo {volume} {5}},\ \bibinfo {pages} {045003} (\bibinfo {year}
  {2022})}\BibitemShut {NoStop}%
\bibitem [{\citenamefont {Maksimov}\ \emph {et~al.}(2022)\citenamefont
  {Maksimov}, \citenamefont {Ushakov}, \citenamefont {Pchelkina}, \citenamefont
  {Li}, \citenamefont {Winter},\ and\ \citenamefont
  {Streltsov}}]{MaksimovPRB2022}%
  \BibitemOpen
  \bibfield  {author} {\bibinfo {author} {\bibfnamefont {P.~A.}\ \bibnamefont
  {Maksimov}}, \bibinfo {author} {\bibfnamefont {A.~V.}\ \bibnamefont
  {Ushakov}}, \bibinfo {author} {\bibfnamefont {Z.~V.}\ \bibnamefont
  {Pchelkina}}, \bibinfo {author} {\bibfnamefont {Y.}~\bibnamefont {Li}},
  \bibinfo {author} {\bibfnamefont {S.~M.}\ \bibnamefont {Winter}},\ and\
  \bibinfo {author} {\bibfnamefont {S.~V.}\ \bibnamefont {Streltsov}},\ }\href
  {https://doi.org/10.1103/PhysRevB.106.165131} {\bibfield  {journal} {\bibinfo
   {journal} {Phys. Rev. B}\ }\textbf {\bibinfo {volume} {106}},\ \bibinfo
  {pages} {165131} (\bibinfo {year} {2022})}\BibitemShut {NoStop}%
\bibitem [{\citenamefont {Pandey}\ and\ \citenamefont
  {Feng}(2022)}]{PandeyPRB2022}%
  \BibitemOpen
  \bibfield  {author} {\bibinfo {author} {\bibfnamefont {S.~K.}\ \bibnamefont
  {Pandey}}\ and\ \bibinfo {author} {\bibfnamefont {J.}~\bibnamefont {Feng}},\
  }\href {https://doi.org/10.1103/PhysRevB.106.174411} {\bibfield  {journal}
  {\bibinfo  {journal} {Phys. Rev. B}\ }\textbf {\bibinfo {volume} {106}},\
  \bibinfo {pages} {174411} (\bibinfo {year} {2022})}\BibitemShut {NoStop}%
\bibitem [{\citenamefont {Lin}\ \emph {et~al.}(2022)\citenamefont {Lin},
  \citenamefont {Zhao}, \citenamefont {Li}, \citenamefont {Shu}, \citenamefont
  {Ma}, \citenamefont {Jiao}, \citenamefont {Huang}, \citenamefont {Sheng},
  \citenamefont {Kolesnikov}, \citenamefont {Li}, \citenamefont {Wu},
  \citenamefont {Wang}, \citenamefont {Zhou}, \citenamefont {Liu},\ and\
  \citenamefont {Ma}}]{LinPreprint2022}%
  \BibitemOpen
  \bibfield  {author} {\bibinfo {author} {\bibfnamefont {G.}~\bibnamefont
  {Lin}}, \bibinfo {author} {\bibfnamefont {Q.}~\bibnamefont {Zhao}}, \bibinfo
  {author} {\bibfnamefont {G.}~\bibnamefont {Li}}, \bibinfo {author}
  {\bibfnamefont {M.}~\bibnamefont {Shu}}, \bibinfo {author} {\bibfnamefont
  {Y.}~\bibnamefont {Ma}}, \bibinfo {author} {\bibfnamefont {J.}~\bibnamefont
  {Jiao}}, \bibinfo {author} {\bibfnamefont {Q.}~\bibnamefont {Huang}},
  \bibinfo {author} {\bibfnamefont {J.}~\bibnamefont {Sheng}}, \bibinfo
  {author} {\bibfnamefont {A.}~\bibnamefont {Kolesnikov}}, \bibinfo {author}
  {\bibfnamefont {L.}~\bibnamefont {Li}}, \bibinfo {author} {\bibfnamefont
  {L.}~\bibnamefont {Wu}}, \bibinfo {author} {\bibfnamefont {X.}~\bibnamefont
  {Wang}}, \bibinfo {author} {\bibfnamefont {H.}~\bibnamefont {Zhou}}, \bibinfo
  {author} {\bibfnamefont {Z.}~\bibnamefont {Liu}},\ and\ \bibinfo {author}
  {\bibfnamefont {J.}~\bibnamefont {Ma}}} (\bibinfo {year} {2022}),\ \bibinfo
  {note} {\url{https://doi.org/10.21203/rs.3.rs-2034295/v1}}\BibitemShut
  {NoStop}%
\bibitem [{\citenamefont {Chen}\ \emph {et~al.}(2021)\citenamefont {Chen},
  \citenamefont {Li}, \citenamefont {Hu}, \citenamefont {Hu}, \citenamefont
  {Yue}, \citenamefont {Sutarto}, \citenamefont {He}, \citenamefont {Iida},
  \citenamefont {Kamazawa}, \citenamefont {Yu}, \citenamefont {Lin},\ and\
  \citenamefont {Li}}]{XPRB2021}%
  \BibitemOpen
  \bibfield  {author} {\bibinfo {author} {\bibfnamefont {W.}~\bibnamefont
  {Chen}}, \bibinfo {author} {\bibfnamefont {X.}~\bibnamefont {Li}}, \bibinfo
  {author} {\bibfnamefont {Z.}~\bibnamefont {Hu}}, \bibinfo {author}
  {\bibfnamefont {Z.}~\bibnamefont {Hu}}, \bibinfo {author} {\bibfnamefont
  {L.}~\bibnamefont {Yue}}, \bibinfo {author} {\bibfnamefont {R.}~\bibnamefont
  {Sutarto}}, \bibinfo {author} {\bibfnamefont {F.}~\bibnamefont {He}},
  \bibinfo {author} {\bibfnamefont {K.}~\bibnamefont {Iida}}, \bibinfo {author}
  {\bibfnamefont {K.}~\bibnamefont {Kamazawa}}, \bibinfo {author}
  {\bibfnamefont {W.}~\bibnamefont {Yu}}, \bibinfo {author} {\bibfnamefont
  {X.}~\bibnamefont {Lin}},\ and\ \bibinfo {author} {\bibfnamefont
  {Y.}~\bibnamefont {Li}},\ }\href
  {https://doi.org/10.1103/PhysRevB.103.L180404} {\bibfield  {journal}
  {\bibinfo  {journal} {Phys. Rev. B}\ }\textbf {\bibinfo {volume} {103}},\
  \bibinfo {pages} {L180404} (\bibinfo {year} {2021})}\BibitemShut {NoStop}%
\bibitem [{\citenamefont {Lee}\ \emph {et~al.}(2021)\citenamefont {Lee},
  \citenamefont {Lee}, \citenamefont {Choi}, \citenamefont {Jang},
  \citenamefont {Kalaivanan}, \citenamefont {Sankar},\ and\ \citenamefont
  {Choi}}]{LeePRB2021}%
  \BibitemOpen
  \bibfield  {author} {\bibinfo {author} {\bibfnamefont {C.~H.}\ \bibnamefont
  {Lee}}, \bibinfo {author} {\bibfnamefont {S.}~\bibnamefont {Lee}}, \bibinfo
  {author} {\bibfnamefont {Y.~S.}\ \bibnamefont {Choi}}, \bibinfo {author}
  {\bibfnamefont {Z.~H.}\ \bibnamefont {Jang}}, \bibinfo {author}
  {\bibfnamefont {R.}~\bibnamefont {Kalaivanan}}, \bibinfo {author}
  {\bibfnamefont {R.}~\bibnamefont {Sankar}},\ and\ \bibinfo {author}
  {\bibfnamefont {K.-Y.}\ \bibnamefont {Choi}},\ }\href
  {https://doi.org/10.1103/PhysRevB.103.214447} {\bibfield  {journal} {\bibinfo
   {journal} {Phys. Rev. B}\ }\textbf {\bibinfo {volume} {103}},\ \bibinfo
  {pages} {214447} (\bibinfo {year} {2021})}\BibitemShut {NoStop}%
\bibitem [{\citenamefont {Kikuchi}\ \emph {et~al.}(2022)\citenamefont
  {Kikuchi}, \citenamefont {Kamoda}, \citenamefont {Mera}, \citenamefont
  {Takahashi}, \citenamefont {Okumura},\ and\ \citenamefont
  {Yasui}}]{KikuchiArxiv2022}%
  \BibitemOpen
  \bibfield  {author} {\bibinfo {author} {\bibfnamefont {J.}~\bibnamefont
  {Kikuchi}}, \bibinfo {author} {\bibfnamefont {T.}~\bibnamefont {Kamoda}},
  \bibinfo {author} {\bibfnamefont {N.}~\bibnamefont {Mera}}, \bibinfo {author}
  {\bibfnamefont {Y.}~\bibnamefont {Takahashi}}, \bibinfo {author}
  {\bibfnamefont {K.}~\bibnamefont {Okumura}},\ and\ \bibinfo {author}
  {\bibfnamefont {Y.}~\bibnamefont {Yasui}},\ }\href
  {https://doi.org/10.48550/arXiv.2206.05409} {\bibfield  {journal} {\bibinfo
  {journal} {arXiv preprint arXiv:2206.05409}\ } (\bibinfo {year}
  {2022})}\BibitemShut {NoStop}%
\bibitem [{\citenamefont {Viciu}\ \emph {et~al.}(2007)\citenamefont {Viciu},
  \citenamefont {Huang}, \citenamefont {Morosan}, \citenamefont {Zandbergen},
  \citenamefont {Greenbaum}, \citenamefont {McQueen},\ and\ \citenamefont
  {Cava}}]{ViciuJSSC2007}%
  \BibitemOpen
  \bibfield  {author} {\bibinfo {author} {\bibfnamefont {L.}~\bibnamefont
  {Viciu}}, \bibinfo {author} {\bibfnamefont {Q.}~\bibnamefont {Huang}},
  \bibinfo {author} {\bibfnamefont {E.}~\bibnamefont {Morosan}}, \bibinfo
  {author} {\bibfnamefont {H.}~\bibnamefont {Zandbergen}}, \bibinfo {author}
  {\bibfnamefont {N.}~\bibnamefont {Greenbaum}}, \bibinfo {author}
  {\bibfnamefont {T.}~\bibnamefont {McQueen}},\ and\ \bibinfo {author}
  {\bibfnamefont {R.}~\bibnamefont {Cava}},\ }\href
  {https://dx.doi.org/10.1016/j.jssc.2007.01.002} {\bibfield  {journal}
  {\bibinfo  {journal} {Journal of Solid State Chemistry}\ }\textbf {\bibinfo
  {volume} {180}},\ \bibinfo {pages} {1060} (\bibinfo {year}
  {2007})}\BibitemShut {NoStop}%
\bibitem [{\citenamefont {Lefran\ifmmode~\mbox{\c{c}}\else \c{c}\fi{}ois}\
  \emph {et~al.}(2016)\citenamefont {Lefran\ifmmode~\mbox{\c{c}}\else
  \c{c}\fi{}ois}, \citenamefont {Songvilay}, \citenamefont {Robert},
  \citenamefont {Nataf}, \citenamefont {Jordan}, \citenamefont {Chaix},
  \citenamefont {Colin}, \citenamefont {Lejay}, \citenamefont {Hadj-Azzem},
  \citenamefont {Ballou},\ and\ \citenamefont {Simonet}}]{LefrancoisPRB2016}%
  \BibitemOpen
  \bibfield  {author} {\bibinfo {author} {\bibfnamefont {E.}~\bibnamefont
  {Lefran\ifmmode~\mbox{\c{c}}\else \c{c}\fi{}ois}}, \bibinfo {author}
  {\bibfnamefont {M.}~\bibnamefont {Songvilay}}, \bibinfo {author}
  {\bibfnamefont {J.}~\bibnamefont {Robert}}, \bibinfo {author} {\bibfnamefont
  {G.}~\bibnamefont {Nataf}}, \bibinfo {author} {\bibfnamefont
  {E.}~\bibnamefont {Jordan}}, \bibinfo {author} {\bibfnamefont
  {L.}~\bibnamefont {Chaix}}, \bibinfo {author} {\bibfnamefont {C.~V.}\
  \bibnamefont {Colin}}, \bibinfo {author} {\bibfnamefont {P.}~\bibnamefont
  {Lejay}}, \bibinfo {author} {\bibfnamefont {A.}~\bibnamefont {Hadj-Azzem}},
  \bibinfo {author} {\bibfnamefont {R.}~\bibnamefont {Ballou}},\ and\ \bibinfo
  {author} {\bibfnamefont {V.}~\bibnamefont {Simonet}},\ }\href
  {https://doi.org/10.1103/PhysRevB.94.214416} {\bibfield  {journal} {\bibinfo
  {journal} {Phys. Rev. B}\ }\textbf {\bibinfo {volume} {94}},\ \bibinfo
  {pages} {214416} (\bibinfo {year} {2016})}\BibitemShut {NoStop}%
\bibitem [{\citenamefont {Bera}\ \emph {et~al.}(2017)\citenamefont {Bera},
  \citenamefont {Yusuf}, \citenamefont {Kumar},\ and\ \citenamefont
  {Ritter}}]{BeraPRB2017}%
  \BibitemOpen
  \bibfield  {author} {\bibinfo {author} {\bibfnamefont {A.~K.}\ \bibnamefont
  {Bera}}, \bibinfo {author} {\bibfnamefont {S.~M.}\ \bibnamefont {Yusuf}},
  \bibinfo {author} {\bibfnamefont {A.}~\bibnamefont {Kumar}},\ and\ \bibinfo
  {author} {\bibfnamefont {C.}~\bibnamefont {Ritter}},\ }\href
  {https://doi.org/10.1103/PhysRevB.95.094424} {\bibfield  {journal} {\bibinfo
  {journal} {Phys. Rev. B}\ }\textbf {\bibinfo {volume} {95}},\ \bibinfo
  {pages} {094424} (\bibinfo {year} {2017})}\BibitemShut {NoStop}%
\bibitem [{\citenamefont {Ye}\ \emph {et~al.}(2012)\citenamefont {Ye},
  \citenamefont {Chi}, \citenamefont {Cao}, \citenamefont {Chakoumakos},
  \citenamefont {Fernandez-Baca}, \citenamefont {Custelcean}, \citenamefont
  {Qi}, \citenamefont {Korneta},\ and\ \citenamefont {Cao}}]{YePRB2012}%
  \BibitemOpen
  \bibfield  {author} {\bibinfo {author} {\bibfnamefont {F.}~\bibnamefont
  {Ye}}, \bibinfo {author} {\bibfnamefont {S.}~\bibnamefont {Chi}}, \bibinfo
  {author} {\bibfnamefont {H.}~\bibnamefont {Cao}}, \bibinfo {author}
  {\bibfnamefont {B.~C.}\ \bibnamefont {Chakoumakos}}, \bibinfo {author}
  {\bibfnamefont {J.~A.}\ \bibnamefont {Fernandez-Baca}}, \bibinfo {author}
  {\bibfnamefont {R.}~\bibnamefont {Custelcean}}, \bibinfo {author}
  {\bibfnamefont {T.~F.}\ \bibnamefont {Qi}}, \bibinfo {author} {\bibfnamefont
  {O.~B.}\ \bibnamefont {Korneta}},\ and\ \bibinfo {author} {\bibfnamefont
  {G.}~\bibnamefont {Cao}},\ }\href
  {https://doi.org/10.1103/PhysRevB.85.180403} {\bibfield  {journal} {\bibinfo
  {journal} {Phys. Rev. B}\ }\textbf {\bibinfo {volume} {85}},\ \bibinfo
  {pages} {180403(R)} (\bibinfo {year} {2012})}\BibitemShut {NoStop}%
\bibitem [{\citenamefont {Sears}\ \emph {et~al.}(2015)\citenamefont {Sears},
  \citenamefont {Songvilay}, \citenamefont {Plumb}, \citenamefont {Clancy},
  \citenamefont {Qiu}, \citenamefont {Zhao}, \citenamefont {Parshall},\ and\
  \citenamefont {Kim}}]{SearsPRB2015}%
  \BibitemOpen
  \bibfield  {author} {\bibinfo {author} {\bibfnamefont {J.~A.}\ \bibnamefont
  {Sears}}, \bibinfo {author} {\bibfnamefont {M.}~\bibnamefont {Songvilay}},
  \bibinfo {author} {\bibfnamefont {K.~W.}\ \bibnamefont {Plumb}}, \bibinfo
  {author} {\bibfnamefont {J.~P.}\ \bibnamefont {Clancy}}, \bibinfo {author}
  {\bibfnamefont {Y.}~\bibnamefont {Qiu}}, \bibinfo {author} {\bibfnamefont
  {Y.}~\bibnamefont {Zhao}}, \bibinfo {author} {\bibfnamefont {D.}~\bibnamefont
  {Parshall}},\ and\ \bibinfo {author} {\bibfnamefont {Y.-J.}\ \bibnamefont
  {Kim}},\ }\href {https://doi.org/10.1103/PhysRevB.91.144420} {\bibfield
  {journal} {\bibinfo  {journal} {Phys. Rev. B}\ }\textbf {\bibinfo {volume}
  {91}},\ \bibinfo {pages} {144420} (\bibinfo {year} {2015})}\BibitemShut
  {NoStop}%
\bibitem [{\citenamefont {Cao}\ \emph {et~al.}(2016)\citenamefont {Cao},
  \citenamefont {Banerjee}, \citenamefont {Yan}, \citenamefont {Bridges},
  \citenamefont {Lumsden}, \citenamefont {Mandrus}, \citenamefont {Tennant},
  \citenamefont {Chakoumakos},\ and\ \citenamefont {Nagler}}]{CaoPRB2016}%
  \BibitemOpen
  \bibfield  {author} {\bibinfo {author} {\bibfnamefont {H.~B.}\ \bibnamefont
  {Cao}}, \bibinfo {author} {\bibfnamefont {A.}~\bibnamefont {Banerjee}},
  \bibinfo {author} {\bibfnamefont {J.-Q.}\ \bibnamefont {Yan}}, \bibinfo
  {author} {\bibfnamefont {C.~A.}\ \bibnamefont {Bridges}}, \bibinfo {author}
  {\bibfnamefont {M.~D.}\ \bibnamefont {Lumsden}}, \bibinfo {author}
  {\bibfnamefont {D.~G.}\ \bibnamefont {Mandrus}}, \bibinfo {author}
  {\bibfnamefont {D.~A.}\ \bibnamefont {Tennant}}, \bibinfo {author}
  {\bibfnamefont {B.~C.}\ \bibnamefont {Chakoumakos}},\ and\ \bibinfo {author}
  {\bibfnamefont {S.~E.}\ \bibnamefont {Nagler}},\ }\href
  {https://doi.org/10.1103/PhysRevB.93.134423} {\bibfield  {journal} {\bibinfo
  {journal} {Phys. Rev. B}\ }\textbf {\bibinfo {volume} {93}},\ \bibinfo
  {pages} {134423} (\bibinfo {year} {2016})}\BibitemShut {NoStop}%
\bibitem [{SM()}]{SM}%
  \BibitemOpen
  \bibinfo {note} {See Supplemental Material at xxx for methods and additional
  data and analyses.}\BibitemShut {Stop}%
\bibitem [{\citenamefont {Balz}\ \emph {et~al.}(2021)\citenamefont {Balz},
  \citenamefont {Janssen}, \citenamefont {Lampen-Kelley}, \citenamefont
  {Banerjee}, \citenamefont {Liu}, \citenamefont {Yan}, \citenamefont
  {Mandrus}, \citenamefont {Vojta},\ and\ \citenamefont
  {Nagler}}]{BalzPRB2021}%
  \BibitemOpen
  \bibfield  {author} {\bibinfo {author} {\bibfnamefont {C.}~\bibnamefont
  {Balz}}, \bibinfo {author} {\bibfnamefont {L.}~\bibnamefont {Janssen}},
  \bibinfo {author} {\bibfnamefont {P.}~\bibnamefont {Lampen-Kelley}}, \bibinfo
  {author} {\bibfnamefont {A.}~\bibnamefont {Banerjee}}, \bibinfo {author}
  {\bibfnamefont {Y.~H.}\ \bibnamefont {Liu}}, \bibinfo {author} {\bibfnamefont
  {J.-Q.}\ \bibnamefont {Yan}}, \bibinfo {author} {\bibfnamefont {D.~G.}\
  \bibnamefont {Mandrus}}, \bibinfo {author} {\bibfnamefont {M.}~\bibnamefont
  {Vojta}},\ and\ \bibinfo {author} {\bibfnamefont {S.~E.}\ \bibnamefont
  {Nagler}},\ }\href {https://doi.org/10.1103/PhysRevB.103.174417} {\bibfield
  {journal} {\bibinfo  {journal} {Phys. Rev. B}\ }\textbf {\bibinfo {volume}
  {103}},\ \bibinfo {pages} {174417} (\bibinfo {year} {2021})}\BibitemShut
  {NoStop}%
\bibitem [{\citenamefont {Zhang}\ \emph {et~al.}(2022)\citenamefont {Zhang},
  \citenamefont {Xu}, \citenamefont {Halloran}, \citenamefont {Zhong},
  \citenamefont {Broholm}, \citenamefont {Cava}, \citenamefont {Drichko},\ and\
  \citenamefont {Armitage}}]{ZhangNM2022}%
  \BibitemOpen
  \bibfield  {author} {\bibinfo {author} {\bibfnamefont {X.}~\bibnamefont
  {Zhang}}, \bibinfo {author} {\bibfnamefont {Y.}~\bibnamefont {Xu}}, \bibinfo
  {author} {\bibfnamefont {T.}~\bibnamefont {Halloran}}, \bibinfo {author}
  {\bibfnamefont {R.}~\bibnamefont {Zhong}}, \bibinfo {author} {\bibfnamefont
  {C.}~\bibnamefont {Broholm}}, \bibinfo {author} {\bibfnamefont
  {R.}~\bibnamefont {Cava}}, \bibinfo {author} {\bibfnamefont {N.}~\bibnamefont
  {Drichko}},\ and\ \bibinfo {author} {\bibfnamefont {N.}~\bibnamefont
  {Armitage}},\ }\href {https://doi.org/10.1038/s41563-022-01403-1} {\bibfield
  {journal} {\bibinfo  {journal} {Nat. Mater.}\ } (\bibinfo {year}
  {2022})}\BibitemShut {NoStop}%
\bibitem [{\citenamefont {Spitz}\ \emph {et~al.}(2022)\citenamefont {Spitz},
  \citenamefont {Nomoto}, \citenamefont {Kitou}, \citenamefont {Nakao},
  \citenamefont {Kikkawa}, \citenamefont {Francoual}, \citenamefont {Taguchi},
  \citenamefont {Arita}, \citenamefont {Tokura}, \citenamefont {Arima},\ and\
  \citenamefont {Hirschberger}}]{SpitzJACS2022}%
  \BibitemOpen
  \bibfield  {author} {\bibinfo {author} {\bibfnamefont {L.}~\bibnamefont
  {Spitz}}, \bibinfo {author} {\bibfnamefont {T.}~\bibnamefont {Nomoto}},
  \bibinfo {author} {\bibfnamefont {S.}~\bibnamefont {Kitou}}, \bibinfo
  {author} {\bibfnamefont {H.}~\bibnamefont {Nakao}}, \bibinfo {author}
  {\bibfnamefont {A.}~\bibnamefont {Kikkawa}}, \bibinfo {author} {\bibfnamefont
  {S.}~\bibnamefont {Francoual}}, \bibinfo {author} {\bibfnamefont
  {Y.}~\bibnamefont {Taguchi}}, \bibinfo {author} {\bibfnamefont
  {R.}~\bibnamefont {Arita}}, \bibinfo {author} {\bibfnamefont
  {Y.}~\bibnamefont {Tokura}}, \bibinfo {author} {\bibfnamefont {T.-h.}\
  \bibnamefont {Arima}},\ and\ \bibinfo {author} {\bibfnamefont
  {M.}~\bibnamefont {Hirschberger}},\ }\href
  {https://doi.org/10.1021/jacs.2c04995} {\bibfield  {journal} {\bibinfo
  {journal} {J. Am. Chem. Soc.}\ }\textbf {\bibinfo {volume} {144}},\ \bibinfo
  {pages} {16866} (\bibinfo {year} {2022})}\BibitemShut {NoStop}%
\bibitem [{\citenamefont {Kr\"{u}ger}\ \emph {et~al.}()\citenamefont
  {Kr\"{u}ger}, \citenamefont {Chen}, \citenamefont {Jin}, \citenamefont {Li},\
  and\ \citenamefont {Janssen}}]{KruegerPreprint}%
  \BibitemOpen
  \bibfield  {author} {\bibinfo {author} {\bibfnamefont {W.~G.~F.}\
  \bibnamefont {Kr\"{u}ger}}, \bibinfo {author} {\bibfnamefont
  {W.}~\bibnamefont {Chen}}, \bibinfo {author} {\bibfnamefont {X.}~\bibnamefont
  {Jin}}, \bibinfo {author} {\bibfnamefont {Y.}~\bibnamefont {Li}},\ and\
  \bibinfo {author} {\bibfnamefont {L.}~\bibnamefont {Janssen}},\ }\bibinfo
  {note} {to appear.}\BibitemShut {Stop}%
\bibitem [{\citenamefont {Jin}\ \emph {et~al.}(2021)\citenamefont {Jin},
  \citenamefont {He}, \citenamefont {Balz}, \citenamefont {Yao}, \citenamefont
  {Li},\ and\ \citenamefont {Li}}]{LET2021}%
  \BibitemOpen
  \bibfield  {author} {\bibinfo {author} {\bibfnamefont {X.}~\bibnamefont
  {Jin}}, \bibinfo {author} {\bibfnamefont {G.}~\bibnamefont {He}}, \bibinfo
  {author} {\bibfnamefont {C.}~\bibnamefont {Balz}}, \bibinfo {author}
  {\bibfnamefont {W.}~\bibnamefont {Yao}}, \bibinfo {author} {\bibfnamefont
  {Y.}~\bibnamefont {Li}},\ and\ \bibinfo {author} {\bibfnamefont
  {X.}~\bibnamefont {Li}}} (\bibinfo {year} {2021}),\ \bibinfo {note}
  {\MakeUppercase{I}nvestigation of spin excitations in Na2Co2TeO6 single
  crystals, STFC ISIS Neutron and Muon Source,
  \url{https://doi.org/10.5286/ISIS.E.RB2010025}}\BibitemShut {NoStop}%
\bibitem [{\citenamefont {Lynn}\ \emph {et~al.}(2012)\citenamefont {Lynn},
  \citenamefont {Chen}, \citenamefont {Chang}, \citenamefont {Zhao},
  \citenamefont {Chi}, \citenamefont {Ratcliff}, \citenamefont {Ueland},\ and\
  \citenamefont {Erwin}}]{LynnNIST2012}%
  \BibitemOpen
  \bibfield  {author} {\bibinfo {author} {\bibfnamefont {J.}~\bibnamefont
  {Lynn}}, \bibinfo {author} {\bibfnamefont {Y.}~\bibnamefont {Chen}}, \bibinfo
  {author} {\bibfnamefont {S.}~\bibnamefont {Chang}}, \bibinfo {author}
  {\bibfnamefont {Y.}~\bibnamefont {Zhao}}, \bibinfo {author} {\bibfnamefont
  {S.}~\bibnamefont {Chi}}, \bibinfo {author} {\bibfnamefont {W.}~\bibnamefont
  {Ratcliff}}, \bibinfo {author} {\bibfnamefont {B.~G.}\ \bibnamefont
  {Ueland}},\ and\ \bibinfo {author} {\bibfnamefont {R.~W.}\ \bibnamefont
  {Erwin}},\ }\href {https://www.ncbi.nlm.nih.gov/pmc/articles/PMC4553874/}
  {\bibfield  {journal} {\bibinfo  {journal} {Journal of research of the
  National Institute of Standards and Technology}\ }\textbf {\bibinfo {volume}
  {117}},\ \bibinfo {pages} {61} (\bibinfo {year} {2012})}\BibitemShut
  {NoStop}%
\bibitem [{\citenamefont {Rodriguez}\ \emph {et~al.}(2008)\citenamefont
  {Rodriguez}, \citenamefont {Adler}, \citenamefont {Brand}, \citenamefont
  {Broholm}, \citenamefont {Cook}, \citenamefont {Brocker}, \citenamefont
  {Hammond}, \citenamefont {Huang}, \citenamefont {Hundertmark}, \citenamefont
  {Lynn}, \citenamefont {Maliszewskyj}, \citenamefont {Moyer}, \citenamefont
  {Orndorff}, \citenamefont {Pierce}, \citenamefont {Pike}, \citenamefont
  {Scharfstein}, \citenamefont {Smee},\ and\ \citenamefont
  {Vilaseca}}]{RodriguezMST2008}%
  \BibitemOpen
  \bibfield  {author} {\bibinfo {author} {\bibfnamefont {J.~A.}\ \bibnamefont
  {Rodriguez}}, \bibinfo {author} {\bibfnamefont {D.~M.}\ \bibnamefont
  {Adler}}, \bibinfo {author} {\bibfnamefont {P.~C.}\ \bibnamefont {Brand}},
  \bibinfo {author} {\bibfnamefont {C.}~\bibnamefont {Broholm}}, \bibinfo
  {author} {\bibfnamefont {J.~C.}\ \bibnamefont {Cook}}, \bibinfo {author}
  {\bibfnamefont {C.}~\bibnamefont {Brocker}}, \bibinfo {author} {\bibfnamefont
  {R.}~\bibnamefont {Hammond}}, \bibinfo {author} {\bibfnamefont
  {Z.}~\bibnamefont {Huang}}, \bibinfo {author} {\bibfnamefont
  {P.}~\bibnamefont {Hundertmark}}, \bibinfo {author} {\bibfnamefont {J.~W.}\
  \bibnamefont {Lynn}}, \bibinfo {author} {\bibfnamefont {N.~C.}\ \bibnamefont
  {Maliszewskyj}}, \bibinfo {author} {\bibfnamefont {J.}~\bibnamefont {Moyer}},
  \bibinfo {author} {\bibfnamefont {J.}~\bibnamefont {Orndorff}}, \bibinfo
  {author} {\bibfnamefont {D.}~\bibnamefont {Pierce}}, \bibinfo {author}
  {\bibfnamefont {T.~D.}\ \bibnamefont {Pike}}, \bibinfo {author}
  {\bibfnamefont {G.}~\bibnamefont {Scharfstein}}, \bibinfo {author}
  {\bibfnamefont {S.~A.}\ \bibnamefont {Smee}},\ and\ \bibinfo {author}
  {\bibfnamefont {R.}~\bibnamefont {Vilaseca}},\ }\href
  {https://doi.org/10.1088/0957-0233/19/3/034023} {\bibfield  {journal}
  {\bibinfo  {journal} {Measurement Science and Technology}\ }\textbf {\bibinfo
  {volume} {19}},\ \bibinfo {pages} {034023} (\bibinfo {year}
  {2008})}\BibitemShut {NoStop}%
\bibitem [{\citenamefont {Bewley}\ \emph {et~al.}(2011)\citenamefont {Bewley},
  \citenamefont {Taylor},\ and\ \citenamefont {Bennington}}]{BewleyLET2011}%
  \BibitemOpen
  \bibfield  {author} {\bibinfo {author} {\bibfnamefont {R.}~\bibnamefont
  {Bewley}}, \bibinfo {author} {\bibfnamefont {J.}~\bibnamefont {Taylor}},\
  and\ \bibinfo {author} {\bibfnamefont {S.}~\bibnamefont {Bennington}},\
  }\href {https://doi.org/10.1016/j.nima.2011.01.173} {\bibfield  {journal}
  {\bibinfo  {journal} {Nuclear Instruments and Methods in Physics Research
  Section A: Accelerators, Spectrometers, Detectors and Associated Equipment}\
  }\textbf {\bibinfo {volume} {637}},\ \bibinfo {pages} {128} (\bibinfo {year}
  {2011})}\BibitemShut {NoStop}%
\bibitem [{\citenamefont {Azuah}\ \emph {et~al.}(2009)\citenamefont {Azuah},
  \citenamefont {Kneller}, \citenamefont {Qiu}, \citenamefont
  {Tregenna-Piggott}, \citenamefont {Brown}, \citenamefont {Copley},\ and\
  \citenamefont {Dimeo}}]{AzuahNIST2009}%
  \BibitemOpen
  \bibfield  {author} {\bibinfo {author} {\bibfnamefont {R.~T.}\ \bibnamefont
  {Azuah}}, \bibinfo {author} {\bibfnamefont {L.~R.}\ \bibnamefont {Kneller}},
  \bibinfo {author} {\bibfnamefont {Y.}~\bibnamefont {Qiu}}, \bibinfo {author}
  {\bibfnamefont {P.~L.}\ \bibnamefont {Tregenna-Piggott}}, \bibinfo {author}
  {\bibfnamefont {C.~M.}\ \bibnamefont {Brown}}, \bibinfo {author}
  {\bibfnamefont {J.~R.}\ \bibnamefont {Copley}},\ and\ \bibinfo {author}
  {\bibfnamefont {R.~M.}\ \bibnamefont {Dimeo}},\ }\href
  {https://www.ncbi.nlm.nih.gov/pmc/articles/PMC4646530/} {\bibfield  {journal}
  {\bibinfo  {journal} {Journal of Research of the National Institute of
  Standards and Technology}\ }\textbf {\bibinfo {volume} {114}},\ \bibinfo
  {pages} {341} (\bibinfo {year} {2009})}\BibitemShut {NoStop}%
\bibitem [{\citenamefont {Ewings}\ \emph {et~al.}(2016)\citenamefont {Ewings},
  \citenamefont {Buts}, \citenamefont {Le}, \citenamefont {Van~Duijn},
  \citenamefont {Bustinduy},\ and\ \citenamefont {Perring}}]{EwingsHorace2016}%
  \BibitemOpen
  \bibfield  {author} {\bibinfo {author} {\bibfnamefont {R.}~\bibnamefont
  {Ewings}}, \bibinfo {author} {\bibfnamefont {A.}~\bibnamefont {Buts}},
  \bibinfo {author} {\bibfnamefont {M.}~\bibnamefont {Le}}, \bibinfo {author}
  {\bibfnamefont {J.}~\bibnamefont {Van~Duijn}}, \bibinfo {author}
  {\bibfnamefont {I.}~\bibnamefont {Bustinduy}},\ and\ \bibinfo {author}
  {\bibfnamefont {T.}~\bibnamefont {Perring}},\ }\href
  {https://doi.org/10.1016/j.nima.2016.07.036} {\bibfield  {journal} {\bibinfo
  {journal} {Nuclear Instruments and Methods in Physics Research Section A:
  Accelerators, Spectrometers, Detectors and Associated Equipment}\ }\textbf
  {\bibinfo {volume} {834}},\ \bibinfo {pages} {132} (\bibinfo {year}
  {2016})}\BibitemShut {NoStop}%
\bibitem [{\citenamefont {Shirane}\ \emph {et~al.}(2002)\citenamefont
  {Shirane}, \citenamefont {Shapiro},\ and\ \citenamefont
  {Tranquada}}]{Shirane2002}%
  \BibitemOpen
  \bibfield  {author} {\bibinfo {author} {\bibfnamefont {G.}~\bibnamefont
  {Shirane}}, \bibinfo {author} {\bibfnamefont {S.~M.}\ \bibnamefont
  {Shapiro}},\ and\ \bibinfo {author} {\bibfnamefont {J.~M.}\ \bibnamefont
  {Tranquada}},\ }\href@noop {} {\emph {\bibinfo {title} {Neutron scattering
  with a triple-axis spectrometer: basic techniques}}}\ (\bibinfo  {publisher}
  {Cambridge University Press},\ \bibinfo {year} {2002})\BibitemShut {NoStop}%
\end{thebibliography}%

\pagebreak
\pagebreak

\widetext
\begin{center}
\textbf{\large Supplemental Material for ``Magnetic ground state of the Kitaev \ch{Na_2Co_2TeO_6} spin liquid candidate''}
\end{center}

\setcounter{equation}{0}
\setcounter{table}{0}
\setcounter{page}{1}
\makeatletter
\renewcommand{\theequation}{S\arabic{equation}}
\renewcommand{\thefigure}{S\arabic{figure}}

\section{Single crystals and neutron diffraction experiments}
\begin{figure}[h]
		\centering{\includegraphics[width=0.48\textwidth]{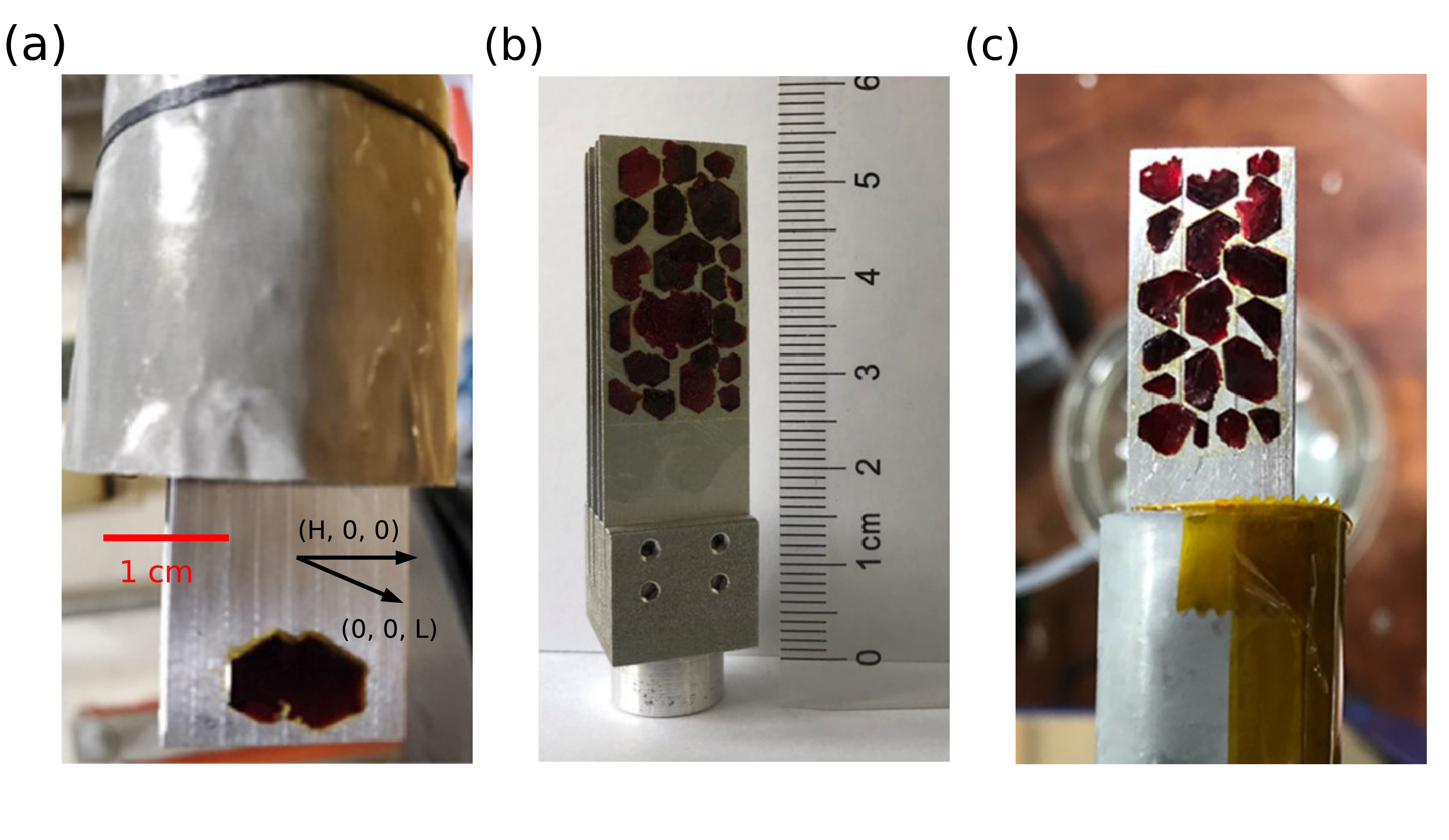}}
		\caption{(a)-(c) Single crystal samples used in BT-7, MACS, and LET experiments, respectively.}
		\label{figS1}
	\end{figure}

	Samples used in three neutron diffraction experiments are shown in Fig. \ref{figS1}, which were prepared with the same method as in \cite{YaoPRB2020} and \cite{YaoPRL2022}. According to previous thermodynamic and diffraction studies on a large number of crystals, we had confirmed that \ch{Na_2Co_2TeO_6} only has one major magnetic ordering transition at $\sim26.5$ K \cite{YaoPRB2020,YaoPRL2022}. This suggests stacking fault is not severe for samples prepared with our method, in comparison with $\alpha$-\ch{RuCl_3} \cite{CaoPRB2016}.
	
	Neutron diffraction measurements were performed in the BT-7 triple-axis spectrometer \cite{LynnNIST2012} with an incident neutron energy $E_i$ = 14.7 meV and the multiaxis crystal spectrometer (MACS) \cite{RodriguezMST2008} with $E_i$ = 5.0 meV, both at NIST Center for Neutron Research (NCNR), USA. Additional neutron diffraction data come from the experiment performed in the time-of-flight spectrometer LET \cite{BewleyLET2011} with $E_i$ = 12.0 meV at the ISIS Spallation Neutron Source, the Rutherford Appleton Laboratory, UK. One single crystal with mass of about 30 mg was used in the BT-7 experiment. Coaligned single crystal arrays of about 800 mg and 750 mg were used in the MACS and LET experiments, respectively. The space group $P6_322$ is used with $a=b=5.28$ $\rm \AA$, $c=11.22$ $\rm \AA$ \cite{BeraPRB2017}. Wave vector is defined as $\textbf{Q} = H\textbf{a*}+K\textbf{b*}+L\textbf{c*}$, with $a^*=b^*=\frac{4\pi}{\sqrt3 a}$, $c^*=\frac{2\pi}{c}$. In all these experiments, the single crystals were aligned with the ($H$, 0, $L$) plane horizontal (Fig. 1 inset and Fig. \ref{figS1}). Magnetic field (up to 10 T in BT-7 and MACS, and up to 8.8 T in LET) was applied vertically, which is parallel to the two-dimensional honeycomb lattice (Fig. 1 inset). Data reductions were performed with DAVE \cite{AzuahNIST2009} for BT-7 and MACS data, and Horace \cite{EwingsHorace2016} for LET data.
	
	In the main text, the data of Fig. 1(c), Fig. 2(a) and (b), and Fig. 3(e) are from BT-7; the data of Fig. 3(a)-(d) are from MACS; the data of Fig. 2(c) are from LET.
	
	\section{Additional Field Dependence Data}
\begin{figure}
		\centering{\includegraphics[width=0.5\textwidth]{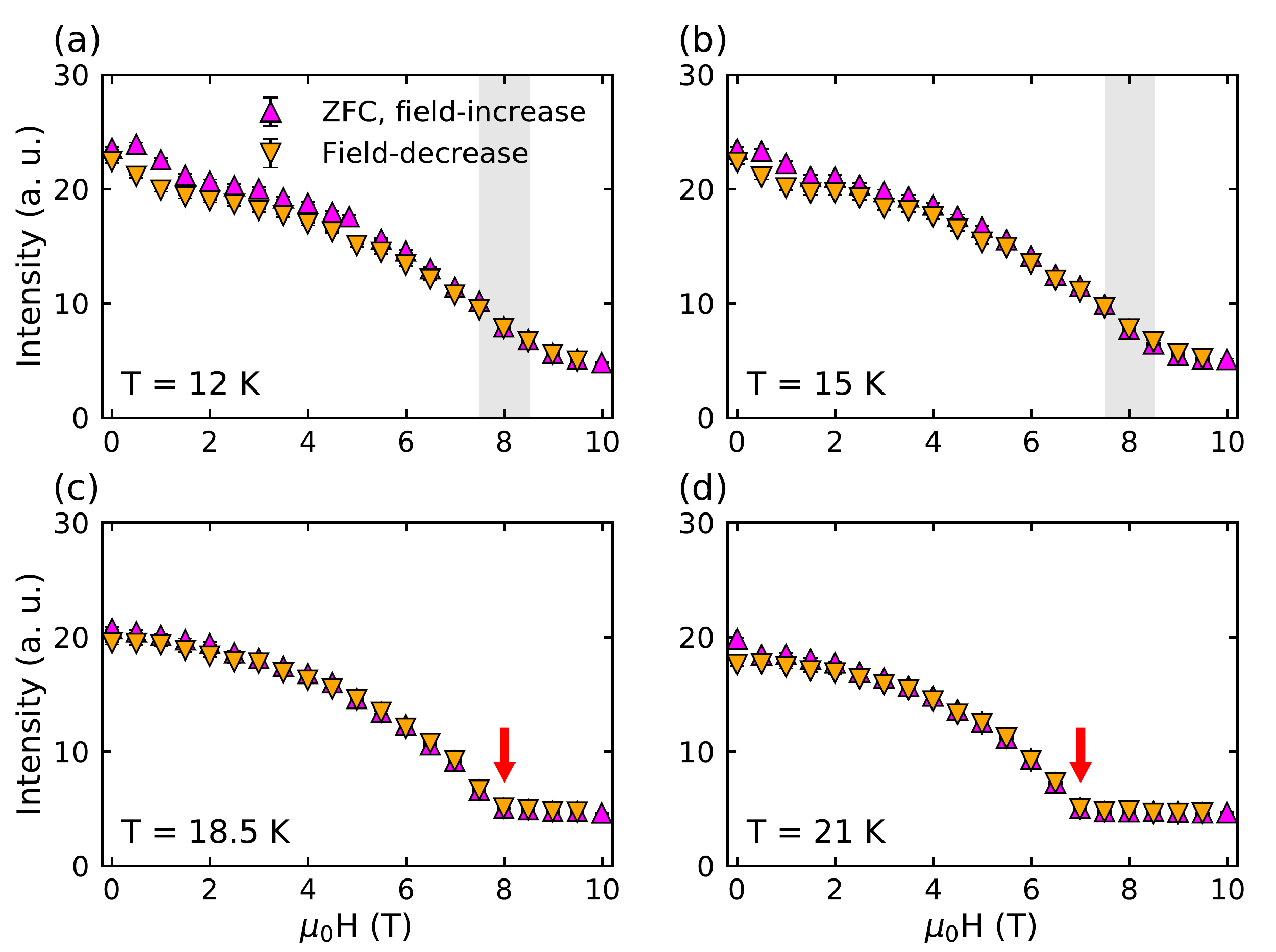}}
		\caption{(a) - (d) Field dependence of the intensity at (0.5, 0, 1) measured with BT-7 at 12 K, 15 K, 18.5 K, and 21 K, respectively. The shaded regions in (a) and (b) indicate the bifurcation of the field increasing and decreasing processes. The red arrows in (c) and (d) indicate the transition to a paramagnetic state.}
		\label{figS2}
	\end{figure}

	Fig. \ref{figS2} presents the field dependence of (0.5, 0, 1) at higher temperatures measured with BT-7. The bifurcation between the data of increasing and decreasing field still persists at 12 K and 15 K, with the anomaly in the field axis basically unchanged. By further increasing temperature, the two sets of data completely overlap with each other and show clear transition to a paramagnetic state at 8 T and 7 T for 18.5 K and 21 K, respectively.
	
\begin{figure}
		\centering{\includegraphics[width=0.4\textwidth]{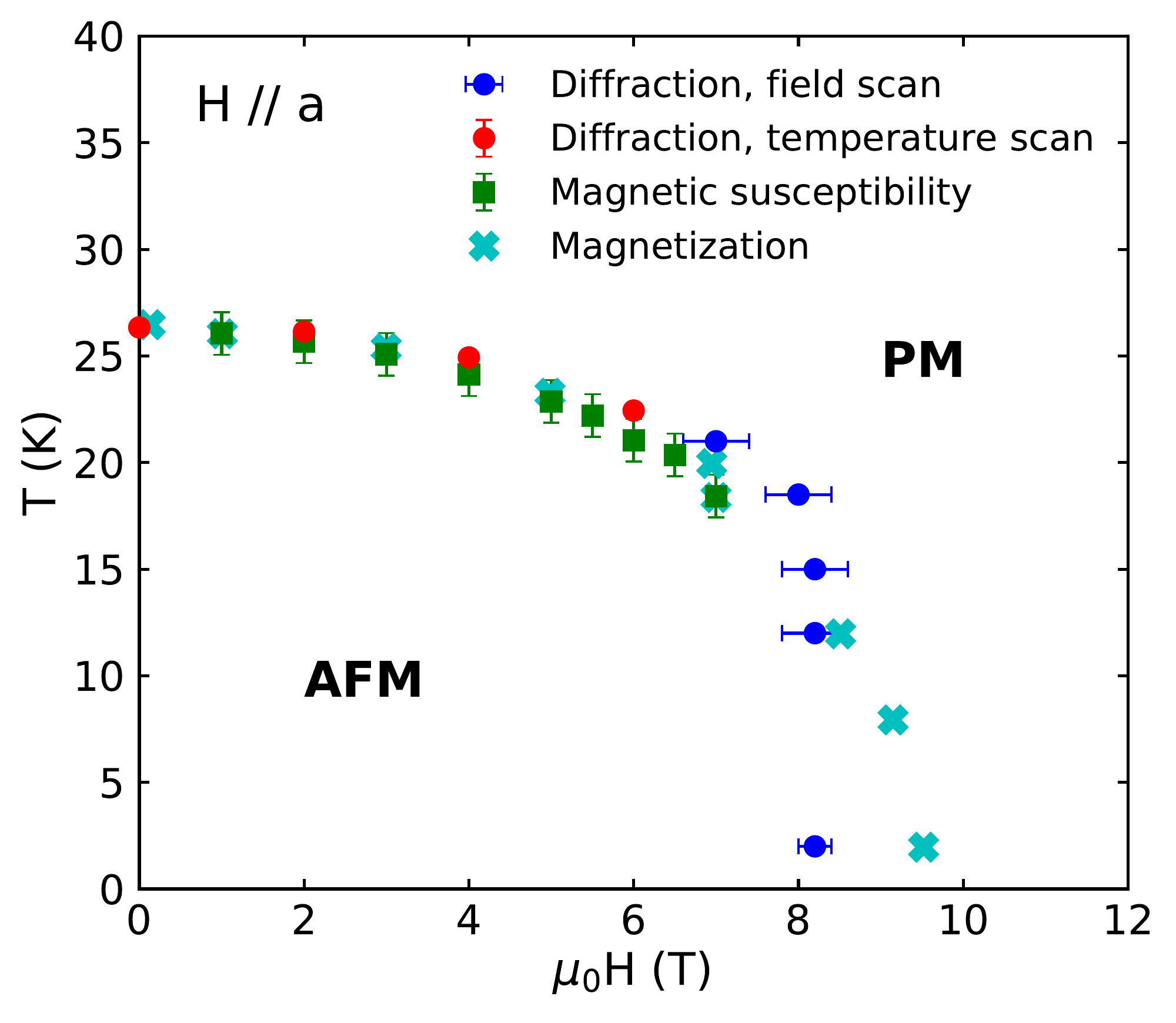}}
		\caption{Phase diagram about in-plane magnetic field ($\textbf{H}$ $\parallel$ $\textbf{a}$) and temperature. The phase boundary determined by magnetic susceptibility and magnetization is adapted from \cite{YaoPRB2020} and \cite{HongPRB2021}, respectively.}
		\label{figS3}
	\end{figure}

	Characteristic temperatures and fields for $\textbf{H}$ $\parallel$ $\textbf{a}$ obtained from this study [Fig. \ref{figS2} and Fig. 1(b) in the main text], as well as those from previous reports \cite{YaoPRB2020,HongPRB2021} are summarized in the phase diagram in Fig. \ref{figS3}. Below $\sim$8 T, the phase boundary determined from neutron diffraction is consistent with magnetic susceptibility and magnetization measurements. A new phase boundary close to 8 T is identified in this study and is related to a first-order phase transition.
	
\begin{figure}[t!]
		\centering{\includegraphics[width=0.5\textwidth]{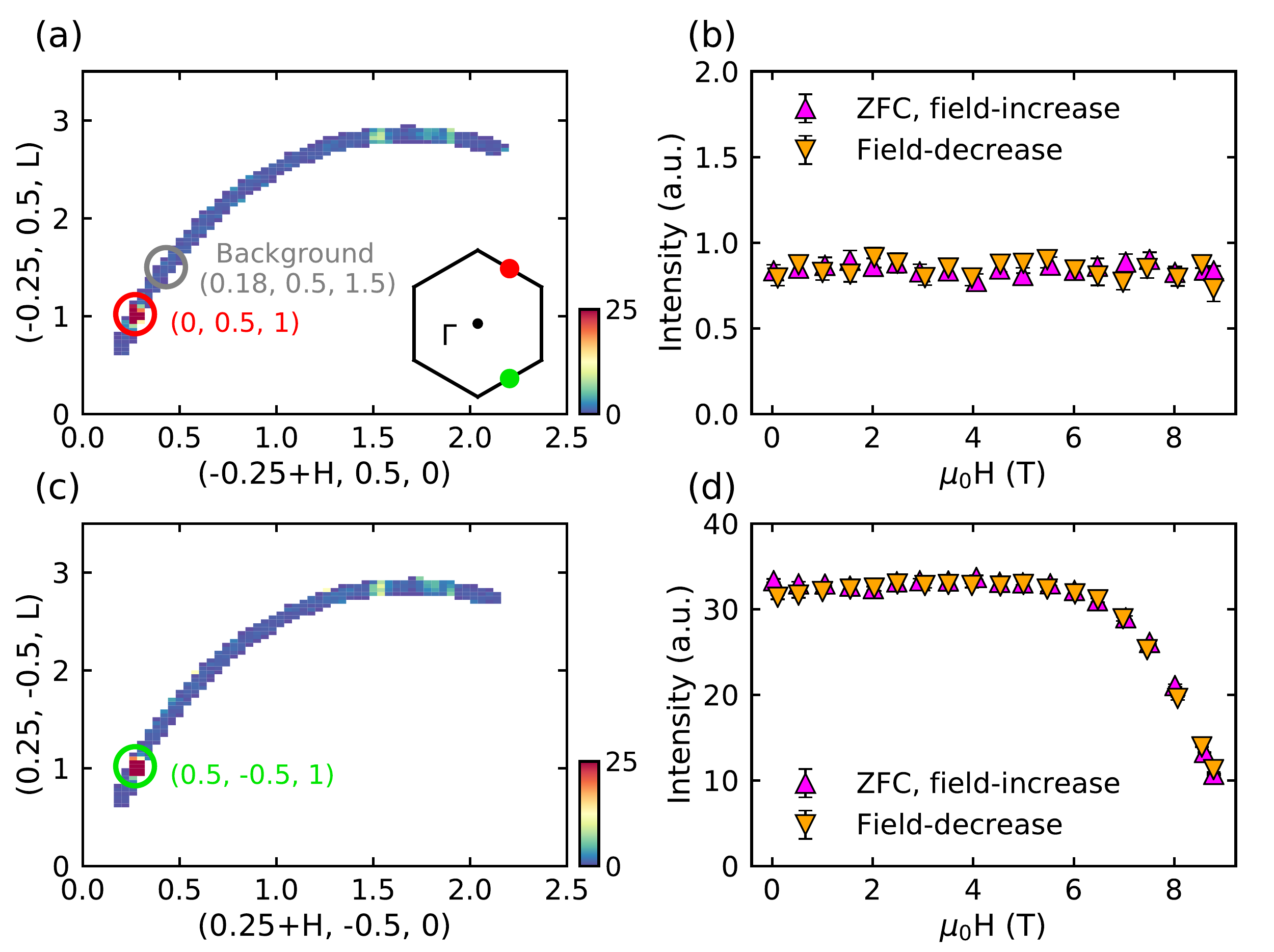}}
		\caption{(a) and (c) The reciprocal space coverage for the field dependence measurements performed with LET at 2 K. The presented data are the constant energy cuts (E = 0 meV, elastic) at 0 T in (-0.25 + $H$, 0.5, $L$) and (0.25 + $H$, -0.5, $L$) planes for (a) and (c), respectively. The inset of (a) shows the positions of two measured magnetic peaks in the Brillouin zone boundary (with $L = 1$). (b) and (d) Field dependence of the background [at (0.18, 0.5, 1.5), grey circle in (a)] and a simultaneously measurable magnetic peak [(0.5, -0.5, 1), green circle in (c)]. The field dependence of (0, 0.5, 1) [red circle in (a)] is presented in the main text [Fig. 2(c)].}
		\label{figS4}
	\end{figure}

	Fig. \ref{figS4}(a) shows reciprocal space coverage of field dependence measurements with LET. In these measurements, we fixed the sample at a specific rotation angle so that the out-of-plane magnetic Bragg peak (0, 0.5, 1) [red circle in Fig. \ref{figS4}(a)] can be covered, then we changed the magnetic field. The measurable trajectory is an arc in the (-0.25+$H$, 0.5, $L$) plane. The field dependence of (0, 0.5, 1) is presented in Fig. 2(c) in the main text. There is no field dependence for the background [grey circle in Fig. \ref{figS4}(a) and Fig. \ref{figS4}(b)], which proves our measurements are reliable. Due to the nearly symmetrical detector distribution vertically, we can simultaneously measure another equivalent magnetic Bragg peak (0.5, -0.5, 1) at the (0.25+$H$, -0.5, $L$) plane [green circle in Fig. \ref{figS4}(c)]. Fig. \ref{figS4}(d) shows its field dependence, which is similar with (0, 0.5, 1). Note that during the interval between field-increase and field-decrease measurements, we made other measurements at 8.8 T, 2 K with sample rotation. So we had corrected the sample movement for the field-decrease measurement of (0.5, -0.5, 1). Its intensity is multiplied by a factor of 1.09.
	
\begin{figure}[h!]
		\centering{\includegraphics[width=0.4\textwidth]{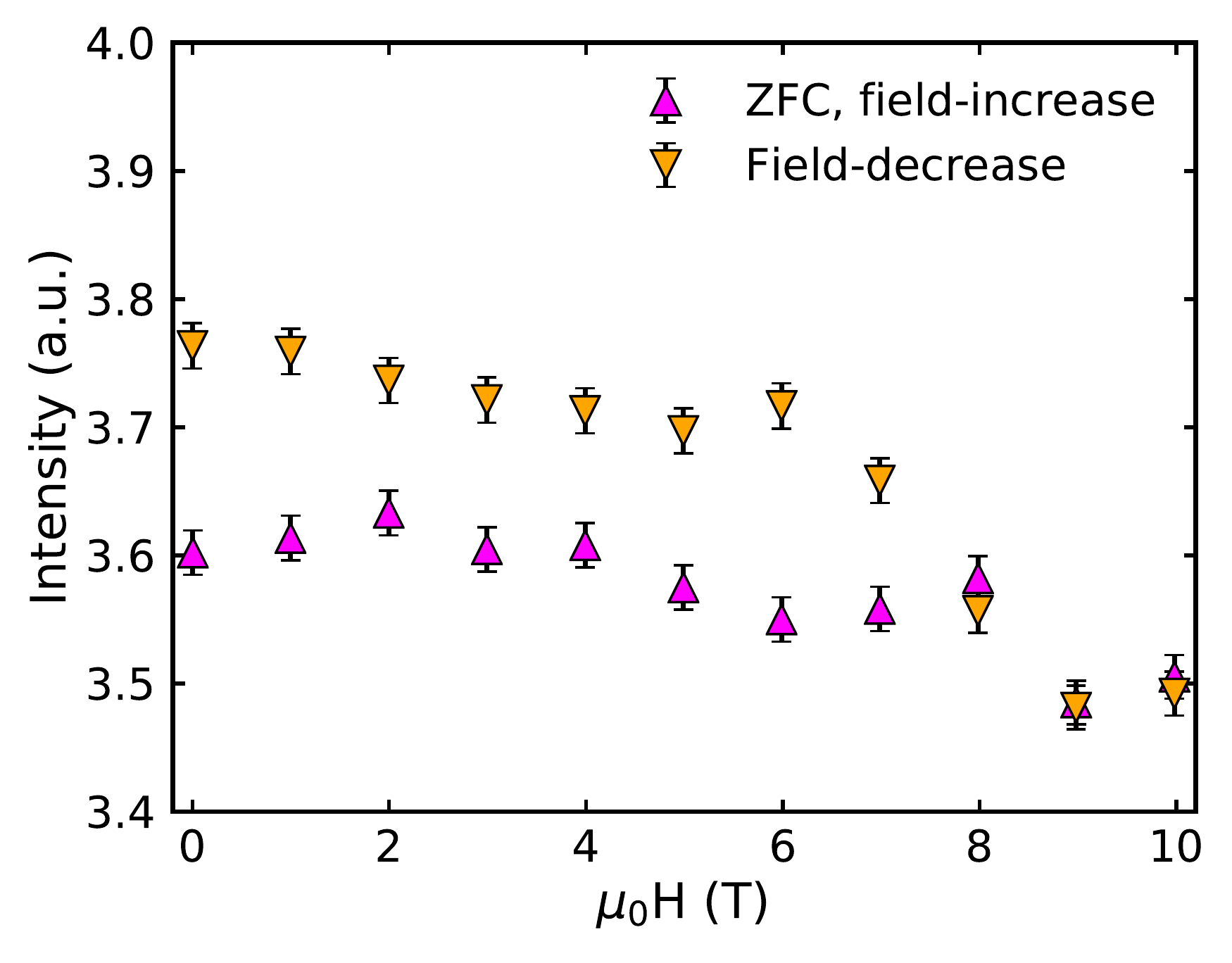}}
		\caption{Field dependence of the intensity of (0.5, 0, 1.5) at 2 K from BT-7 measurements.}
		\label{figS5}
	\end{figure}

	Fig. \ref{figS5} shows the field dependence of the intensity at (0.5, 0, 1.5), which shows opposite behavior to the intensity at (0.5, 0, 1). The intensity at (0.5, 0, 1.5) gets enhanced after field training, proving that the lost intensity at (0.5, 0, 1) goes into non-integer-$L$ positions.
	
	\section{Additional Momentum-Scan Data}
	Fig. \ref{figS6} shows $H$-scans at selected structural Bragg peaks, based on which the field-induced moment is estimated (Section IV). The non-magnetic nature of these peaks at 0 T can be confirmed by the measurement above T$\rm_N$, as showed in Fig. \ref{figS7}. Fig. \ref{figS8} compares $H$-scans at 10 T, 2 K and 0 T, 35 K. It demonstrates that the remaining intensity at 10 T, 2 K is from magnetic scattering. Fig. \ref{figS9} shows long $L$-scans from BT-7 measurements at 0 T before and after field training. The intensity at (0.5, 0, 1) goes into non-integer-$L$ positions, which is consistent with Fig. 3(d) in the main text (from MACS measurements).
	
	\section{Evaluation of Field-induced Moment}
	For a structural Bragg peak, the integrated intensity is proportional to the modulus square of the structure factor $F_{N}(\textbf{Q})$:
	\begin{equation}
		I_{N}(\textbf{Q})=A|F_{N}(\textbf{Q})|^2,
	\end{equation}
	where
	\begin{equation}
		F_{N}(\textbf{Q})=\sum_{j}b_{j}e^{i\textbf{Q}\cdot\textbf{r}_j}e^{W_j}.
	\end{equation}
	
	For a magnetic Bragg peak, the integrated intensity is proportional to the modulus square of the the magnetic structure factor $\textbf{F}_{M}(\textbf{Q})$:
	\begin{equation}
		I_{M}(\textbf{Q})=B|\textbf{F}_{M}(\textbf{Q})|^2,
	\end{equation}
	where
	\begin{equation}
		\textbf{F}_{M}(\textbf{Q})=\sum_{j}\frac{\gamma r_0}{2} g_jf_j(Q)\textbf{S}_{\perp j}e^{i\textbf{Q}\cdot\textbf{r}_j}e^{W_j}.
	\end{equation}
	$\textbf{S}_{\perp j}$ is the spin size at site $j$ that is detectable by neutrons:
	\begin{equation}
		\textbf{S}_{\perp j}=\textbf{S}_j-\hat{\textbf{Q}}(\hat{\textbf{Q}}\cdot\textbf{S}_j),
	\end{equation}
	where $\hat{\textbf{Q}}$ is the unit vector of $\textbf{Q}$ and $\textbf{S}_j$ is the spin vector at site $j$. The magnetic moment at site $j$ in Bohr magneton is $g_jS_j$. The term $\frac{\gamma r_0}{2}$ in (4) containing the classical electron radius ($r_0$) and gyromagnetic ratio ($\gamma$) acts as an effective scattering length of per Bohr magneton, which is 2.695 fm.
	
	The full integrated intensity in the magnetic fields is the sum of $I_N$ and $I_M$:
	\begin{equation}
		I(\textbf{Q})=I_N(\textbf{Q})+I_M(\textbf{Q}).
	\end{equation}
	Since we measure at low temperature, we approximate the Debye–Waller factor ($e^{W_j}$) to be unity in (2) and (4).
	
	The factors A and B in (1) and (3) contain information about the number density of the structural or magnetic unit cells. A and B also take account for influences of resolution, geometrical, and absorption factors in a real scattering process. These factors will be the same for structural and magnetic reflections at a specific position in the reciprocal space.
	
	For an estimation of the field induced moment, we assume all spins are pointing along the field (within the honeycomb plane), so that they can be fully detected by neutrons. Thus spin vectors defined in (4) are parallel with each other. This assumption is reasonable as the intensity of AFM peaks is already quite weak at 10 T (see Fig. 2 and 3 in the main text). Next we assume the moment sizes in two cobalt sites are the same (they are different but close according to previous neutron diffraction studies \cite{LefrancoisPRB2016,BeraPRB2017,SamarakoonPRB2021}).With these assumptions, we can reduce (4) into
	\begin{equation}
		\textbf{F}_{M}(\textbf{Q})=\frac{\gamma r_0}{2} g\textbf{S}f(Q)\sum_{j}e^{i\textbf{Q}\cdot\textbf{r}_j},
	\end{equation}
	where $g\textbf{S}$ is the moment size (in $\mu_B$) along the field and is to be calculated. $f(Q)$ is the magnetic form factor of \ch{Co^{2+}} ion. Therefore, for a specific position [\textit{e.g.} (1, 0, 1)] $A = B$ in (1) and (3). $F_N(\textbf{Q})$ can be calculated directly based on reported crystal structure and also the summation in (7). Finally, we can solve (1), (3), and (6) to get the moment size in (7).
	
\begin{figure}[h!]
		\centering{\includegraphics[width=0.55\textwidth]{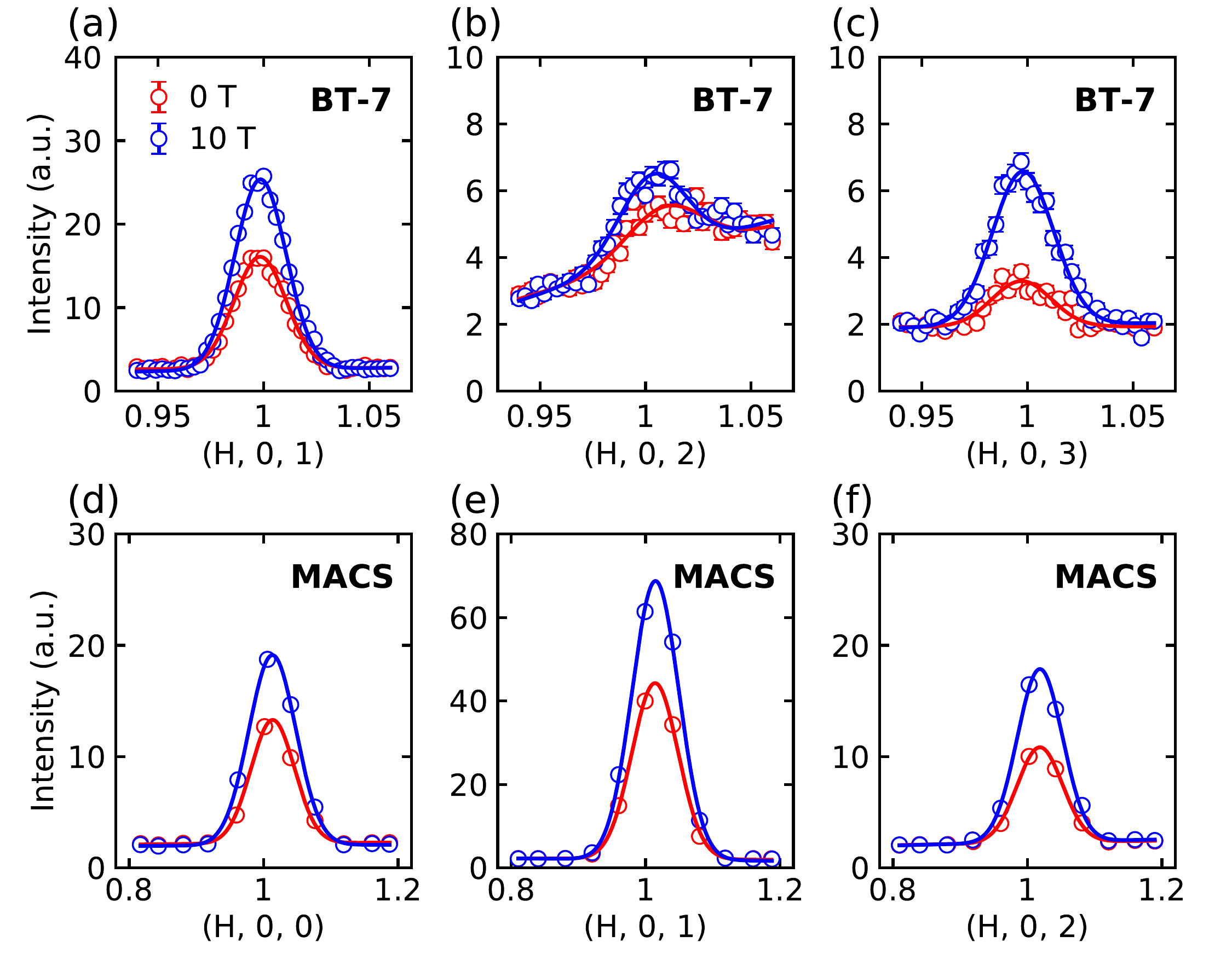}}
		\caption{(a)-(c) $H$-scans through (1, 0, 1), (1, 0, 2), and (1, 0, 3) at 0 T (red) and 10 T (blue) from BT-7 measurements. (d)-(f) $H$-scans through (1, 0, 0), (1, 0, 1), and (1, 0, 2) at 0 T (red) and 10 T (blue) from MACS measurements. The solid lines are fits with Gaussian profiles.}
		\label{figS6}
	\end{figure}

\begin{figure}
		\centering{\includegraphics[width=0.35\textwidth]{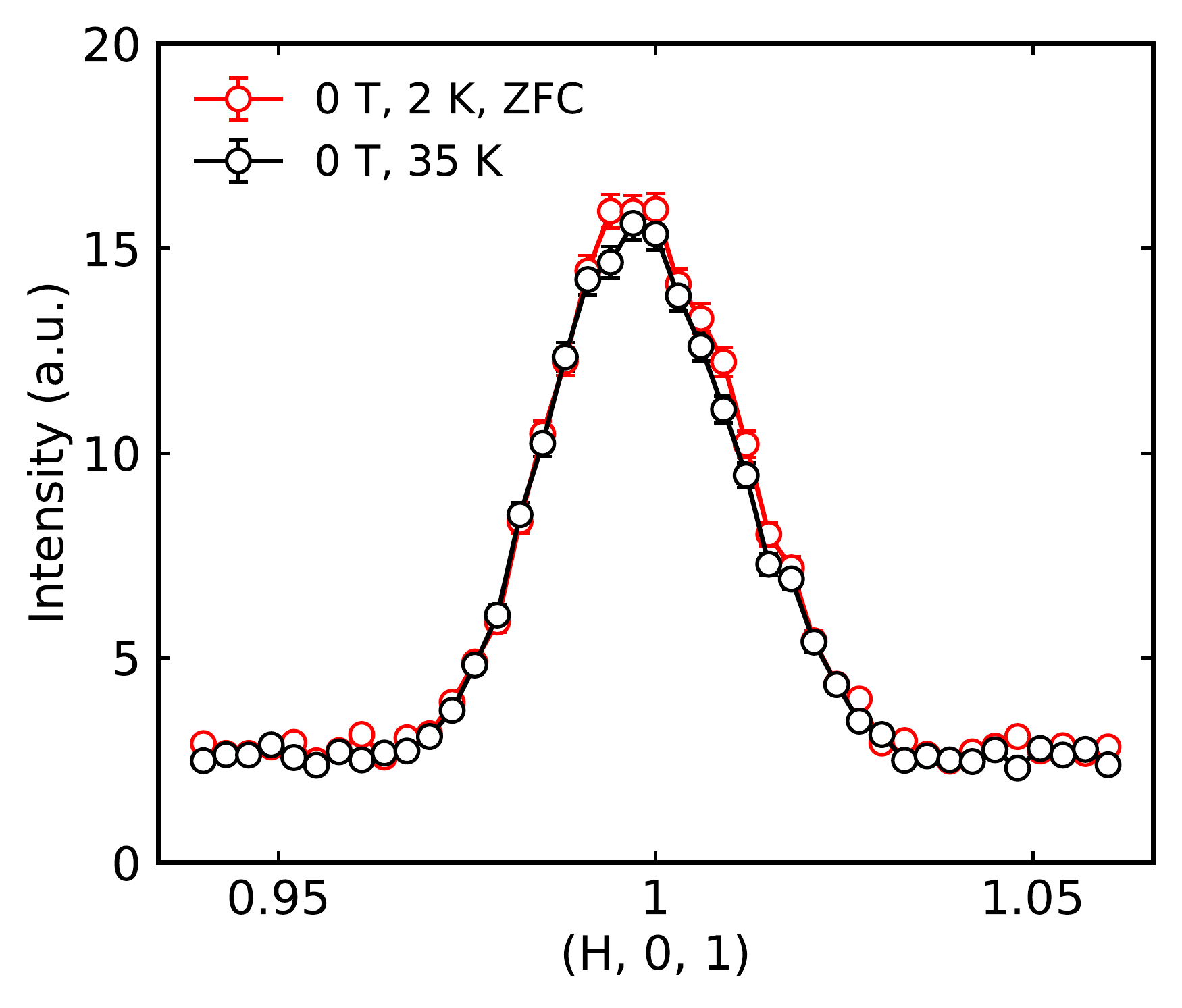}}
		\caption{$H$-scans through (1, 0, 1) at 0 T, 2 K and 0 T, 35 K from BT-7 measurements.}
		\label{figS7}
	\end{figure}

\begin{figure}
		\centering{\includegraphics[width=0.8\textwidth]{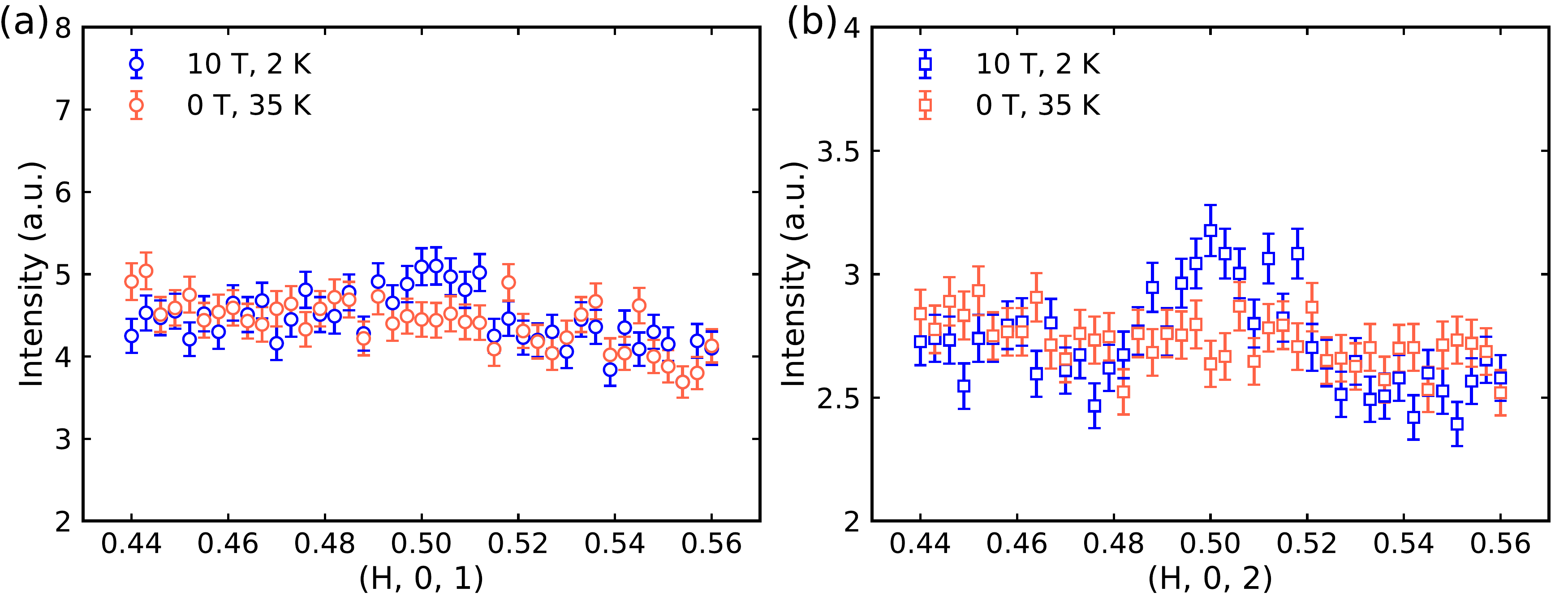}}
		\caption{$H$-scans through (0.5, 0, 1) (a) and (0.5, 0, 2) (b) at 10 T, 2 K and 0 T, 35 K from BT-7 measurements.}
		\label{figS8}
	\end{figure}

\begin{figure}[h!]
		\centering{\includegraphics[width=0.8\textwidth]{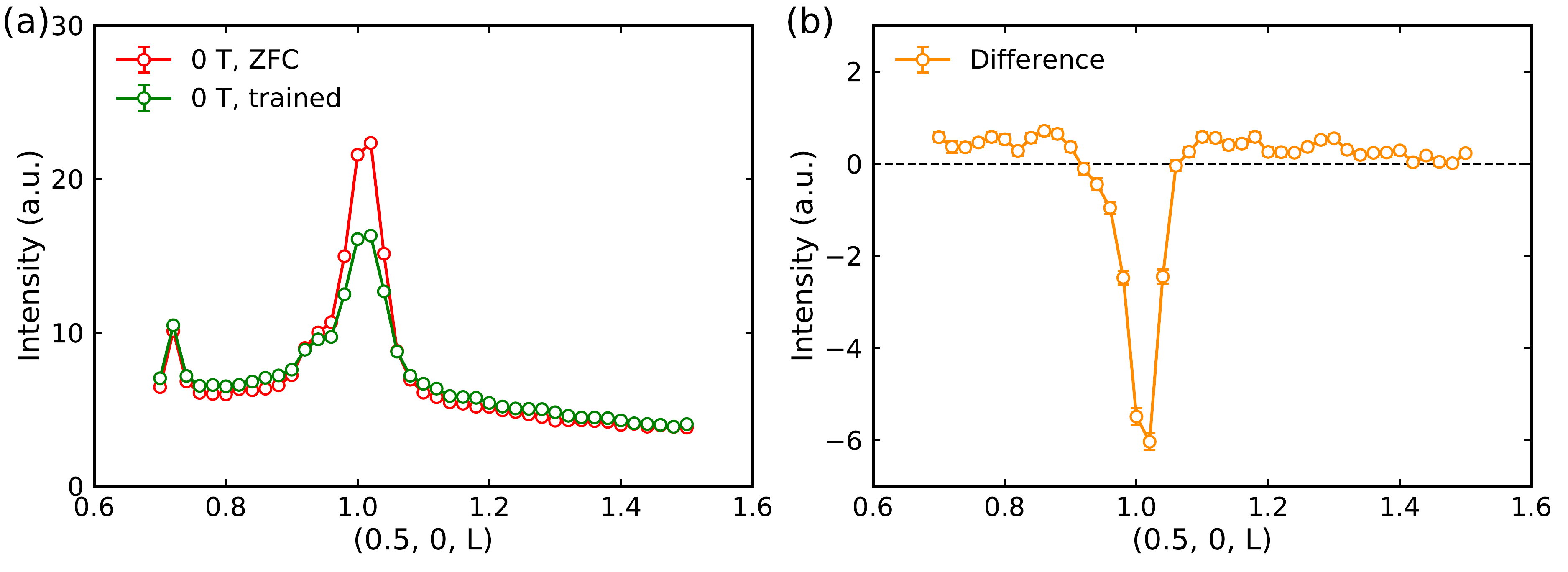}}
		\caption{(a) $L$-scans performed on BT-7 through (0.5, 0, 1) at 0 T after ZFC and field training. (b) Intensity difference (0 T, trained - 0 T, ZFC) between the two measurements in (a).}
		\label{figS9}
	\end{figure}

	The calculation is based on $H$-scans at structural Bragg peaks as showed in Fig. \ref{figS6}. The results are listed in Table \ref{tb1}. The average of five trustable measurements gives a magnetic moment of 2.05(3) $\mu_B$/\ch{Co^{2+}}. We had referred to \cite{Shirane2002} in the above calculation procedure.
	
	\begin{table}[h]
		\caption{\label{tab:table1}
			Field induced magnetic moments (in $\mu_{\rm{B}}$/\ch{Co^{2+}}) deduced from structural Bragg peaks in two diffraction experiments.
		}
		\begin{ruledtabular}
			\begin{tabular}{ccccc}
				&(1, 0, 0)&(1, 0, 1)&(1, 0, 2)&(1, 0, 3)\\
				\midrule
				BT-7& -& 2.08(6)& -\footnote{This moment could not be determined reliably due to background problem, see Fig. \ref{figS6}(b).} & 2.01(7)\\
				MACS& 2.06(7)& 1.97(8)& 2.13(8)& -\\
			\end{tabular}	
		\end{ruledtabular}
		\label{tb1}
	\end{table}
	
	\section{Temperature dependence behavior expected from the ``Triple-q'' order after field training}
	
\begin{figure}
		\centering{\includegraphics[width=0.5\textwidth]{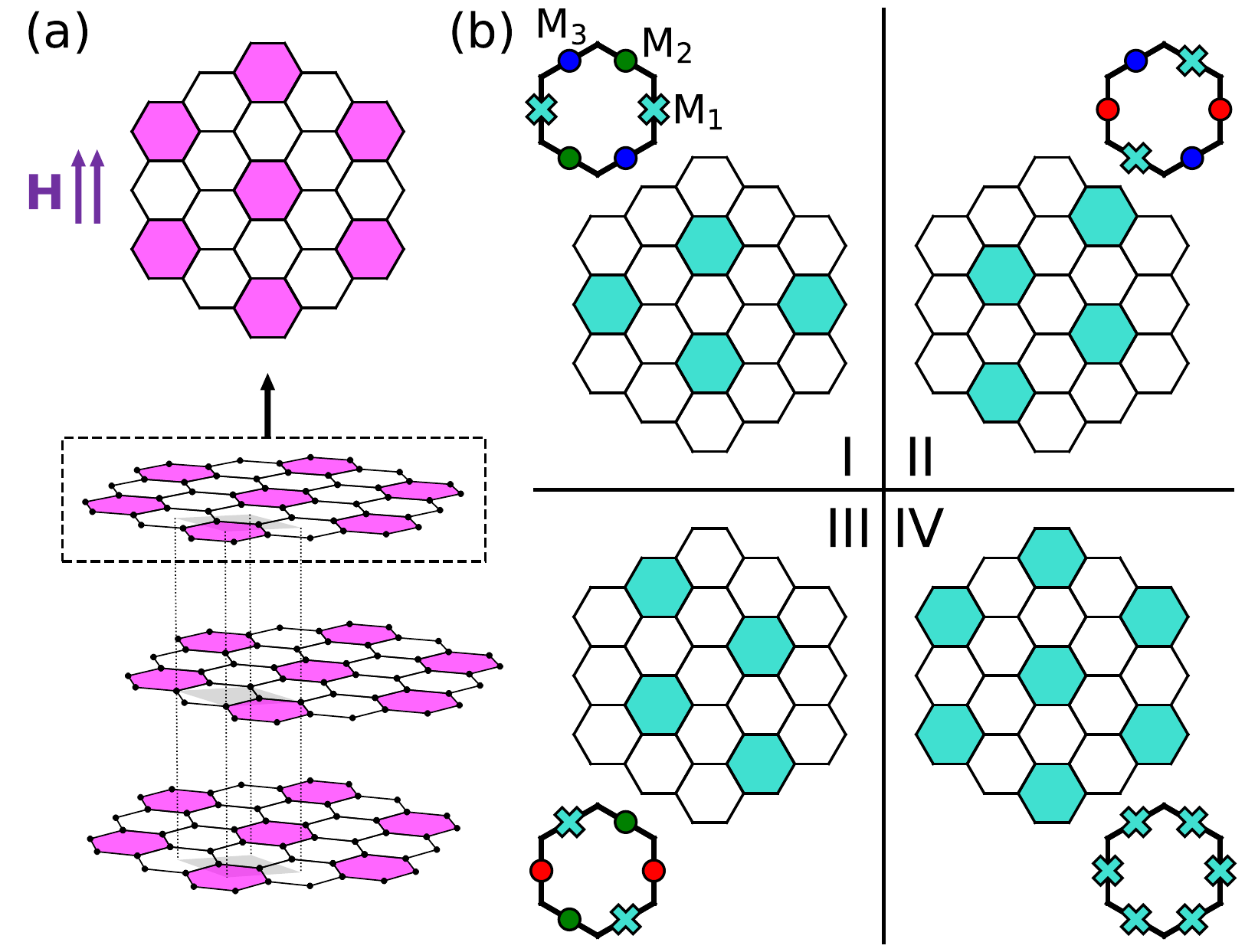}}
		\caption{(a) $c$-axis stacking of the ``triple-q'' order after ZFC. The upper part shows the spin chirality distribution in the honeycomb lattice, which follows Fig. 1(b) in the main text. (b) Four kinds of (opposite) spin chirality distributions (I - IV) in the honeycomb lattice. Magnetic Bragg peaks in the Brillouin zone are displayed at the corner of each panel. The colored dots follow the inset of Fig. 1(c) in the main text. The crosses show the Bragg peaks where broadening happens due to the loss of partial $c$-axis correlation.}
		\label{figS10}
	\end{figure}

	After ZFC, the chirality distribution is shown in Fig. \ref{figS10}(a), where we have assumed a uniform chirality for simplicity. With the application of in-plane magnetic field, the opposite spin chirality is introduced in certain layers, which causes partial loss of $c$-axis correlation. For the opposite spin chirality, we have four kinds of distributions with respect to the initial one [upper part of Fig. \ref{figS10}(a)], as showed in Fig. \ref{figS10}(b). We note that these four distributions affect the in-plane Bragg peaks differently. The pattern I in Fig. \ref{figS10}(b) will lead to peak broadening for M$_1$, while the M$_2$ and M$_3$ are intact. In the same vein, patterns II, III, and IV will respectively lead to peak broadening for M$_2$, M$_3$, and all these three kinds of peaks. From the experiment, we know that M$_2$ and M$_3$ do not get broadened after field training, therefore pattern I conforms to our case. We expect these four patterns are the same in energy. Therefore, when warming up from the field trained state, pattern I would be activated to other three patterns. As discussed above, the pattern II and III will not cause peak broadening for M$_1$, but for M$_2$ and M$_3$, respectively. So, peak of M$_1$ can be narrowed along $c$-axis, which is exactly what we have observed. As a consequence, peaks of M$_2$ and M$_3$ are expected to lose intensity faster than directly warming from a ZFC state [Fig. 3(e) in the main text]. When the temperature is higher enough, the four patterns occur with the same probability, which will lead to the converge of the temperature dependence behaviors for three kinds of peaks. This process amounts to thermal redistribution of field induced opposite spin chirality.

\end{document}